\documentclass[useAMS,amssymb,usenatbib,letterpaper]{mn2e}
\usepackage{epsfig}
\usepackage{color}
\usepackage{graphicx}
\usepackage{tabularx}
\usepackage{longtable}

\newcommand{\Ms}{\mathrm{M}_\odot}

\title[]{Local Group Dwarf Galaxies: Nature {\it And} Nurture}

\author[Sawala et al.] {Till Sawala$^{1,3}$\thanks{E-Mail:
    till.sawala@durham.ac.uk}, Cecilia Scannapieco$^{2}$, and Simon
  White$^{3}$ \\ $^{1}$Institute for Computational Cosmology,
  Department of Physics, University of Durham, South Road, Durham DH1
  3LE, UK \\ $^{2}$Astrophysikalisches Institut Potsdam, An der
  Sternwarte 16, 14482 Potsdam, Germany \\ $^{3}$Max-Planck Institute
  for Astrophysics, Karl-Schwarzschild-Strasse 1, 85748 Garching,
  Germany}

\begin{document}

\date{Accepted ***. Received ***; in original form 09 March 2011}

\pagerange{\pageref{firstpage}--\pageref{lastpage}} \pubyear{2011}

\maketitle

\label{firstpage}

\begin{abstract}
We investigate the formation and evolution of dwarf galaxies in a high
resolution, hydrodynamical cosmological simulation of a Milky Way
sized halo and its environment. Our simulation includes gas cooling,
star formation, supernova feedback, metal enrichment and UV
heating. In total, 90 satellites and more than 400 isolated dwarf
galaxies are formed in the simulation, allowing a systematic study of
the internal and environmental processes that determine their
evolution. We find that $95\%$ of satellite galaxies are gas-free at
$z=0$, and identify three mechanisms for gas loss: supernova feedback,
tidal stripping, and photo-evaporation due to re-ionization. Gas-rich
satellite galaxies are only found with total masses above $\sim5\times
10^9\Ms$. In contrast, for isolated dwarf galaxies, a total mass of
$\sim10^9\Ms$ constitutes a sharp transition; less massive galaxies
are predominantly gas-free at $z=0$, more massive, isolated dwarf
galaxies are often able to retain their gas. In general, we find that
the total mass of a dwarf galaxy is the main factor which determines
its star formation, metal enrichment, and its gas content, but that
stripping may explain the observed difference in gas content between
field dwarf galaxies and satellites with total masses close to
$10^9\Ms$. We also find that a morphological transformation via tidal
stripping of infalling, luminous dwarf galaxies whose dark matter is
less concentrated than their stars, cannot explain the high total
mass-light ratios of the faint dwarf spheroidal galaxies.
\end{abstract}


\section{Introduction} \label{introduction}
The Milky Way galaxy is believed to be surrounded by hundreds of dwarf
galaxies \citep{Koposov-2008, Tollerud-2008}. Most of the known
satellites are faint, early type dwarf spheroidal galaxies, with
stellar masses ranging from less than $10^3$ up to $10^7\Ms$,
predominantly old and metal-poor stellar populations, no detectable
interstellar gas, and eponymous, spheroidal morphologies. Stellar
kinematics suggest that dwarf spheroidal galaxies are dominated by
dark matter \citep{Faber-Lin-1983}, with estimates for the total
dynamical masses in the range of $\sim 2\times10^8$~ --~$10^9\Ms$
\citep{Walker-2007}.

Dwarf galaxies of such low total mass are believed to be highly
susceptible to both external and internal effects: Supernova feedback
can establish a self-regulation of star formation, and also lead to an
ejection of gas, while the interaction with the environment, and with
neighbouring galaxies, can potentially strip their interstellar gas,
and may also change their morphology. The effect of the cosmological
UV-background, whilst external, is assumed to be independent of
environment, and therefore identical for satellites and for isolated
dwarf galaxies.

Since the correlation of morphological type with proximity to larger
galaxies was first demonstrated by \cite{Einasto-1974}, the role of
{\it Nature} versus {\it Nurture} in the formation of early-type dwarf
galaxies has been the cause of debate \citep[e.g.][]{Mateo-1998,
  Grebel-2003, Tolstoy-2009, Annibali-2010}. Evidence for
environmental effects on galaxy formation in general is well known,
and expressed in the morphology-density relation
\citep[e.g.][]{Davis-1976}. Two studies comparing early type dwarf
galaxies in the field to early type dwarf galaxies in clusters
(\cite{Michielsen-2008} for the Virgo cluster and \cite{Koleva-2009}
for the Fornax cluster), found no clear difference between the
environments. However, comparing Coma cluster dwarf galaxies to dwarf
galaxies in poor groups, \cite{Annibali-2010} reported that dwarf
galaxies in low-density environments may experience more prolonged
star formation; evidence that the morphology-density relation also
extends to dwarf galaxies. It should also be emphasized that even if
the environmental correlation is weak, external effects could still
play an important role if, for dwarfs, the responsible mechanisms are
also efficient in low-density environments.

In a sample of SDSS galaxies with stellar masses in the range of
$2\times 10^7-3\times10^8\Ms$, \cite{Geha-2006} report that the gas
fraction of dwarf galaxies increases with distance to their nearest
luminous neighbour. Within the Local Group, \cite{Einasto-1974} show
that the HI mass in dwarf galaxies follow a similar trend with
distance to M31 or the Milky Way. \cite{Grebel-2003} also report a
morphology-distance relation: late-type, dwarf irregular galaxies are
found predominantly at the outskirts of the Local Group; early type
dwarf galaxies are mostly satellites; so-called transition type
galaxies, which share some of the characteristics of early type dwarf
galaxies whilst also containing some gas, are found at intermediate
distances. \cite{Grebel-1997} also find a correlation between
galactocentric distance and duration of star formation in Milky Way
satellites.

On the other hand, there is also evidence for supernova-driven
outflows in starburst dwarf galaxies \citep[e.g.][]{Meurer-1992,
  Schwartz-2004}. \cite{Sanchez-2010} conclude that low-mass galaxies
tend to be less strongly affected by tidal forces, and that the
morphological trend of increasing thickness found in fainter galaxies,
is related to the increasing importance of feedback mechanisms, rather
than environmental effects.

Future studies of isolated dwarf galaxies in the Local Group may
answer the question whether internal processes suffice to expel gas
from dwarf galaxies, or whether interactions are a necessary
factor. Two isolated dwarf spheroidal galaxies, Cetus
\citep{Whiting-1999} and Tucana \citep{Lavery-1992}, have been
discovered in the Local Group, with distances of $780 \pm 40$~kpc and
$890\pm50$~kpc, respectively \citep{Bernard-2009}. Deep observations
and modeling of the stellar populations by \cite{Monelli-2010}
suggests that these presently isolated objects are similar to the
oldest Milky Way dSph satellites; this runs counter to the morphology
density relation and suggests a formation mechanism independent of
environment. However, it is not known with certainty whether present
isolation also implies isolated evolution.

Numerical simulations of isolated dwarf
galaxies~\citep[e.g.][]{Stinson-2007, Valcke-2008, Revaz-2009,
  Sawala-2010} have shown that dark matter-dominated, early-type dwarf
galaxies comparable to the Local Group dwarf spheroidals {\it can}
form in isolation through the processes of supernova feedback and UV
radiation, which regulate and eventually quench star formation, and
eject most of the interstellar gas. In particular, this work showed
that the dependence of the efficiency of the two effects on the depth
of the gravitational potential can explain the observed scaling
relations between stellar mass and mass-to-light ratio
\citep[e.g.][]{Mateo-1998, Gilmore-2007, Strigari-2008}, and between
stellar mass and metallicity \citep{Grebel-2003, Tolstoy-2009}.

Simulations of interacting dwarfs \citep{Mayer-2001, Mayer-2006,
  Klimentowski-2009} have indicated that evolved, disk-like galaxies
can also be transformed into fainter, early-type dwarf galaxies in the
environment of the Milky-Way. Two mechanisms that can lead to such a
transformation are ram pressure and tidal interactions.

Ram pressure refers to the pressure exerted on the interstellar gas of
a satellite galaxy as it passes through the hot intergalactic medium
in a dense environment {\citep{Einasto-1974}}. Ram-pressure
stripping affects the interstellar gas, but not the stars or
dark matter. While this mechanism is observable in galaxy clusters,
estimates of the gas density of the Milky-Way halo
\citep[e.g.][]{Murali-2000} show it to be several orders of magnitude
too low for ram pressure stripping to be efficient \citep{Mayer-2001}.

Tidal perturbations are caused by the differential gravitational
acceleration across the diameter of a satellite as it orbits the main
halo. They affect gas, stars and dark matter alike, and can alter the
internal kinematics and morphology of the satellite (``tidal
distortion''), as well as remove mass from the objects (``tidal
stripping''). Tidal stripping can still change the composition of
objects, if the spatial distribution of the components differ. Clear
observational evidence for tidal stripping within the Milky-Way halo
is provided by the presence of tidal streams \citep[e.g.][now also
  observed around other galaxies]{Johnston-1999}. These elongated
substructures consist of stars that were tidally unbound from
satellite galaxies, and deposited along their orbits.

Most present day dwarf spheroidals, however, show no sign of tidal
distortion in their stellar kinematics \citep{Walker-2007}. Models in
which similar-mass late type progenitors are transformed into
different-mass dwarf spheroidals based on orbital parameters alone,
may have also difficulty in explaining the strong-mass metallicity
scaling relation \citep{Grebel-2003}. Most dwarf spheroidals also show
metallicity gradients, which are reproduced in simulations where star
formation is regulated by internal processes \citep{Revaz-2009}, but
are not typically found in dwarf irregulars, and may not be preserved
during strong tidal stirring.

So far, simulations have largely separated internal and external
mechanisms: by investigating either the formation of isolated dwarf
galaxies \citep{Stinson-2007, Valcke-2008, Revaz-2009, Sawala-2010},
or the transformation of evolved objects in an external gravitational
potential \citep{Mayer-2001, Mayer-2006, Klimentowski-2009}. In
reality, all effects are simultaneously present in the Local Group. In
this work, we present results from the {\it Aquila} Simulation
\citep{Scannapieco-2009}, which follows the formation of a Milky Way
sized galaxy and its environment in the fully cosmological context of
a $\Lambda$CDM universe. The satellites that grow and evolve are thus
subject to tidal forces, but also all the astrophysical processes
associated with cooling, star formation, supernova feedback and UV
heating. This simulation allows us to study all these effects in a
consistent manner, and to compare their relative importance for the
evolution of each object individually, as well as for the ensemble of
the Milky Way satellites.

This paper is organized as follows: Section~\ref{aquila_methods}
contains a description of the initial conditions for the simulation,
the numerical methods, and the method for the identification of
substructure. Section~\ref{aquila_evolution} describes the formation
and time-evolution of the halo along with its substructures. In
Section~\ref{aquila:gas_loss} we take a closer look at the different
mechanisms for gas removal in individual
satellites. Section~\ref{aquila_relations} summarizes the statistical
properties of the present-day satellite population and the derived
scaling relations. In Section~\ref{aquila_outside}, we also compare
the population of satellites to the population of isolated dwarf
galaxies formed in the same simulation. We conclude with a summary in
Section~\ref{aquila_summary}.

\section{The Aquila Simulation} \label{aquila_methods}
The simulation was set up with the principal aim of studying the
formation of disk galaxies in a $\Lambda$CDM universe. The results
presented here are based on simulation ``AQ-C-5'' of
\cite{Scannapieco-2009}, which resulted in the formation of a
Milky-Way sized disk galaxy. In this paper, the focus is on the
formation and evolution of the $\sim$90 satellites of the central
galaxy, and the isolated dwarf galaxies that form in its environment.

\subsection{Initial Conditions}
The initial conditions used here are based on one of several haloes
(labeled halo ``C''), which were extracted from the Millennium~II
Simulation, and resimulated with pure dark matter in the {\it
  Aquarius} project \citep{Springel-2008}. The cosmological parameters
are identical to those of the Millennium Simulations, $\Omega_m=0.25,
\Omega_\Lambda = 0.75, h=0.73$ and $\sigma_8=0.9$, consistent with
WMAP-1 cosmology. The simulation is performed with periodic boundary
conditions in a box of side length 137 Mpc ($100 h^{-1}$~Mpc in
comoving coordinates). The central Lagrangian region is filled with an
equal number of high resolution dark matter and gas particles, at a
mass ratio of $\Omega_{DM} = 0.21$ to $\Omega_b=0.04$. The particle
masses in the level 5 {\it Aquila} simulation are $2.6\times10^6\Ms$
for dark matter and 2--4$\times10^5\Ms$ for gas particles. Star
particles which form have a mass of 1--2$\times10^5\Ms$. The
gravitational softening parameter is fixed to 1.37 (1$ h^{-1}$ kpc) in
comoving coordinates, and equal for all particle types. Throughout
this paper, masses and distances will be stated in physical units of
$\Ms$ and kpc. Note that despite the limited accuracy of gravitational
forces at the softening scale, tidal accelerations due to an external
potential are almost unaffected by the softening.

\subsection{Computational Methods}
The simulations are performed with the smoothed particle hydrodynamics
code \textsc{GADGET-3} \citep{Springel-2005, Springel-2008}, together
with the star formation, feedback and multiphase ISM model of
\cite{Scannapieco-2005, Scannapieco-2006}. The same model is also used
in the higher resolution simulations of isolated dwarf galaxies
presented in \cite{Sawala-2010, Sawala-2011}. Specific to this
simulation, a star formation parameter of $c_\star=0.1$ is used,
whereas most simulations presented in \cite{Sawala-2010}, and all
simulations of \cite{Sawala-2011} use a value of
$c_\star=0.05$. \cite{Sawala-2010} also contains a discussion of the
effect of varying $c_\star$ for dwarf galaxies. The energy per
supernova is set to $0.7 \times 10^{51}$~ergs, identical to the value
of \cite{Sawala-2011}, but slightly lower than the value of
$1.0\times10^{51}$~ergs used in \cite{Sawala-2010}. The energy and
metals are distributed equally between the hot and cold gas
phases. The cooling model is based on \cite{Sutherland-1993}, with no
cooling below $10^4K$, and the UV heating mechanism is implemented
following \cite{Haardt-1996}, assuming a cosmic UV background active
from $z=6$.

The mass resolution of the {\it Aquila} simulation is several hundred
times lower than can be achieved in simulations of individual dwarf
galaxies. This reflects primarily the much larger Lagrangian volume,
but the presence of a large galaxy with its high cold gas fraction and
specific star formation rate also increases the computational cost per
particle. As a result, the properties of individual dwarf galaxies are
not resolved with the same level of detail as in simulations of
isolated objects; dwarf galaxies with stellar masses of $\sim10^6 \Ms$
are clearly stretching the resolution limit. Still, the large number
of satellites that form allow a number of statistical comparisons. In
addition, the fact that, apart from the differences stated above, the
same astrophysical and numerical models are used in the high
resolution simulations of \cite{Sawala-2010, Sawala-2011}, allows a
direct comparison and an estimation of the effects of resolution. We
include the results of our high-resolution simulations in
Figure~\ref{fig:relation_mdm-ms}.

The same initial conditions have also been used in other studies
\citep{Okamoto-2009, Okamoto-2010, Wadepuhl-2010}, with different
physical astrophysical models and numerical codes. The star formation
and feedback model of \citeauthor{Wadepuhl-2010} is based on
\cite{Springel-2003}, combined with a black hole wind model and a
model for cosmic rays. They conclude that cosmic rays, or some other
mechanism in addition to thermal supernova feedback and UV heating is
necessary in order to bring the satellite luminosity function in
agreement with observations. Our results agree qualitatively with
their simulations, despite the fact that we do not include cosmic
rays, indicating that supernova feedback produces similar effects in
our ISM model to cosmic ray pressure in theirs.
\citeauthor{Okamoto-2010} investigate how varying the treatment of
feedback affects the properties of dwarfs. They show that stellar mass
and metallicity scale strongly with circular velocity (i.e. subhalo
mass), and conclude that a threshold established by re-ionization
results in the small fraction of visible satellite galaxies populating
a much larger number of satellite subhaloes. Neither study
investigates the {\it isolated} dwarfs and compares them to the
satellites; this is a major part of our work.

\subsection{Identification of Substructure} \label{aquila:substructure}
In each snapshot, haloes are identified in a two step process using a
Friend-of-Friend (FoF) algorithm. The FoF groups are defined first by
linking only the dark matter particles. In a second step, star and gas
particles are linked to the particles already belonging to these
groups in the same way. In the Aquila simulation, the larger haloes,
and in particular, the main halo that will become the host of the
``Milky Way'' galaxy, also contain a number of gravitationally bound,
over-dense substructures called {\it subhaloes}, which are identified
using the {\sc subfind} algorithm of \cite{Springel-2001}. For each
FoF halo, {\sc subfind} begins by computing a smoothed local density
by an SPH-interpolation over all particles of the halo. A potential
substructure is first identified as a local overdensity with respect
to this smooth background density. It is then subjected to
gravitational unbinding, whereby all unbound particles are iteratively
removed, until the substructure either vanishes, falling below the
threshold of 20 particles, or is identified as a genuine self-bound
subhalo.

In order to trace subhaloes over time between different snapshots, the
20 most bound dark matter particles of each subhalo in a given
snapshot were compared to the list of particles in all subhaloes of
the previous snapshot, identifying as the progenitor the subhalo that
contained at least 11 of the 20 particles amongst its 20\% of most
bound particles. The process is repeated until a progenitor is no
longer found, and we define this as the time of {\it formation}. If
the subhalo can be linked to a subhalo in a previous snapshot, but not
to its most bound particles, we consider the progenitor to have
fragmented, and define this as the time of {\it
  fragmentation}. Fragmentation can be due to the breaking-up of
larger haloes, or to the amplification of substructures above the
particle threshold, without the newly identified subhaloes hosting
galaxies. In Table~\ref{table:aquila-satellites}, fragmentation is
indicated with an asterisk next to the formation redshift. We note
that a large majority of subhaloes of satellite galaxies can be traced
back as independent objects well beyond $z = 6$.

Most of the subhaloes belonging to the main halo at $z=0$ belong to
different groups at earlier times. We call the time when a subhalo is
{\it first} identified as a subhalo of the main halo the time of {\it
  infall}, noting that some subhaloes subsequently become transitorily
isolated, and fall in again at a later time.

\begin{figure*}
  \begin{center}
    \begin{tabular}{lll}                     
      \hspace{-3mm} \includegraphics*[trim = 0mm 0mm 0mm 0mm, clip, scale= .64]{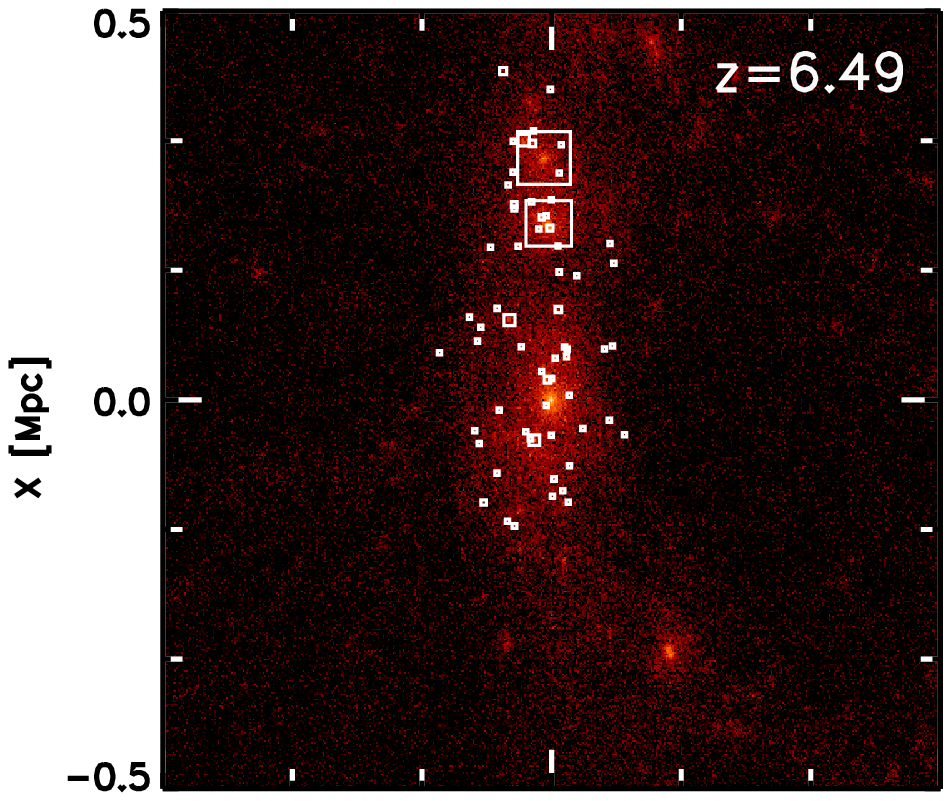} &
      \hspace{-4mm} \includegraphics*[trim = 10mm 0mm 0mm 0mm, clip, scale = .64]{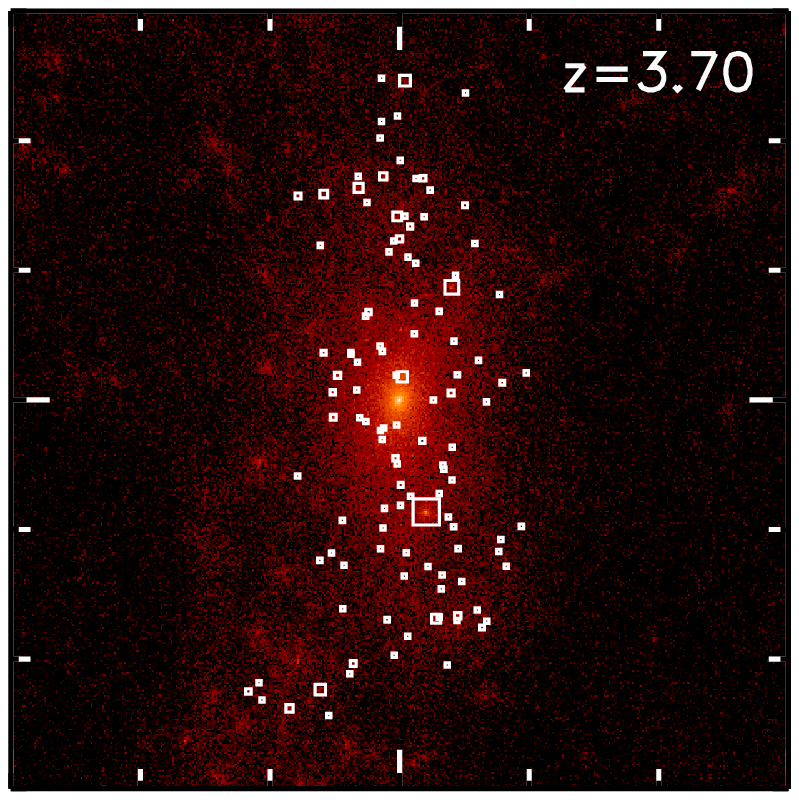} &
      \hspace{-4mm} \includegraphics*[trim = 10mm 0mm 0mm 0mm, clip, scale = .64]{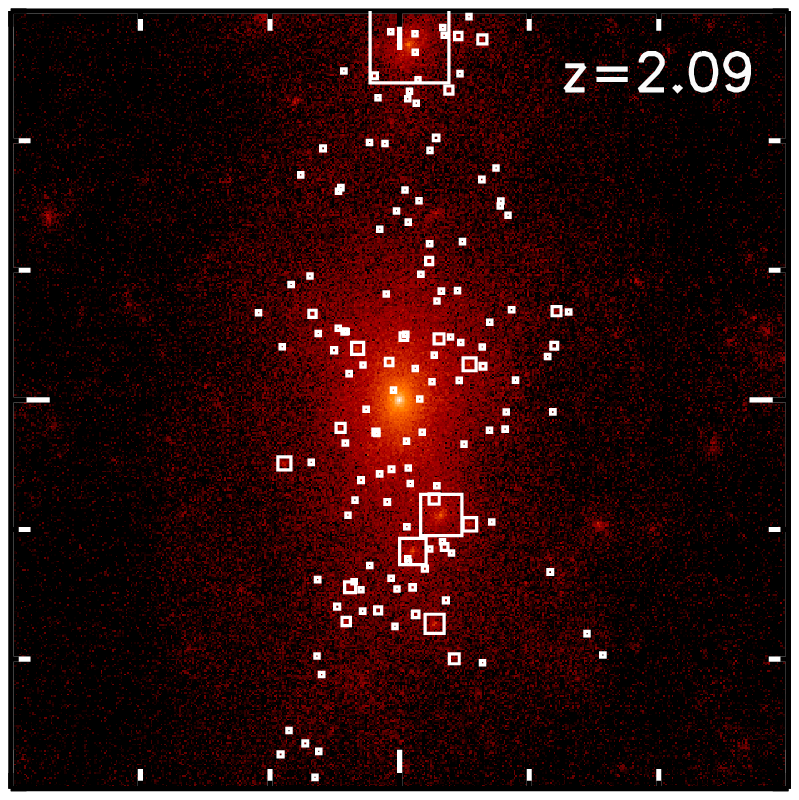} \vspace{-8mm}\\  
      \hspace{-3mm} \includegraphics*[trim = 0mm 0mm 0mm 0mm, clip, scale= .64]{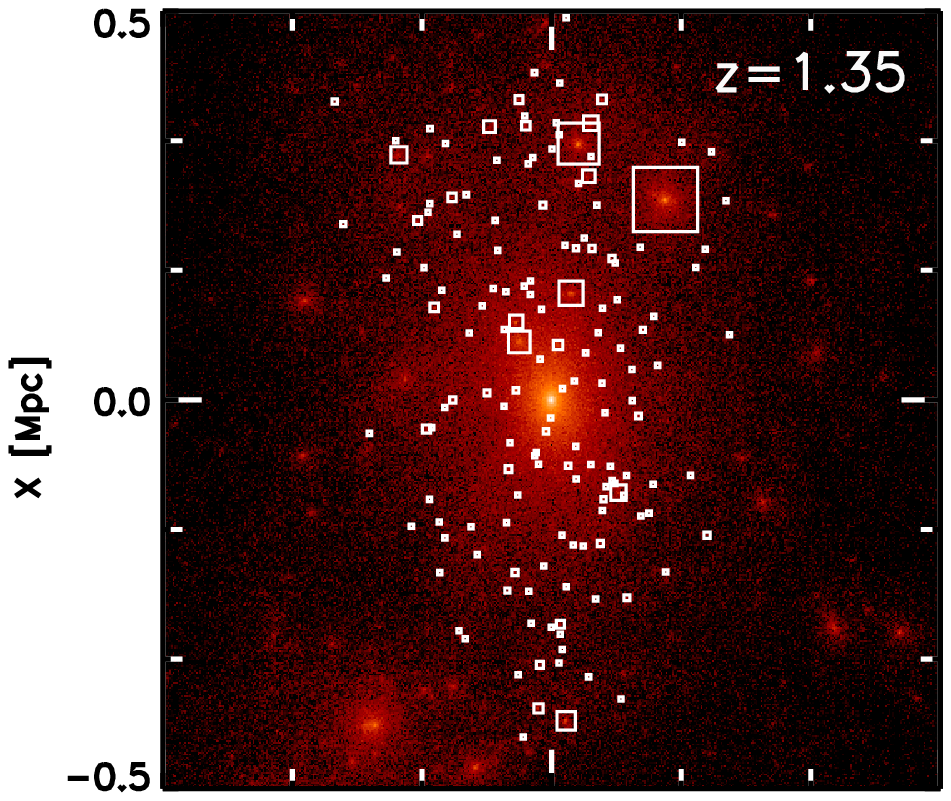} &
      \hspace{-4mm} \includegraphics*[trim = 10mm 0mm 0mm 0mm, clip, scale = .64]{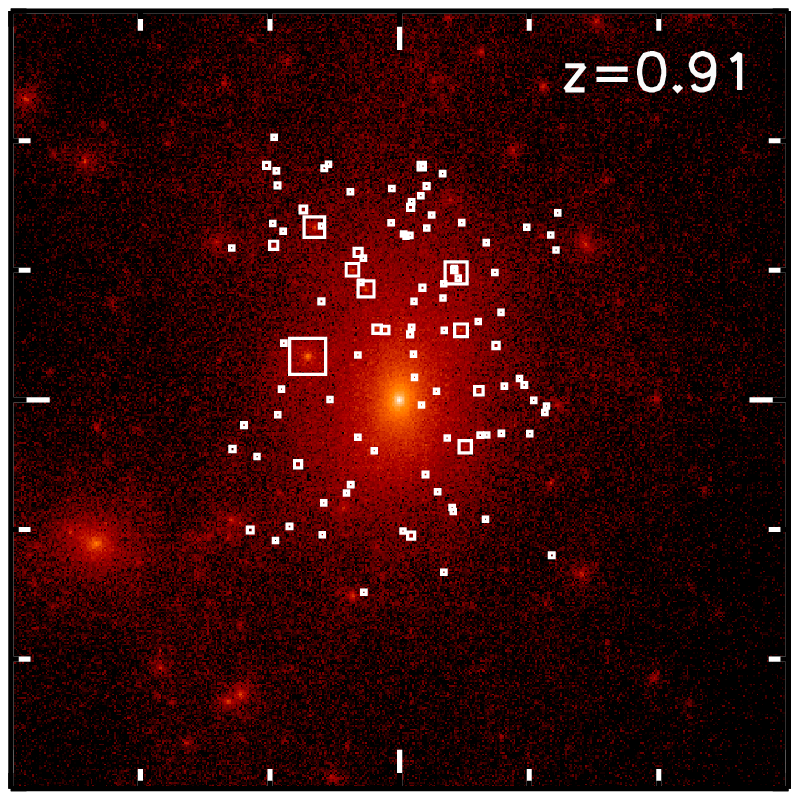} &
      \hspace{-4mm} \includegraphics*[trim = 10mm 0mm 0mm 0mm, clip, scale = .64]{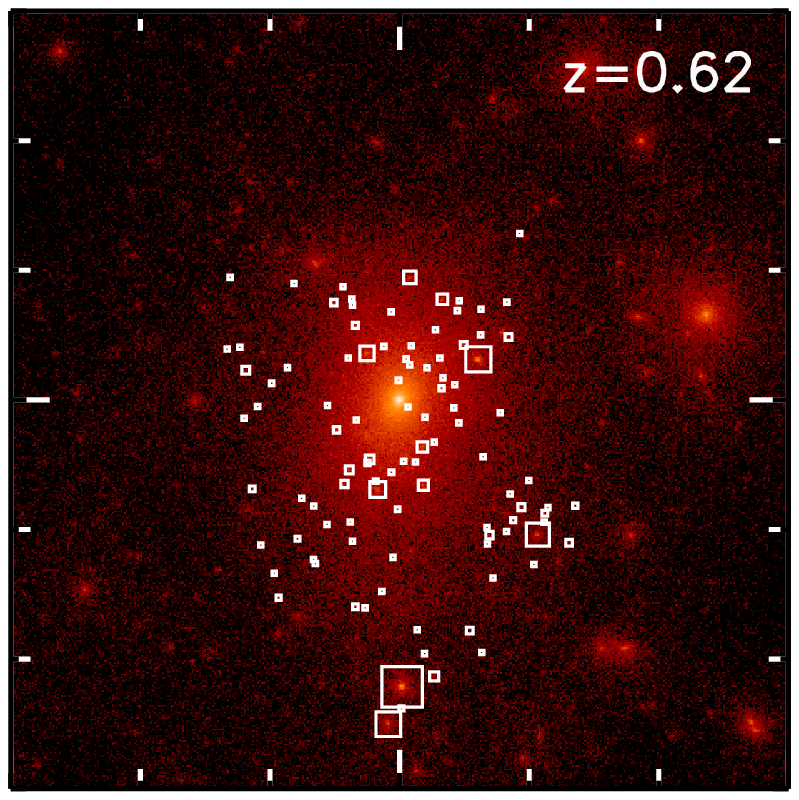} \vspace{-8mm}\\  
      \hspace{-3mm} \includegraphics*[trim = 0mm 0mm 0mm 0mm, clip, scale= .64]{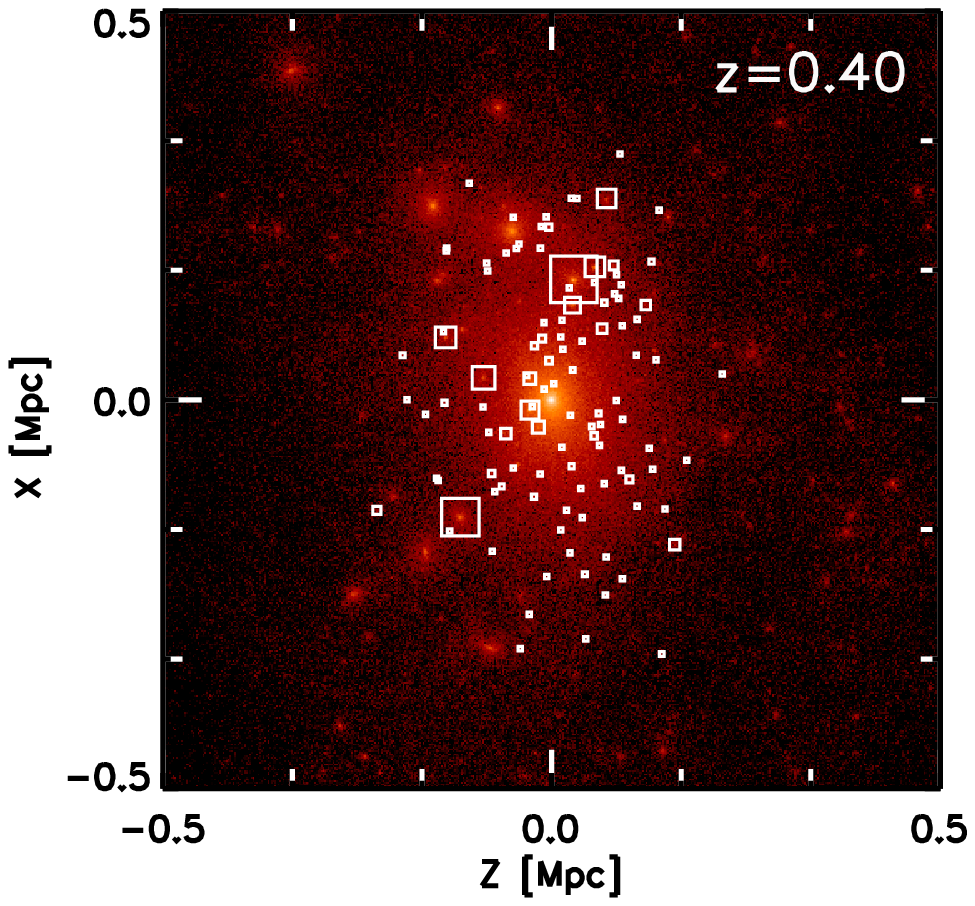} &
      \hspace{-4mm} \includegraphics*[trim = 10mm 0mm 0mm 0mm, clip, scale = .64]{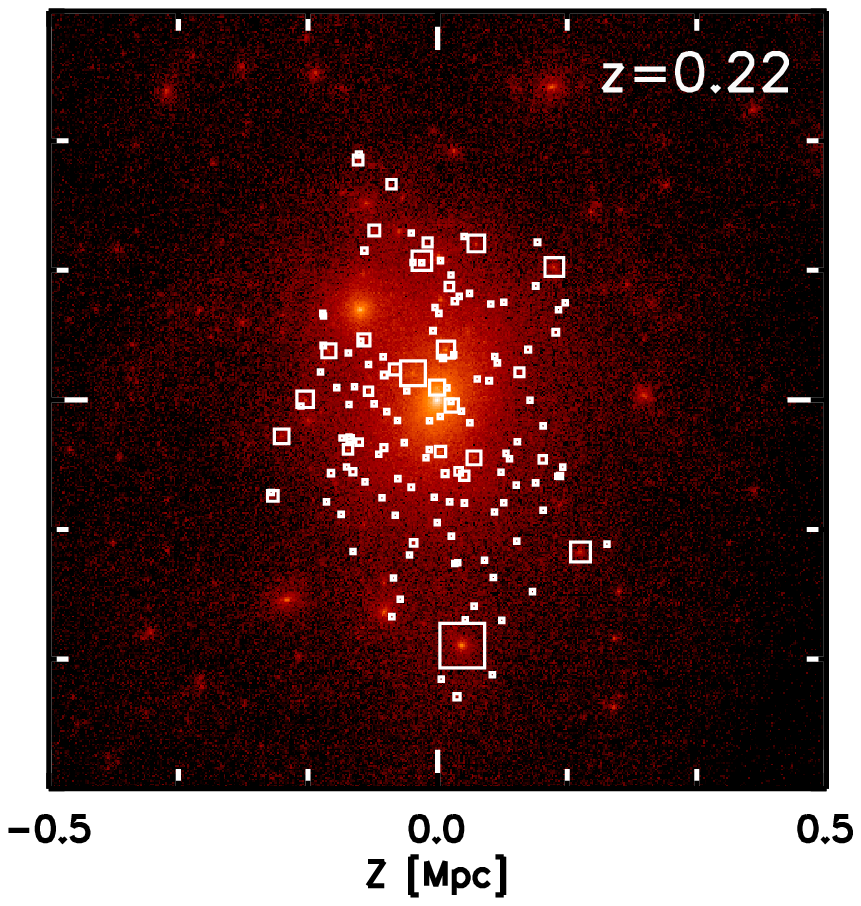} &
      \hspace{-4mm} \includegraphics*[trim = 10mm 0mm 0mm 0mm, clip, scale = .64]{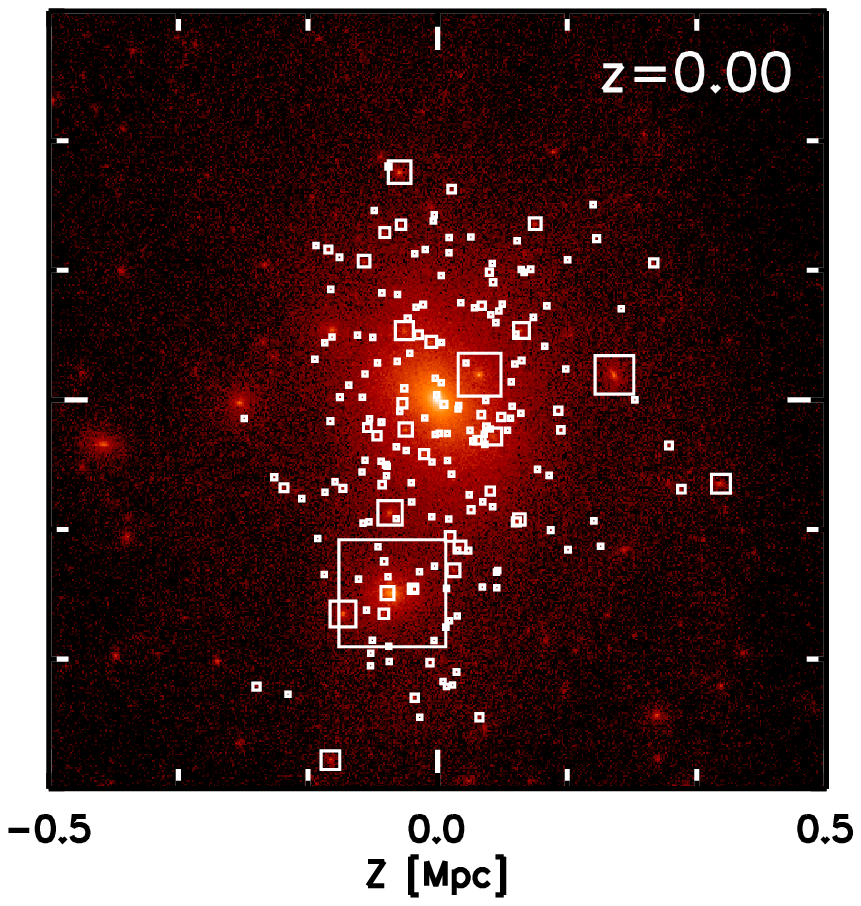} 
    \end{tabular}
  \end{center}
  \caption{Evolution of the dark matter distribution in the central
    region of the Aquila simulation. Each panel shows a box of
    sidelength 1~Mpc (comoving), centred on the central subhalo, and
    oriented along the major (X) and minor (Z) component of the
    inertia tensor of the main halo. The squares show the position of
    identified subhaloes belonging to the FoF-group of the main halo
    present at each snapshot, with the area proportional to the
    subhalo mass. Over time, the distribution of mass and of subhaloes
    changes from an elongated distribution at high redshift to a more
    rounded distribution at lower redshift (different projections of
    the subhaloes at $z=0$ are shown in
    Figure~\ref{fig:aquila_projections}). \label{fig:aquila_dm}}
\end{figure*}

\begin{figure*}
  \begin{center}
    \begin{tabular}{lll}                     
      \hspace{-3mm} \includegraphics*[trim = 0mm 0mm 0mm 0mm, clip, scale= .64]{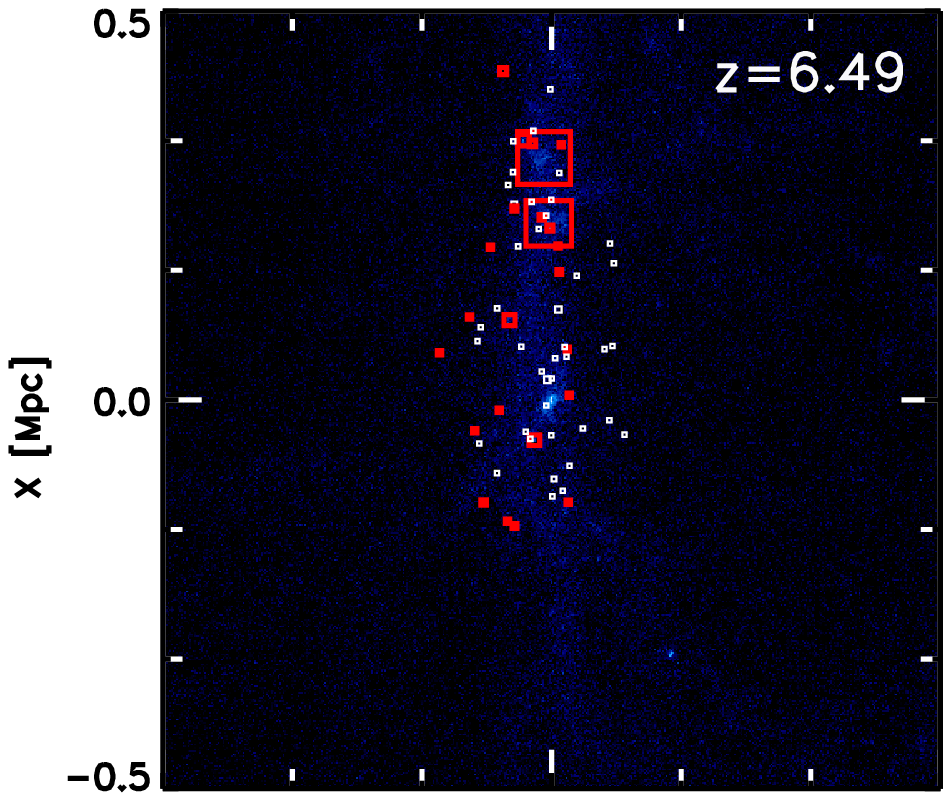} &
      \hspace{-4mm} \includegraphics*[trim = 10mm 0mm 0mm 0mm, clip, scale = .64]{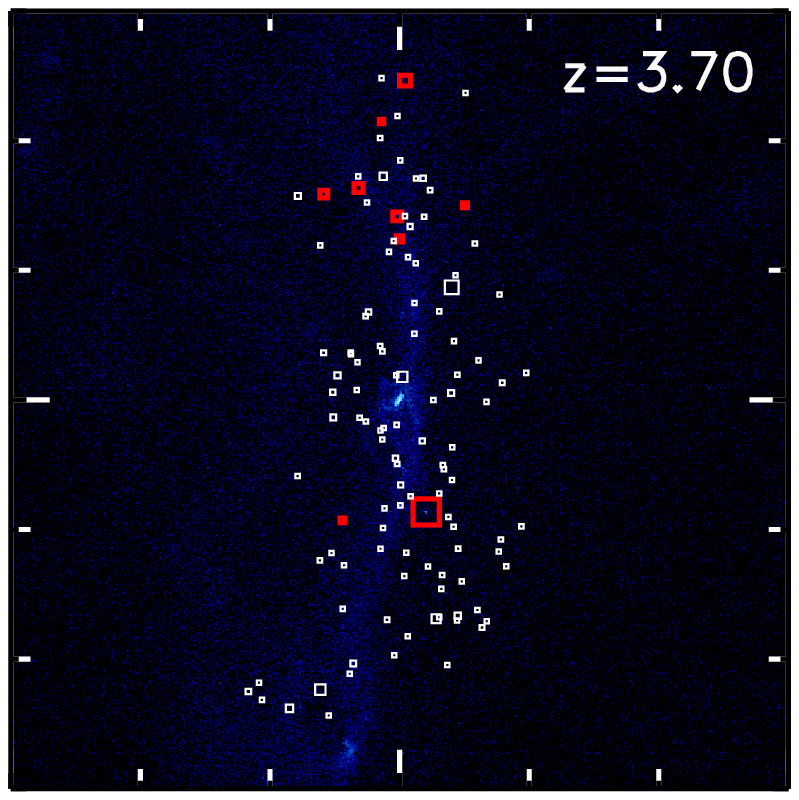} &
      \hspace{-4mm} \includegraphics*[trim = 10mm 0mm 0mm 0mm, clip, scale = .64]{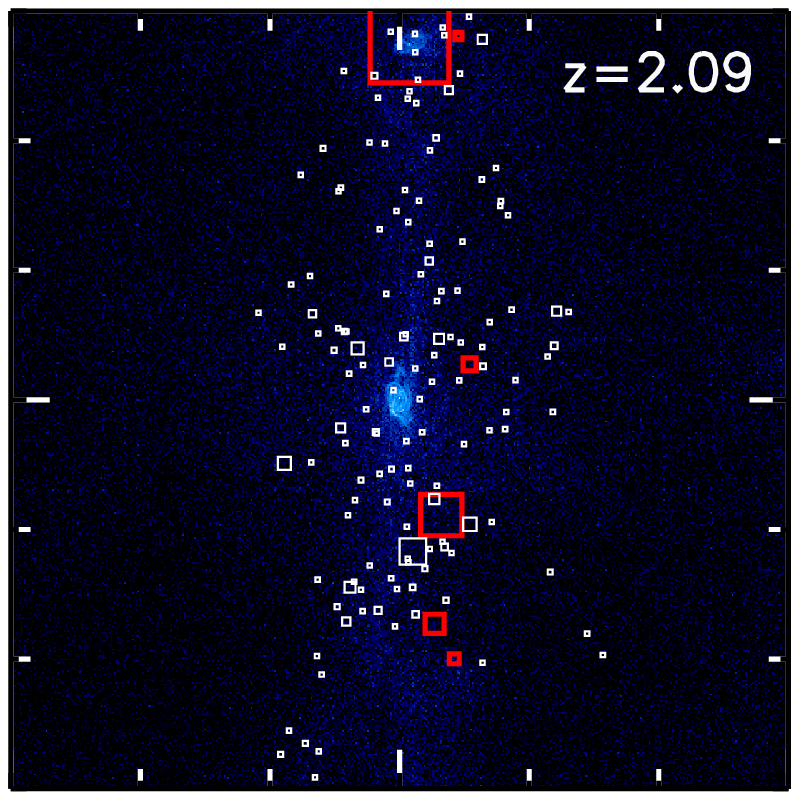} \vspace{-8mm}\\  
      \hspace{-3mm} \includegraphics*[trim = 0mm 0mm 0mm 0mm, clip, scale= .64]{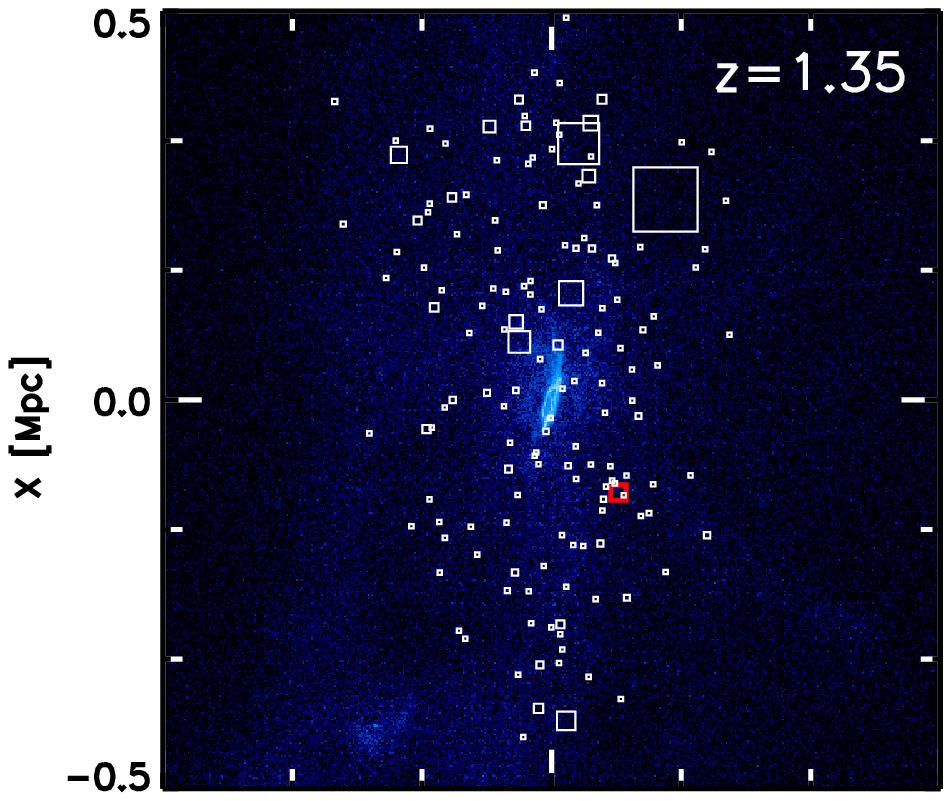} &
      \hspace{-4mm} \includegraphics*[trim = 10mm 0mm 0mm 0mm, clip, scale = .64]{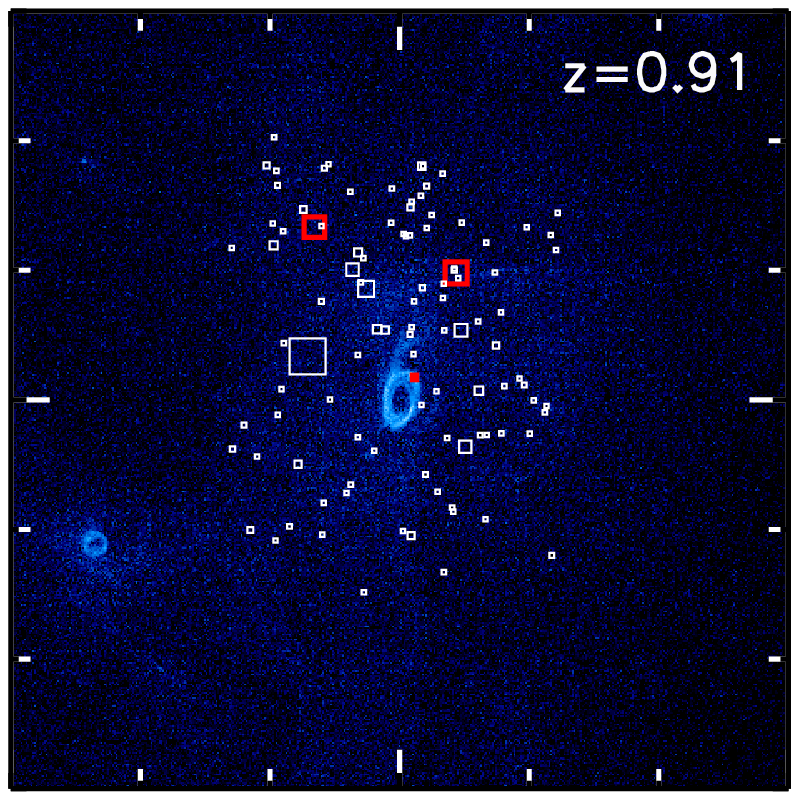} &
      \hspace{-4mm} \includegraphics*[trim = 10mm 0mm 0mm 0mm, clip, scale = .64]{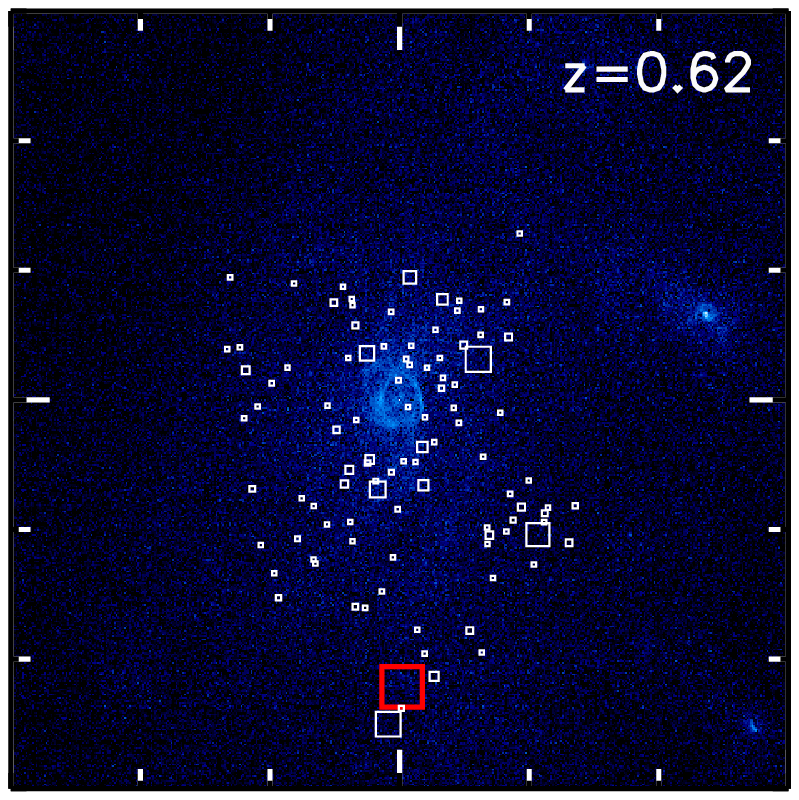} \vspace{-8mm}\\  
      \hspace{-3mm} \includegraphics*[trim = 0mm 0mm 0mm 0mm, clip, scale= .64]{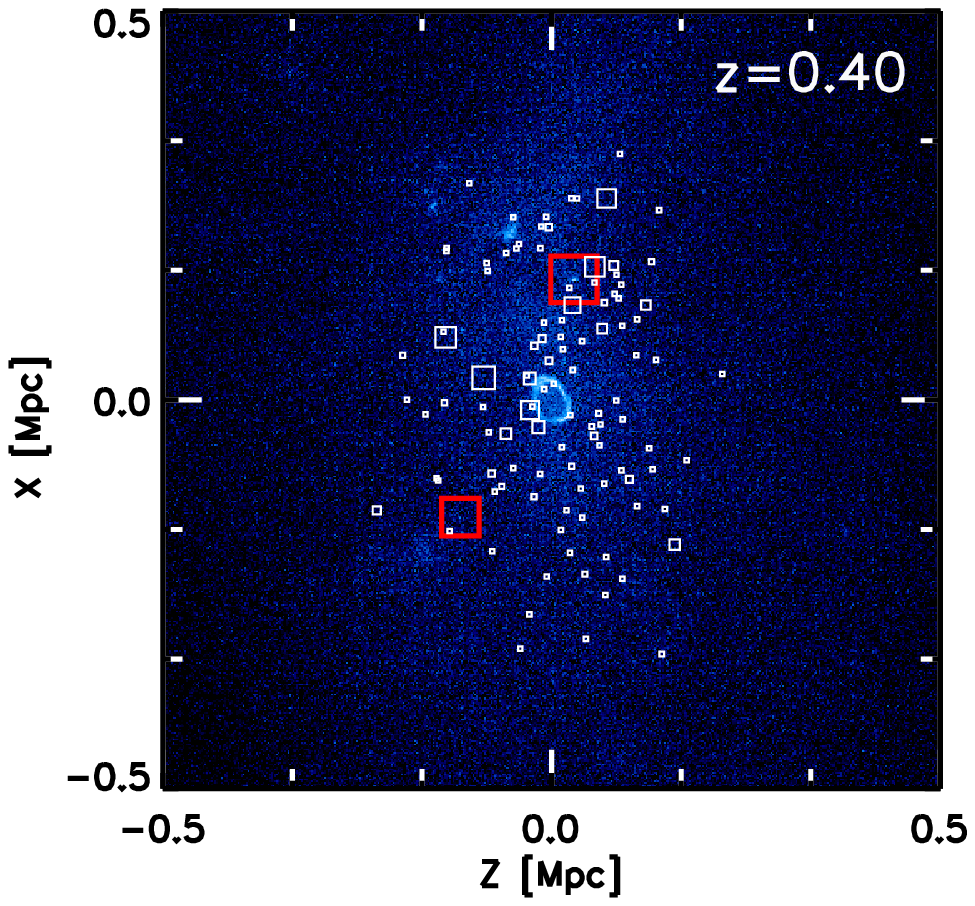} &
      \hspace{-4mm} \includegraphics*[trim = 10mm 0mm 0mm 0mm, clip, scale = .64]{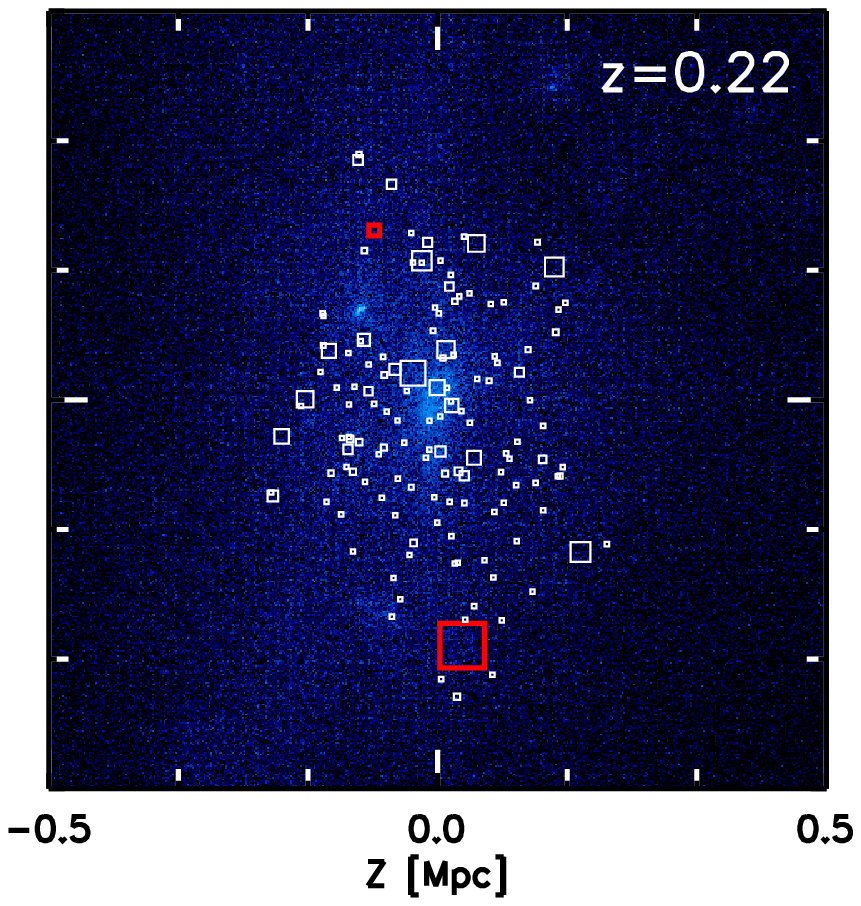} &
      \hspace{-4mm} \includegraphics*[trim = 10mm 0mm 0mm 0mm, clip, scale = .64]{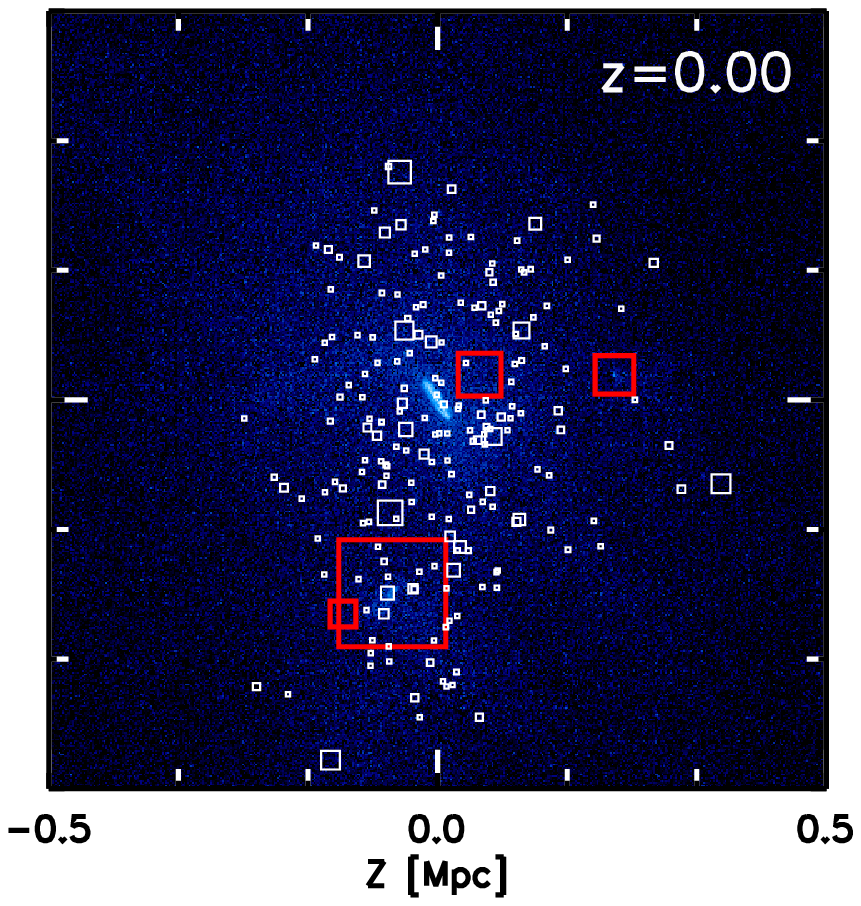} 
    \end{tabular}
  \end{center}
  \caption{Gas distribution in the central region of the Aquila
    simulation at different redshifts, in volumes identical to
    Figures~\ref{fig:aquila_dm} and Figures~\ref{fig:aquila_stars}.
    Subhaloes of the main FoF halo with gas are shown as red squares,
    Gas-free subhaloes are shown in white. While most satellites
    contain gas at $z=7$, this fraction drops significantly, and only
    four of the most massive satellites are not gas-free at
    $z=0$.\label{fig:aquila_gas}}
\end{figure*}

\begin{figure*}
  \begin{center}
    \begin{tabular}{lll}                     
      \hspace{-3mm} \includegraphics*[trim = 0mm 0mm 0mm 0mm, clip, scale= .64]{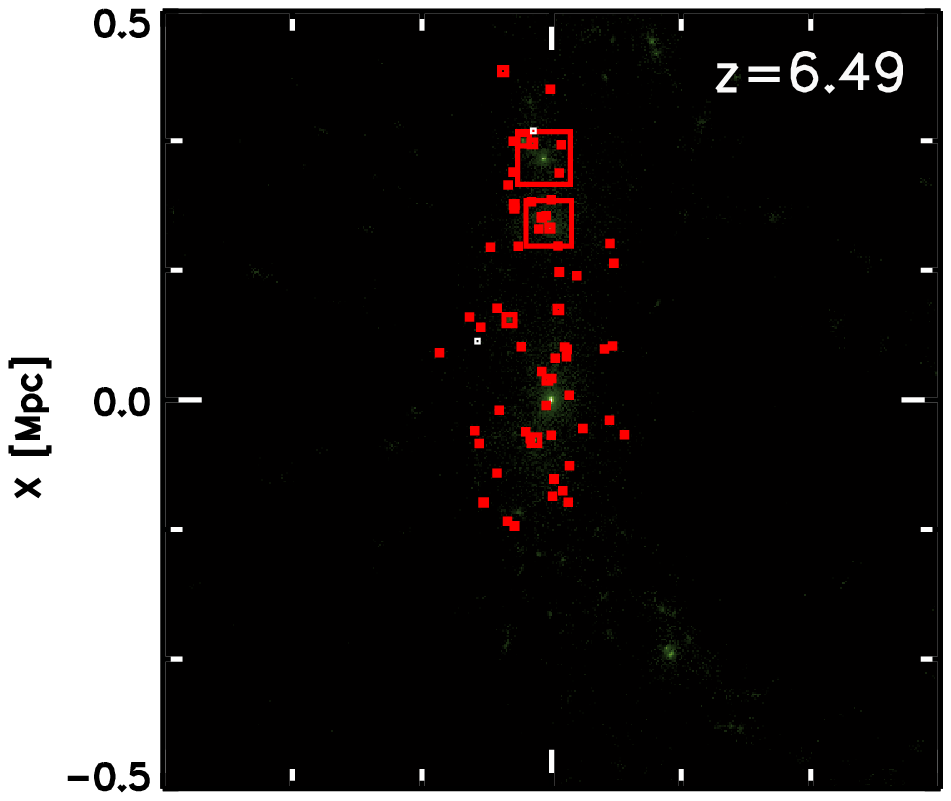} &
      \hspace{-4mm} \includegraphics*[trim = 10mm 0mm 0mm 0mm, clip, scale = .64]{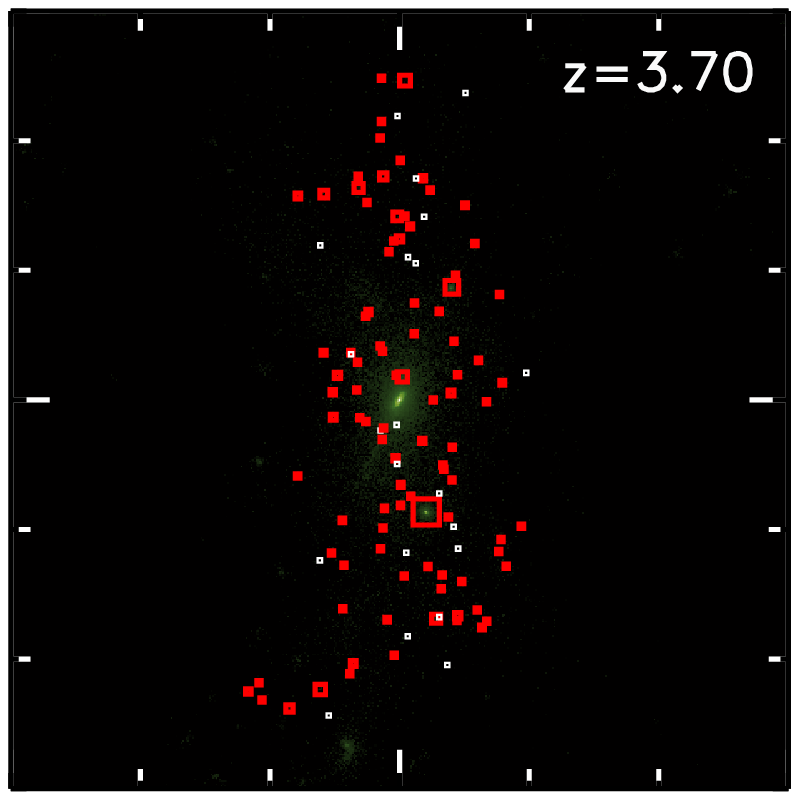} &
      \hspace{-4mm} \includegraphics*[trim = 10mm 0mm 0mm 0mm, clip, scale = .64]{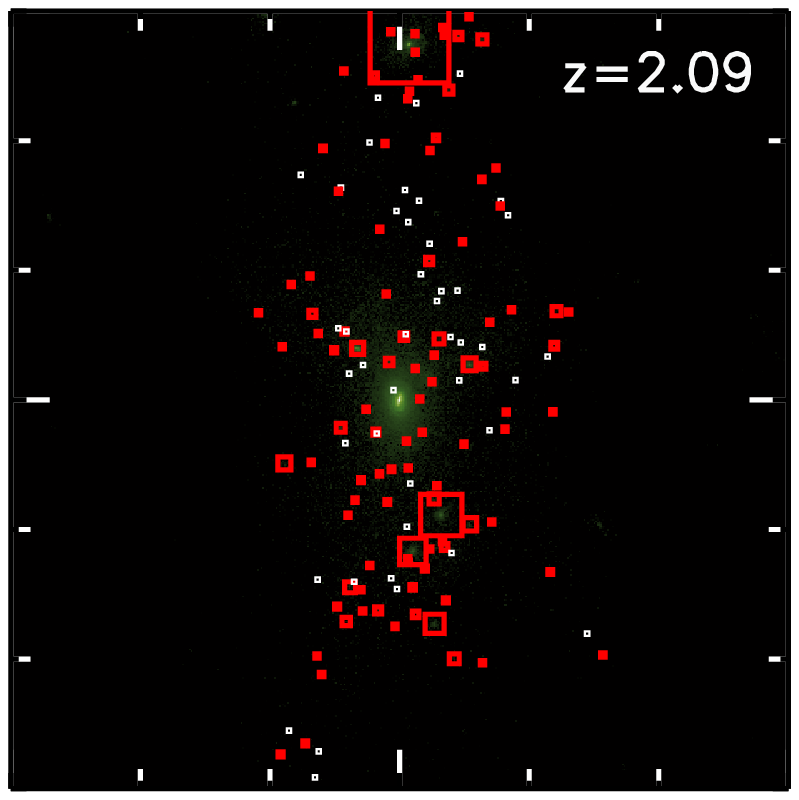} \vspace{-8mm}\\  
      \hspace{-3mm} \includegraphics*[trim = 0mm 0mm 0mm 0mm, clip, scale= .64]{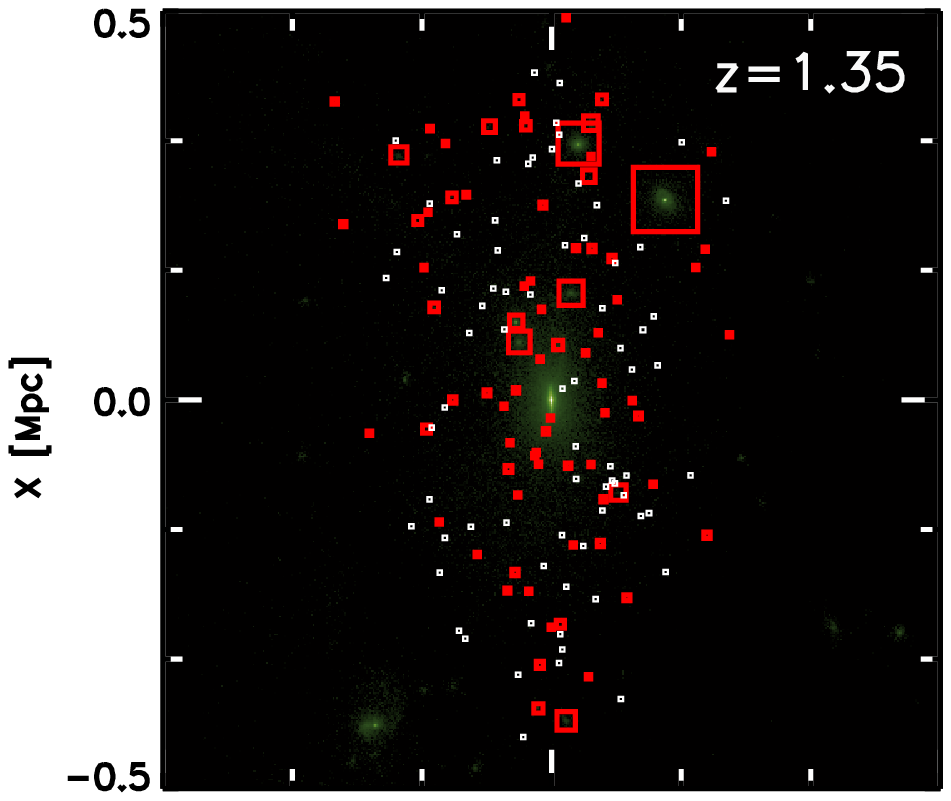} &
      \hspace{-4mm} \includegraphics*[trim = 10mm 0mm 0mm 0mm, clip, scale = .64]{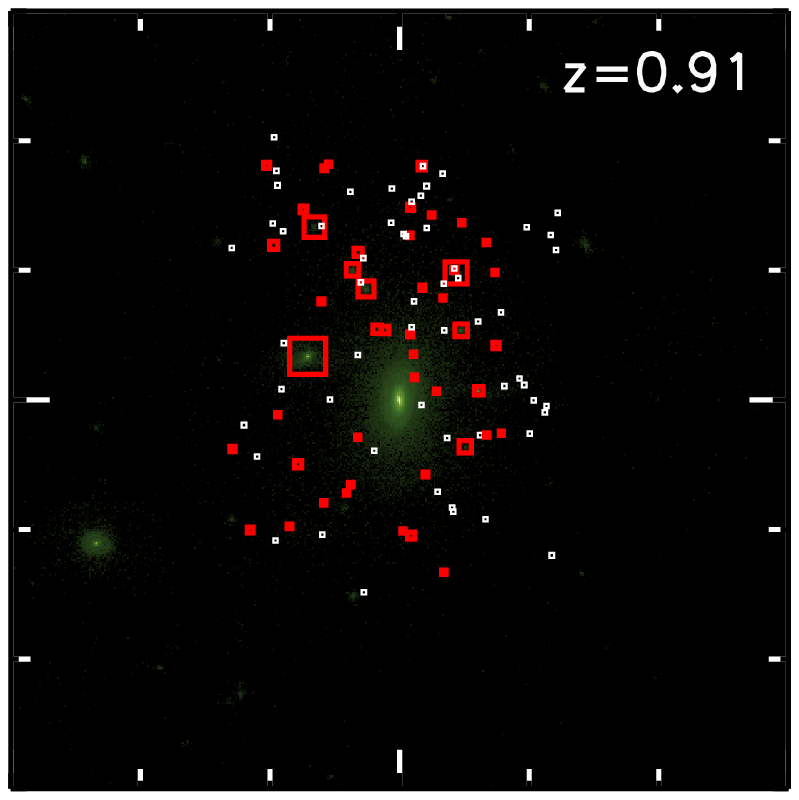} &
      \hspace{-4mm} \includegraphics*[trim = 10mm 0mm 0mm 0mm, clip, scale = .64]{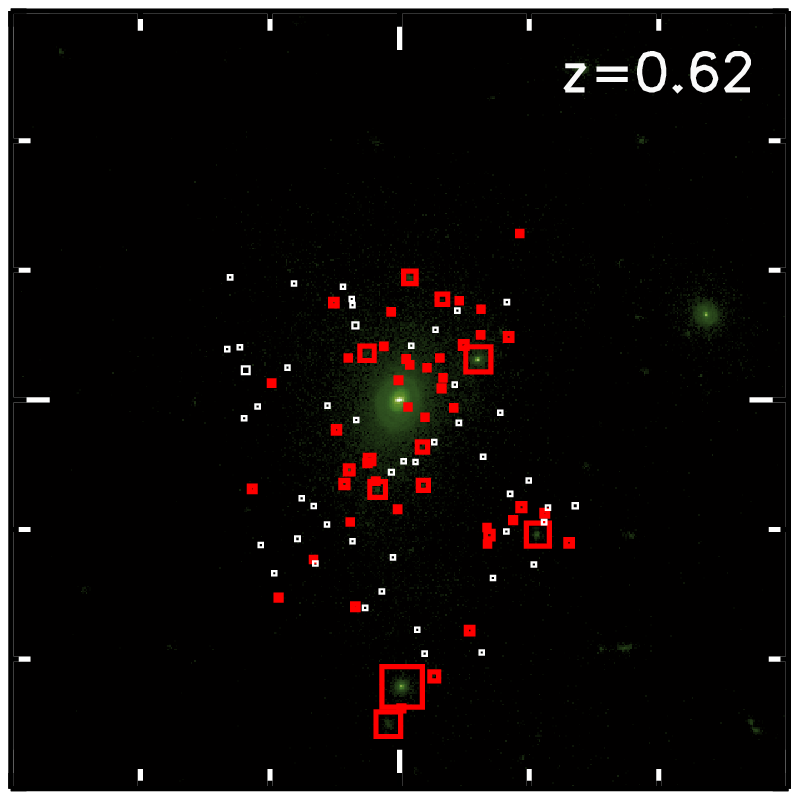} \vspace{-8mm}\\  
      \hspace{-3mm} \includegraphics*[trim = 0mm 0mm 0mm 0mm, clip, scale= .64]{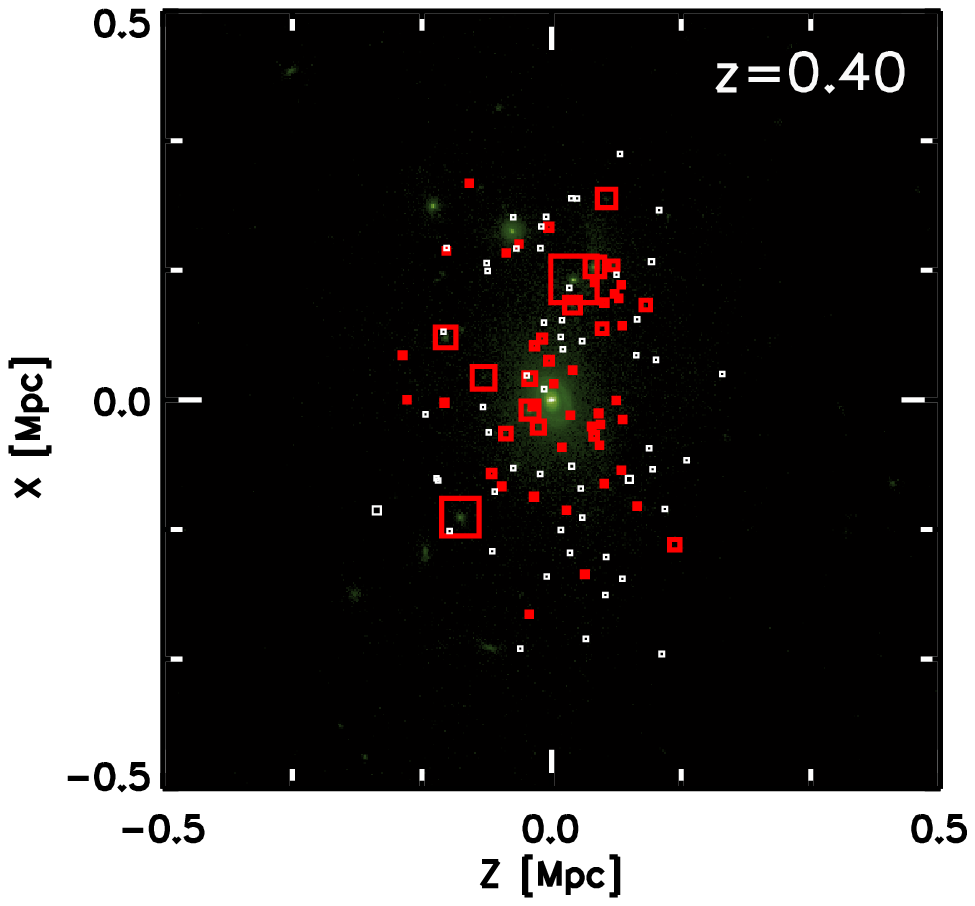} &
      \hspace{-4mm} \includegraphics*[trim = 10mm 0mm 0mm 0mm, clip, scale = .64]{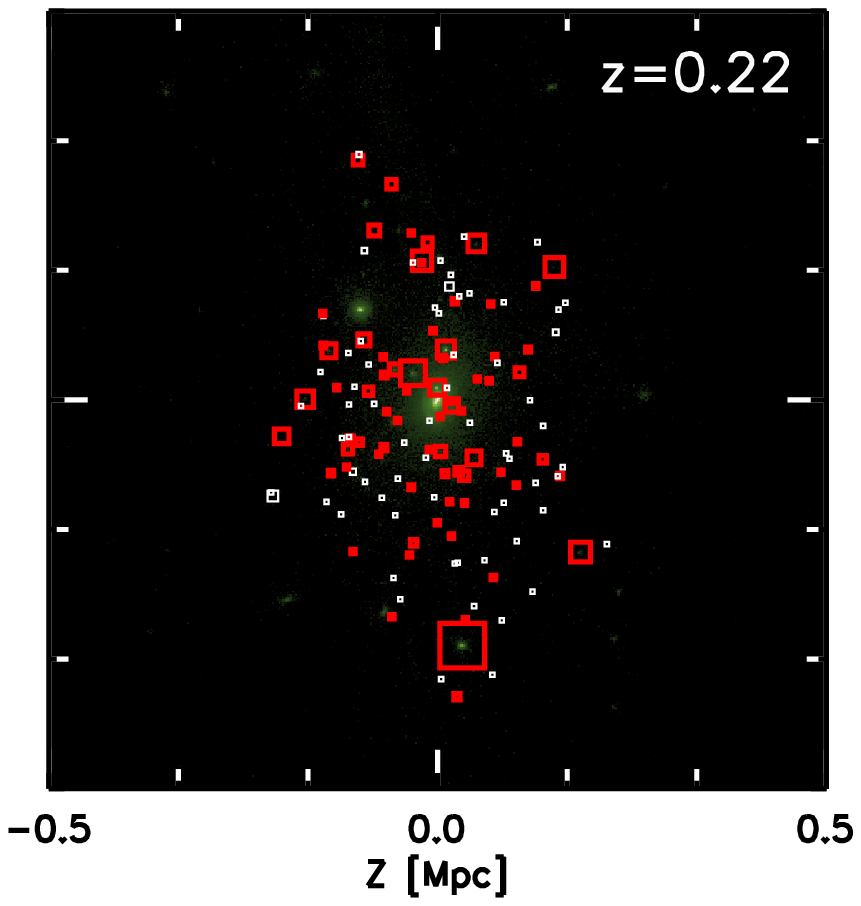} &
      \hspace{-4mm} \includegraphics*[trim = 10mm 0mm 0mm 0mm, clip, scale = .64]{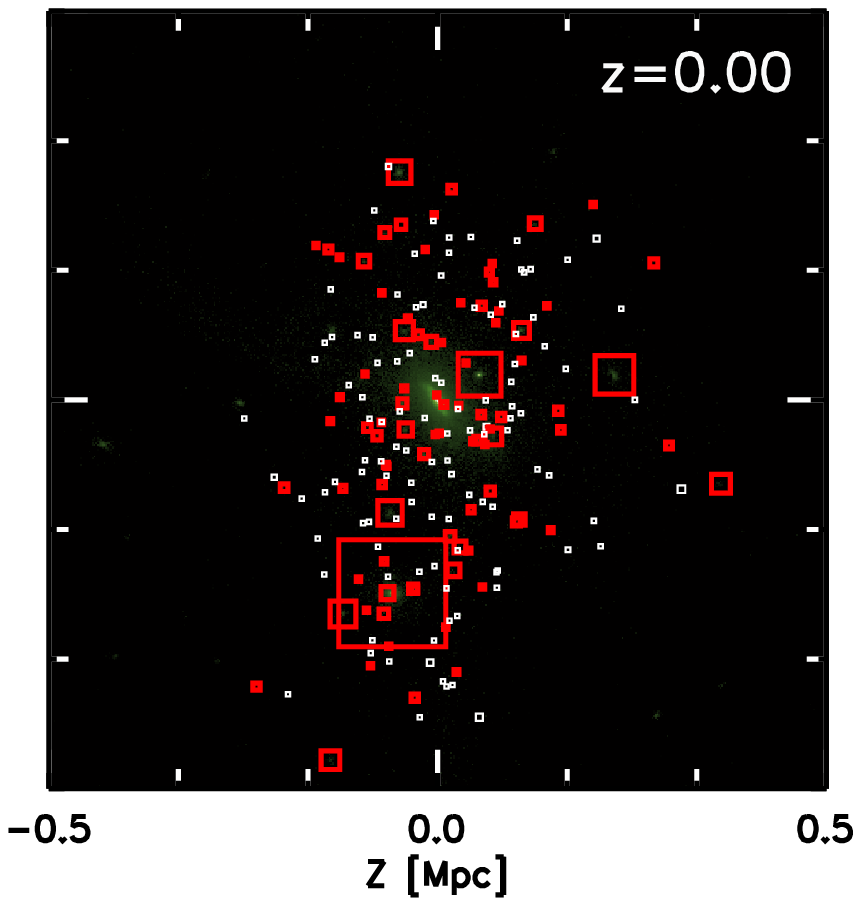} 
    \end{tabular}
  \end{center}
  \vspace{-.1in}
  \caption{Stellar mass distribution in n the central region of the
    Aquila simulation at different redshifts, in volumes identical to
    Figures~\ref{fig:aquila_dm} and Figures~\ref{fig:aquila_gas}. The
    central, Milky Way type galaxy dominates the total stellar mass at
    every redshift. The positions of satellite galaxies are shown in
    red, while dark subhaloes are shown as white squares. The size of
    each square corresponds to the dark matter mass of each
    subhalo. While nearly all subhaloes present at $z=7$ also contain
    stars, the fraction drops to $\sim45\%$ at $z=0$, with more
    massive subhaloes more likely to contain
    stars. \label{fig:aquila_stars}}
\end{figure*} 

\begin{figure*}
  \begin{center}
    \begin{tabular}{lll}
      \hspace{-.15in} \includegraphics*[width = .34 \textwidth]{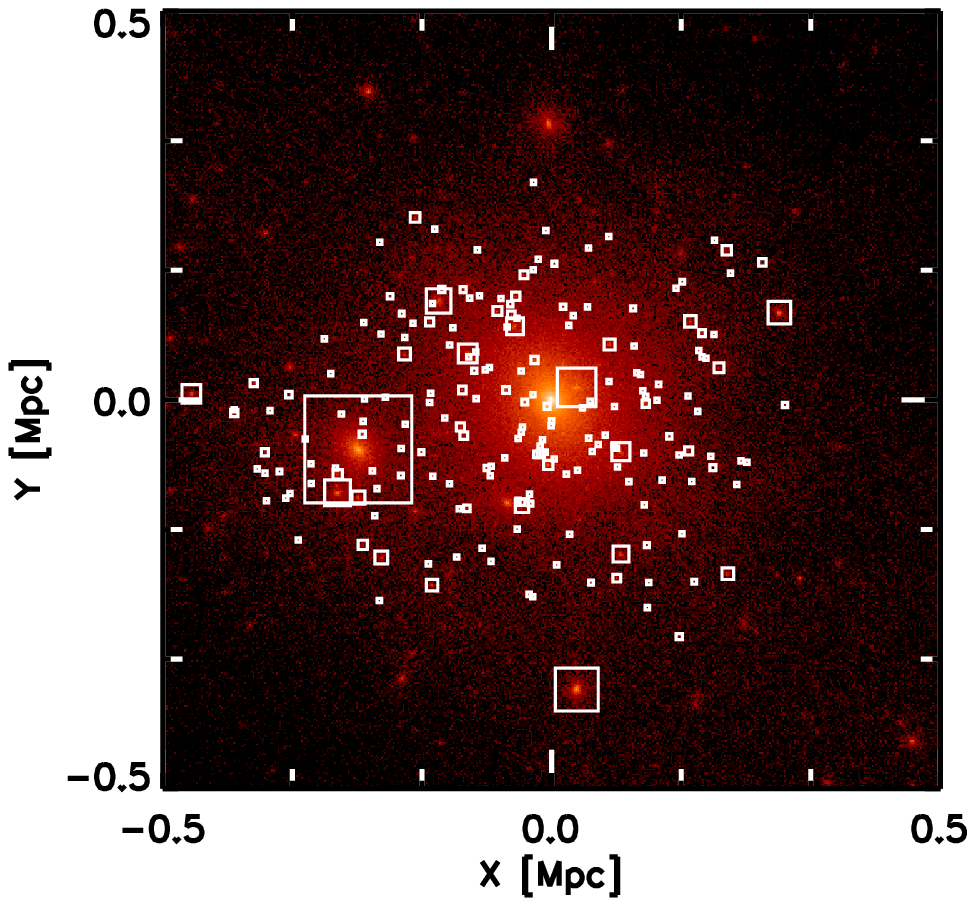} &
      \hspace{-.22in} \includegraphics*[width = .34 \textwidth]{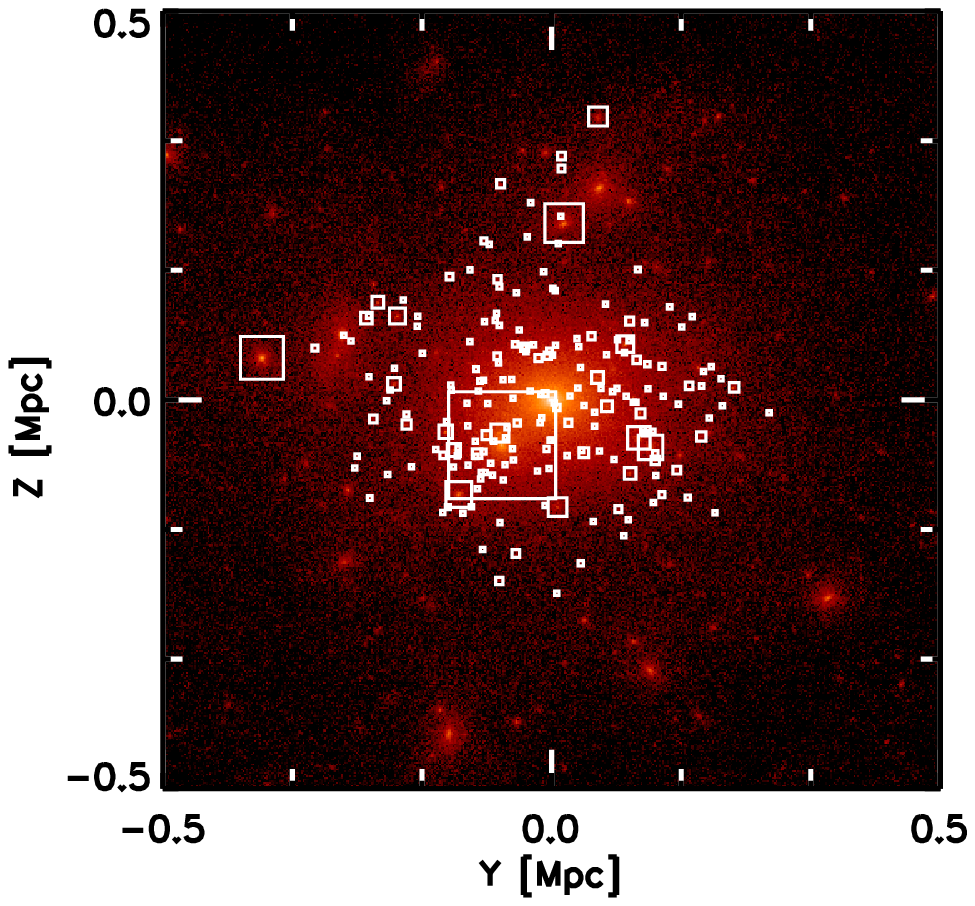} &
      \hspace{-.22in} \includegraphics*[width = .34 \textwidth]{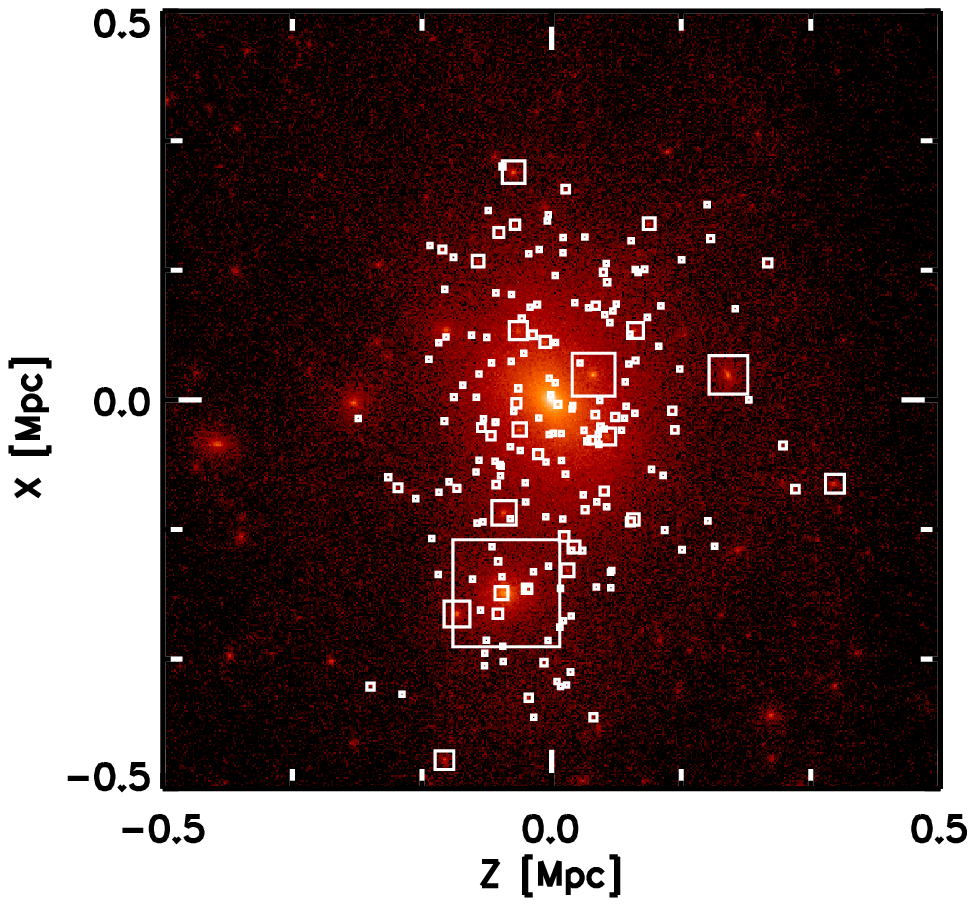} \vspace{-.1in}\\
      \hspace{-.15in} \includegraphics*[width = .34 \textwidth]{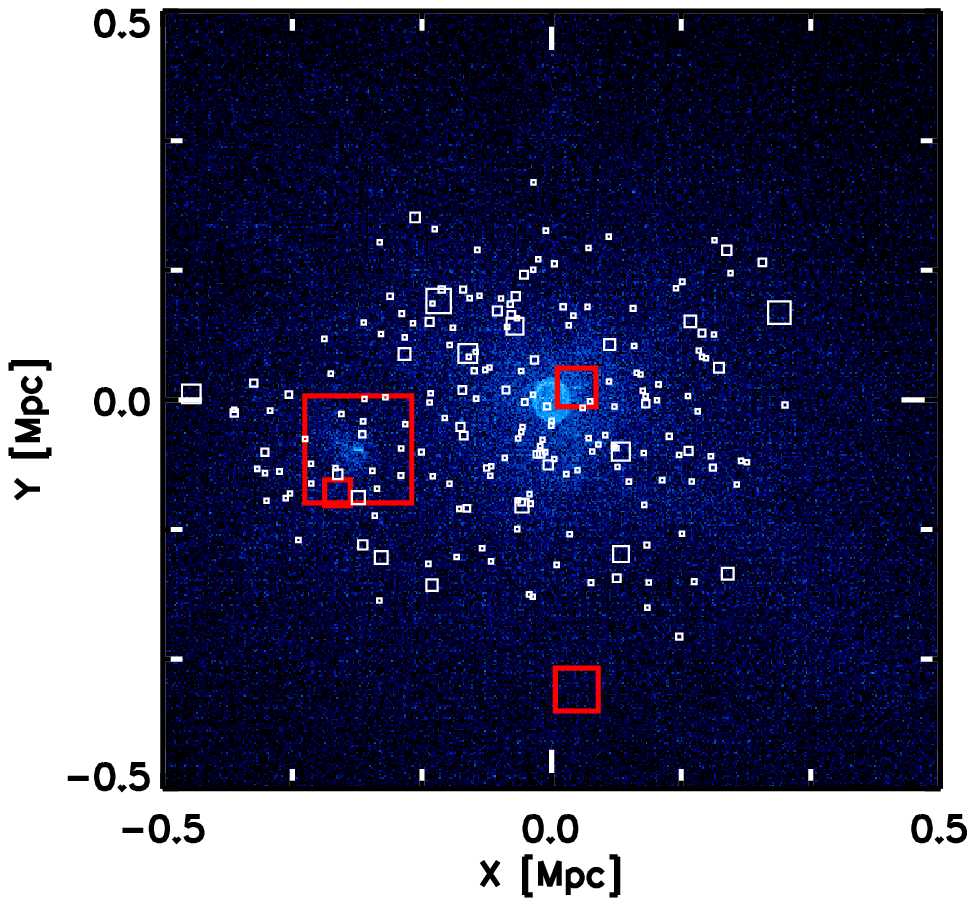} &
      \hspace{-.22in} \includegraphics*[width = .34 \textwidth]{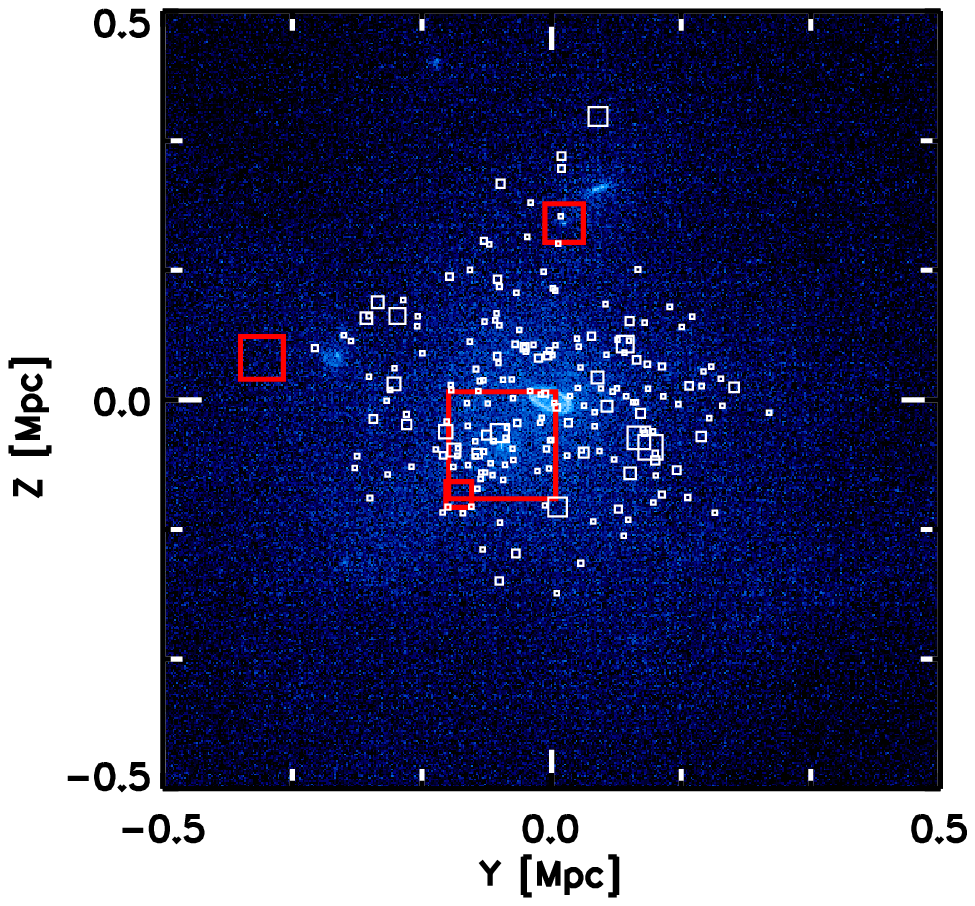} &
      \hspace{-.22in} \includegraphics*[width = .34 \textwidth]{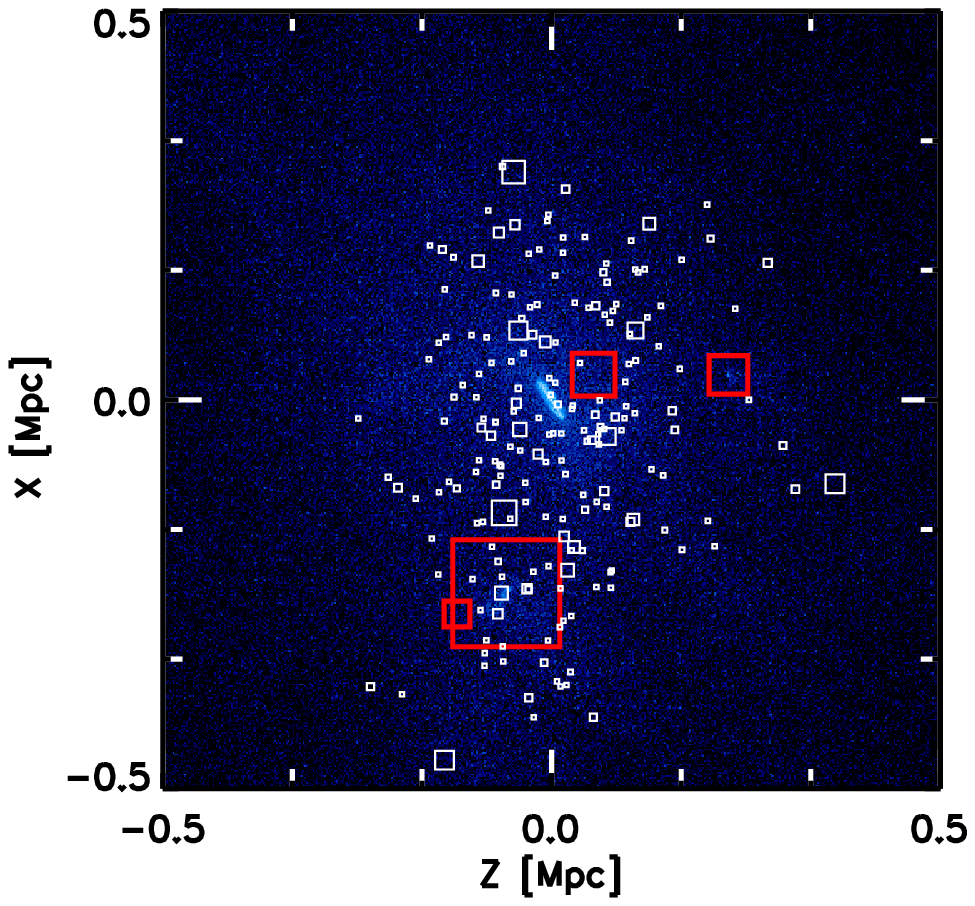} \vspace{-.1in}\\
      \hspace{-.15in} \includegraphics*[width = .34 \textwidth]{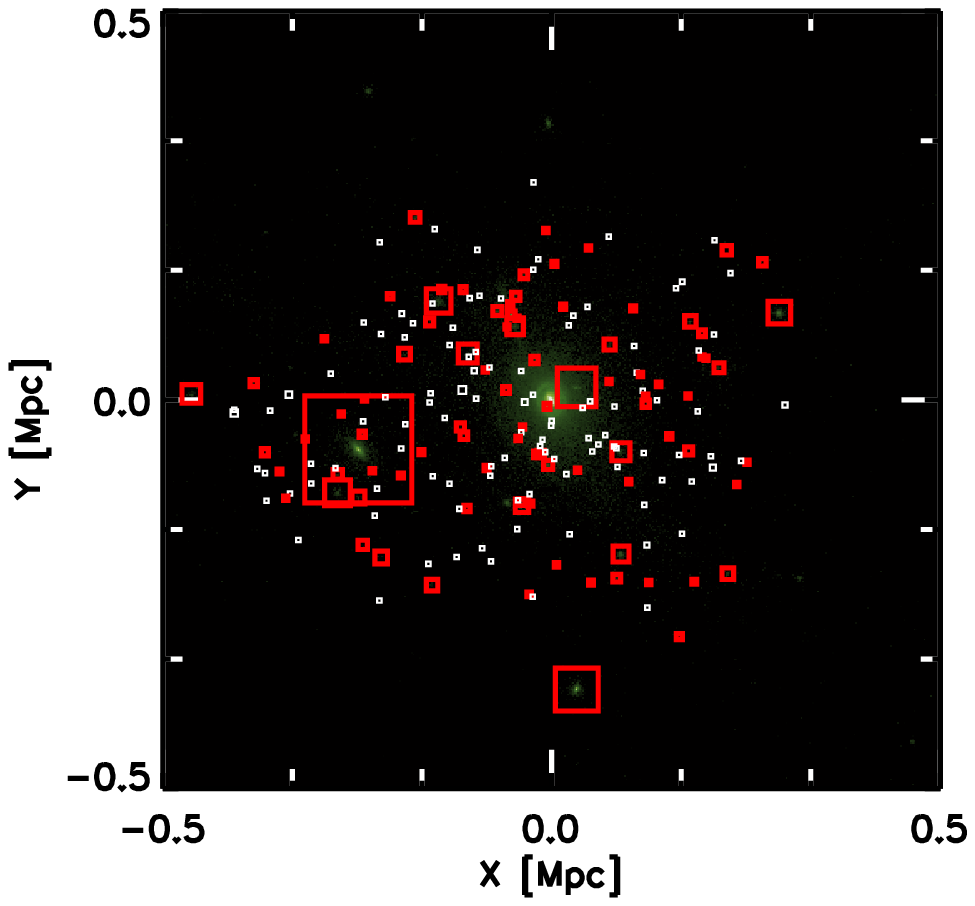} &
      \hspace{-.22in} \includegraphics*[width = .34 \textwidth]{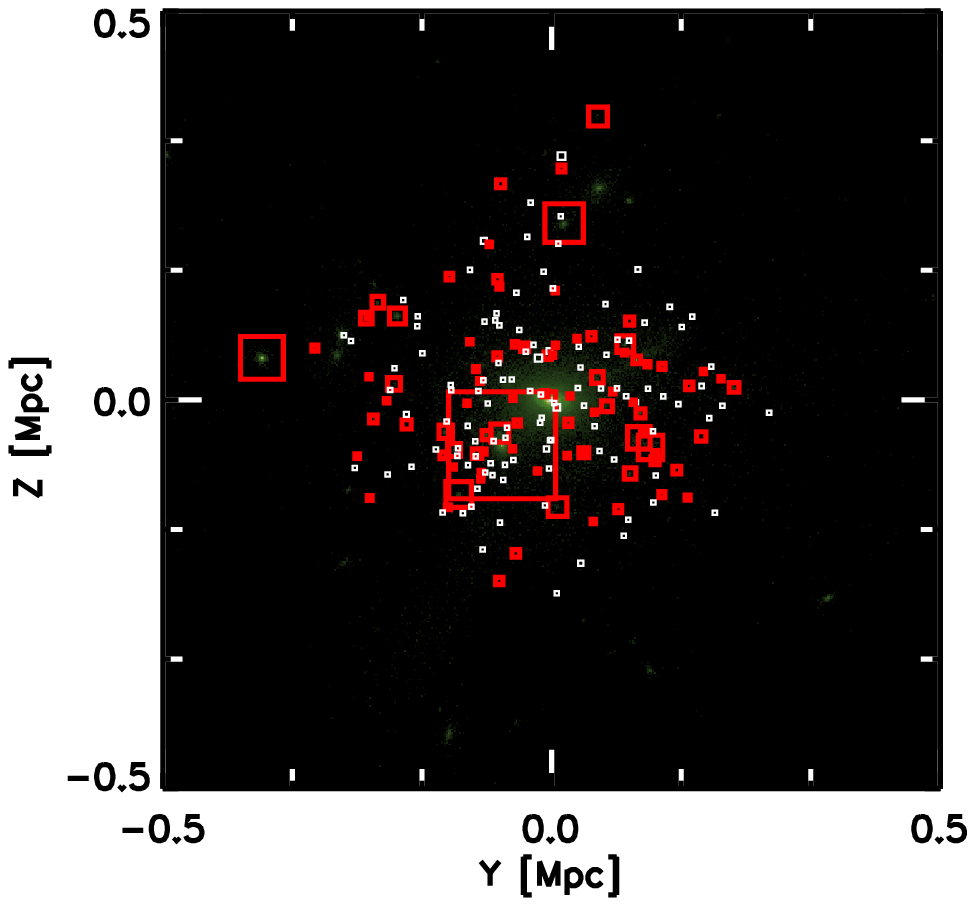} &
      \hspace{-.22in} \includegraphics*[width = .34 \textwidth]{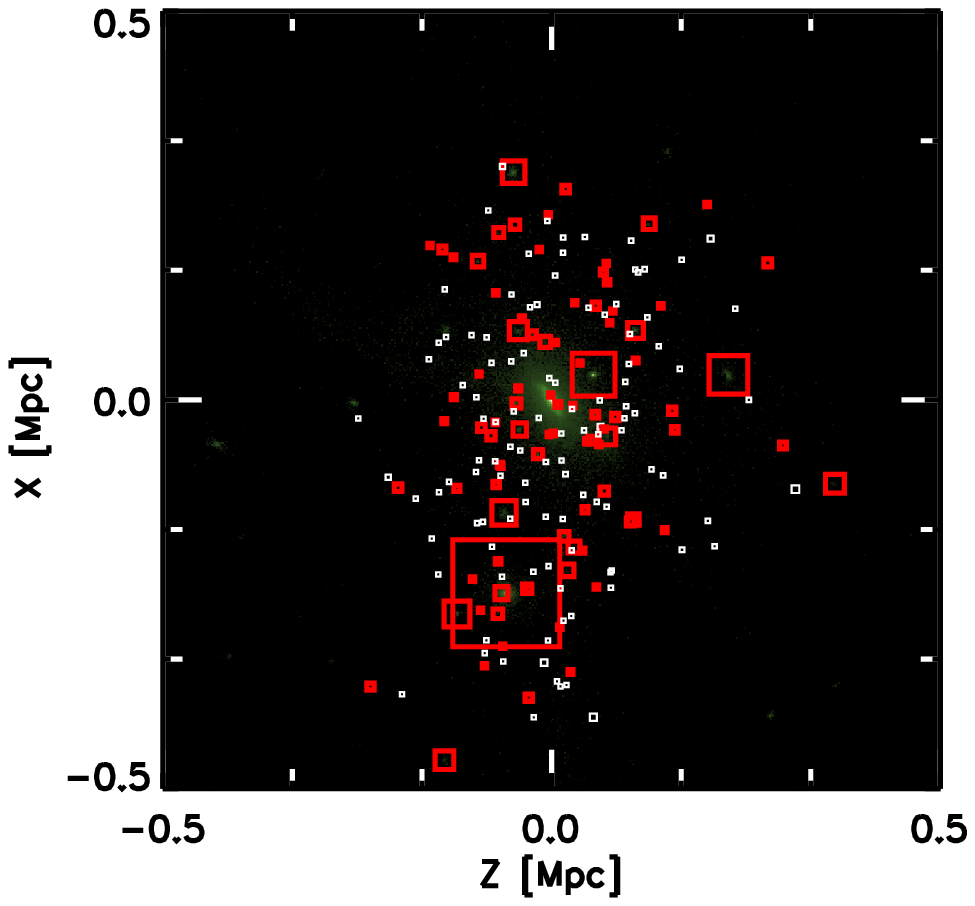}
    \end{tabular}
  \end{center}
  \vspace{-.1in}
  \caption{Projections of the dark matter (top), gas mass (middle) and
    stellar mass (bottom) distributions at $z=0$, with the location of
    subhaloes overplotted as squares. As in
    Figures~\ref{fig:aquila_dm}~--~\ref{fig:aquila_stars}, the size of
    the squares indicate the dark matter mass of each subhalo.
    Analogous to Figure~\ref{fig:aquila_gas}
    and~\ref{fig:aquila_stars}, in the middle row, red and white
    squares distinguish satellites which have gas from those that are
    gas-free, while in the bottom row, the distinction is between
    subhaloes with and without stars. Notable from the middle row is
    the pair of late-infalling, gas-rich satellites 1 and 7, as
    described in Section 4.3.
\label{fig:aquila_projections}}
\end{figure*}

\section{Time Evolution of the Aquila Simulation} \label{aquila_evolution}
Figures \ref{fig:aquila_dm}, \ref{fig:aquila_gas} and
\ref{fig:aquila_stars} respectively show the time-evolution of the
projected dark matter, gas and stellar mass distributions in the
central region of the simulation. Each panel is centred on the
position of the main subhalo (which is the host of the ``Milky Way''),
and shows all particles within a cubic volume of side length 1~Mpc At
each redshift, the $X,Y$ and $Z$-coordinates are defined parallel to
the principal moments of the inertia tensor of the halo, with
eigenvalues $I_x > I_y > I_z$. Keeping the volume fixed in comoving
coordinates corresponds to a zoom-out in physical coordinates by a
factor of 7.5 as the universe grows with time from $z=6.5$ to
$z=0$. The squares indicate the position of all satellite subhaloes
belonging to the main FoF halo identified at the time of the snapshot,
with the size of the squares in all figures indicative of (but not
strictly proportional to) the dark matter mass of the subhalo.

In Figure~\ref{fig:aquila_gas}, where the blue colour indicates gas
density, red boxes denote the subhaloes that contain gas, while white
boxes denote subhaloes that are gas-free. It can be seen that even at
high redshifts, the majority of subhaloes are gas-free, and only four
relatively massive satellites contain gas at $z=0$. The smallest of
these four satellites, which are the subject of
Section~\ref{aquila:massive}, has a total mass of $\sim 5\times10^9
\Ms$. All lower mass satellites, many of which formed stars, have lost
their gas during their evolution. The different mechanisms of gas
loss, internal and environmental, are discussed in
Section~\ref{aquila:gas_loss}.

In Figure~\ref{fig:aquila_stars}, the green colour shows the stellar
density, which is clearly dominated by the central object and its
halo. Here, red boxes show subhaloes that contain stars, while white
boxes show haloes that are essentially dark. At $z=0$, there are 90
subhaloes containing stars, including the four which contain gas, as
shown in Figure~\ref{fig:aquila_gas}. The highest mass subhalo that
does not contain any stars at $z=0$ has a mass of
$4.3\times10^8\Ms$. The lower mass limit for star formation becomes
difficult to assess, because of the limited resolution of our
simulation. The total number of subhaloes with stars is comparable,
however, with observational estimates of the number of dwarf
satellites around the Milky Way.

The halo reaches a final virial mass of $1.6 \times 10^{12} \Ms$,
comparable to recent observational estimates of the Milky Way halo,
for example $10^{12}\Ms$ \citep{Xue-2008}, $1.4\times10^{12}\Ms$
\citep{Smith-2007}, $1.6\times10^{12}\Ms$ \citep{Gnedin-2010} and
$2.4\times10^{12}\Ms$ \citep{Li-2008}. The corresponding spherical
virial radius is $\sim 250$~kpc, but we include as satellites all
subhaloes within the FoF group. 40\% of the satellites are presently
located outside of $r_{vir}$, with the furthest satellite at a
galactocentric distance of $490$~kpc. The central galaxy reaches a
stellar mass of $10.8 \times10^{10}\Ms$, higher than current
observational estimates for the Milky Way, for example $5.5
\times10^{10}\Ms$ from \citep{Flynn-2006}. Distributions of the
different mass components, and the positions of all satellites at
$z=0$ are shown in three orthogonal projections along the principal
axes of the inertia tensor of the halo in
Figure~\ref{fig:aquila_projections}. By comparison with the flattened
distribution seen at high redshift, the final shape of the halo
appears round and largely featureless. This transformation from a
triaxial mass distribution, expected from purely gravitational
assembly, to an oblate halo similar to that observed, is also studied
in \cite{Tissera-2010}, who attribute the difference to baryonic
effects.

\section{Formation and Evolution of Satellites} \label{aquila:gas_loss}
When the environment is included, different mechanisms of gas loss can
play a role, and all are observed in the simulation. Internal and
external mechanisms often act simultaneously, and are not always easy
to disentangle. Just as supernova heating aids in the gas removal by
UV radiation \citep{Sawala-2010}, the thermal expansion caused by the
energy input can also make it easier for tidal interactions to remove
gas. In section~\ref{aquila:examples}, we discuss four exemplary cases
of gas loss which are representative of the total subhalo population
in terms of their final properties, but where the different mechanisms
are relatively easily identified.  Section~\ref{aquila:extremes}
considers two extreme cases of subhaloes very heavily affected by
stripping. Section~\ref{aquila:massive} describes four satellites that
still contain gas at $z=0$, and contrasts them to the many gas-free
satellites.

\subsection{Gas Loss by Example}\label{aquila:examples}
\begin{figure*}
  \begin{center}
    \begin{tabular}{ll}
      \hspace{-.42in} \includegraphics*[trim = 00mm 12mm 00mm 6mm, clip, scale= .5]{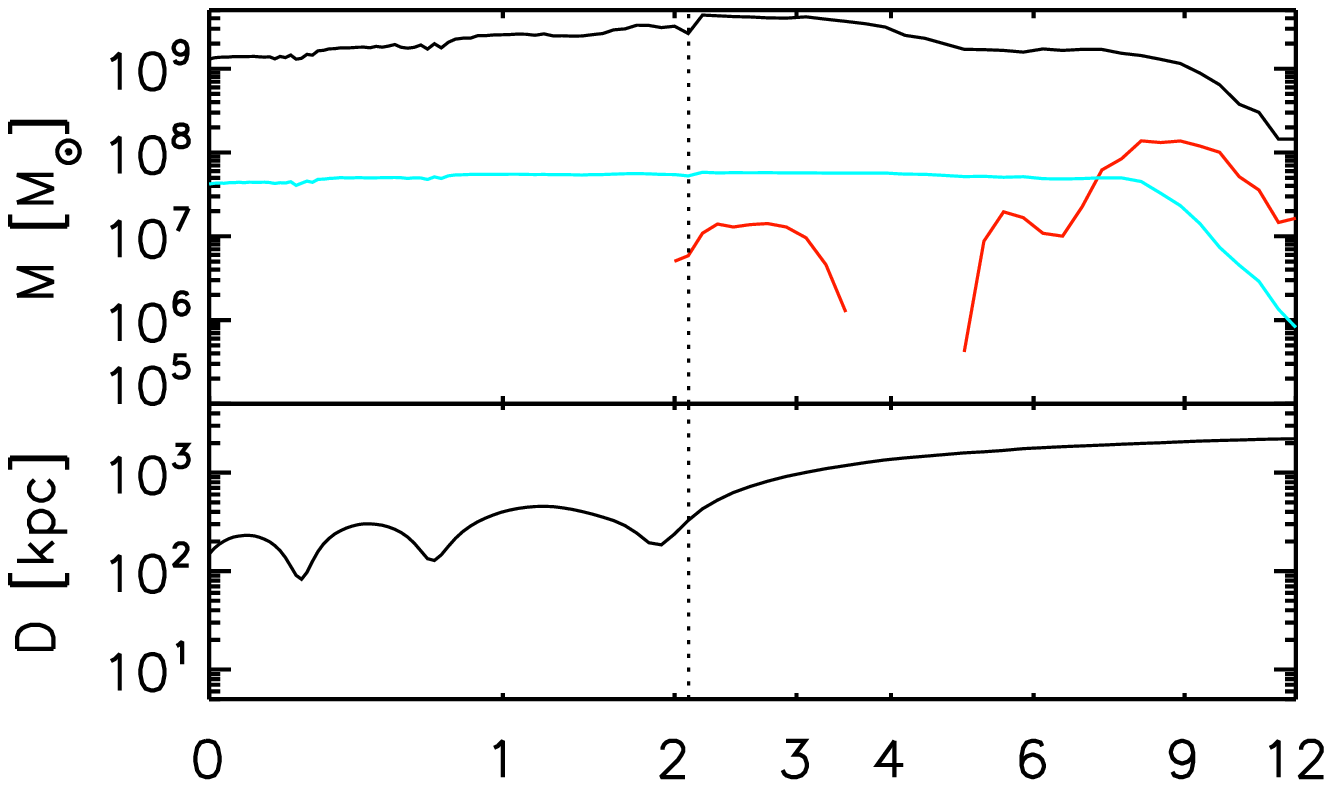} &
      \hspace{-.30in} \includegraphics*[trim = 00mm 12mm 00mm 6mm, clip, scale= .5]{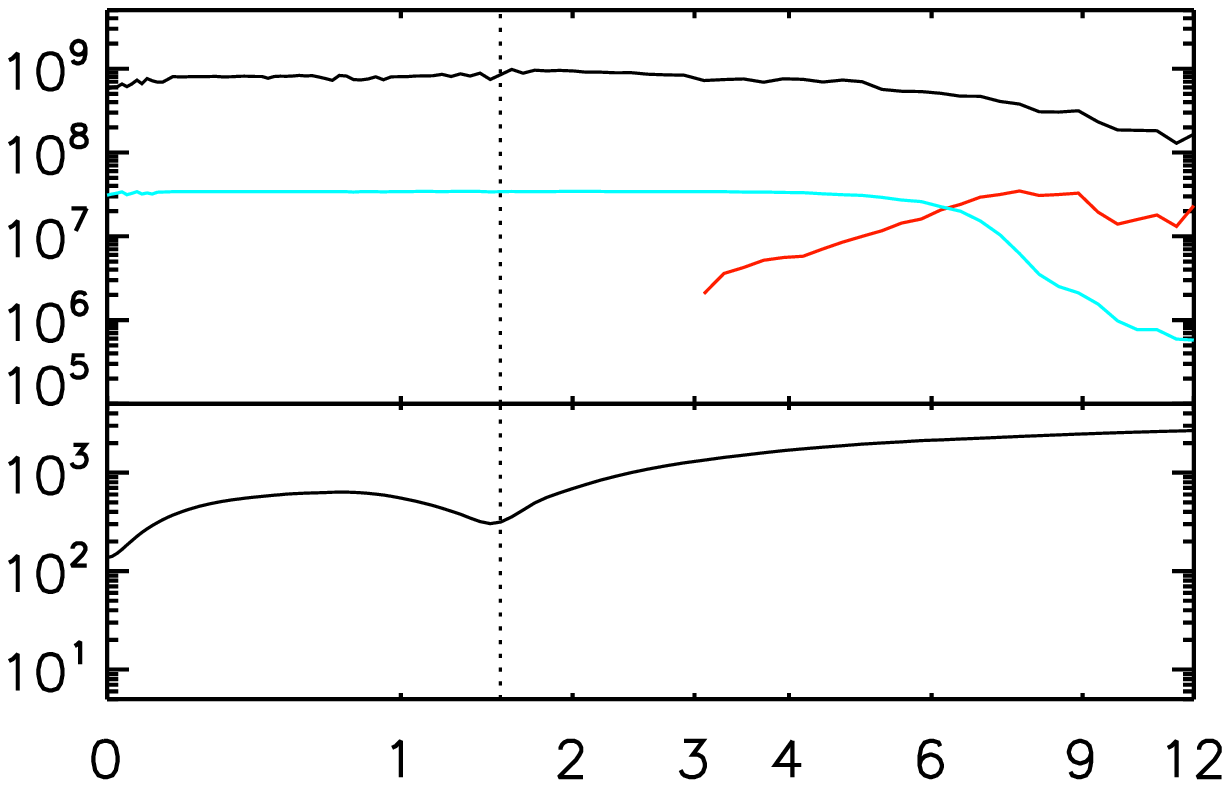} \\
      \hspace{-.42in} \includegraphics*[trim = 00mm 0mm 00mm 6mm, clip, scale= .5]{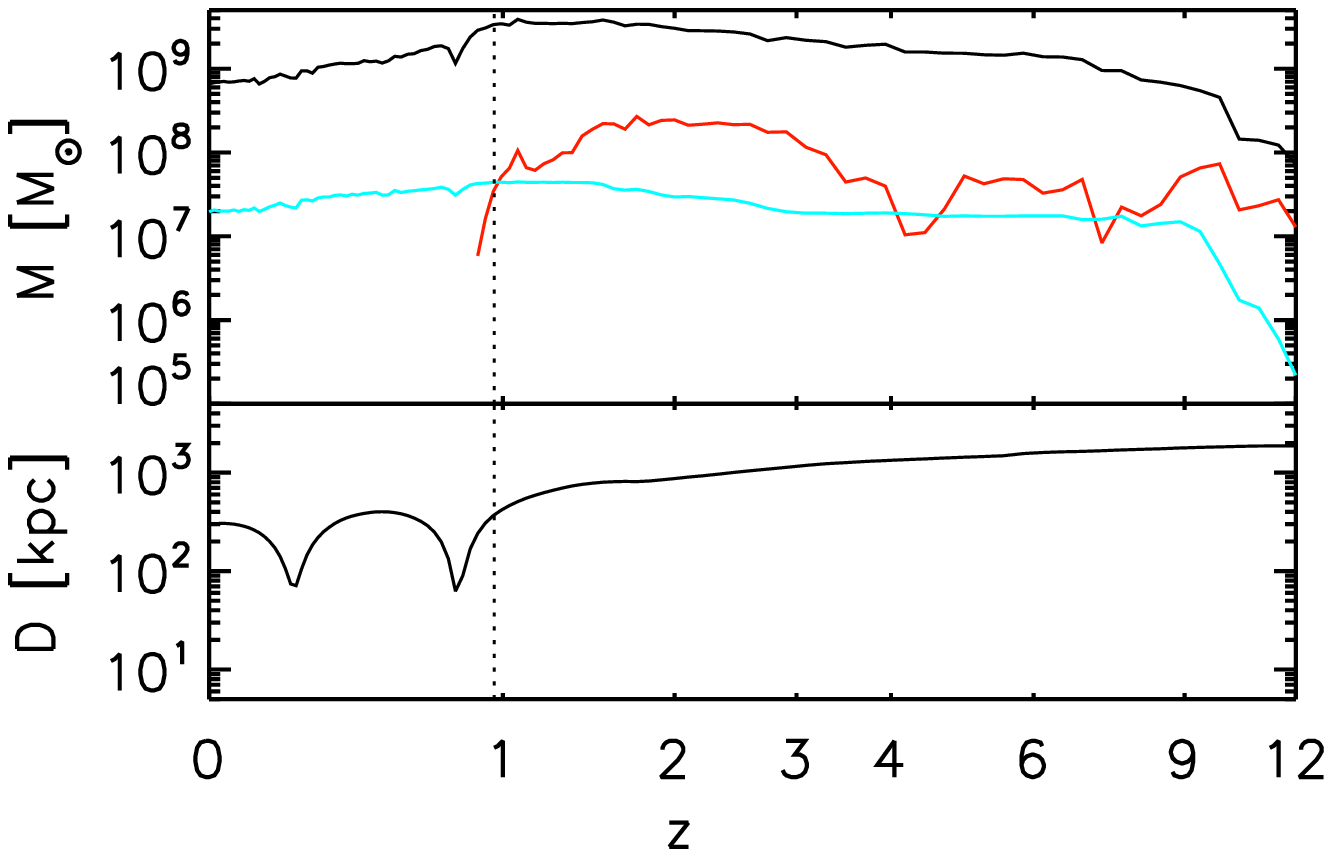} &
      \hspace{-.30in} \includegraphics*[trim = 00mm 0mm 00mm 6mm, clip, scale= .5]{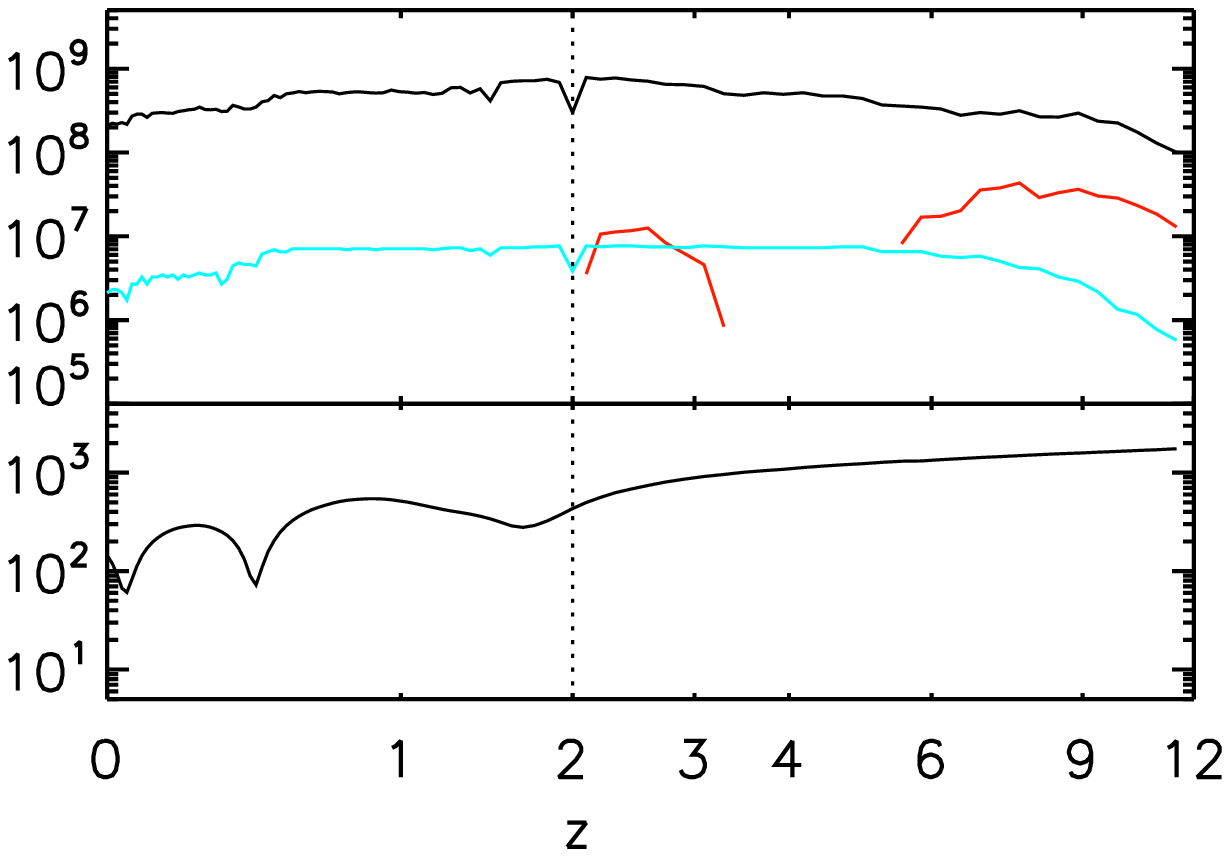}
    \end{tabular}
  \end{center}
  \vspace{-.1in}
  \caption{Evolution of the mass components (top) and distance to the
    centre of the central galaxy (bottom) for four subhaloes: 14 (top
    left), 23 (top right), 25 (bottom left) and 56 (bottom right) as a
    function of redshift. In each of the top panels, the black line
    denotes the dark matter mass, the red line denotes the gas mass,
    and the blue line denotes the stellar mass. The dotted lines
    denote the time of infall. Before outflow and/or stripping, the
    galaxies reach peak baryon masses of $\sim 5 \times 10^7$--$3
    \times 10^8 \Ms$, corresponding to $\sim 2 \times 10^2$--$10^3$
    SPH particles. \label{fig:selected_evolution}}
\end{figure*}

\begin{figure*}
  \begin{center}
    \begin{tabular}{lll}
       \hspace{-.2in} \includegraphics*[trim = 10mm 0mm 0mm 0mm, clip, scale= .40]{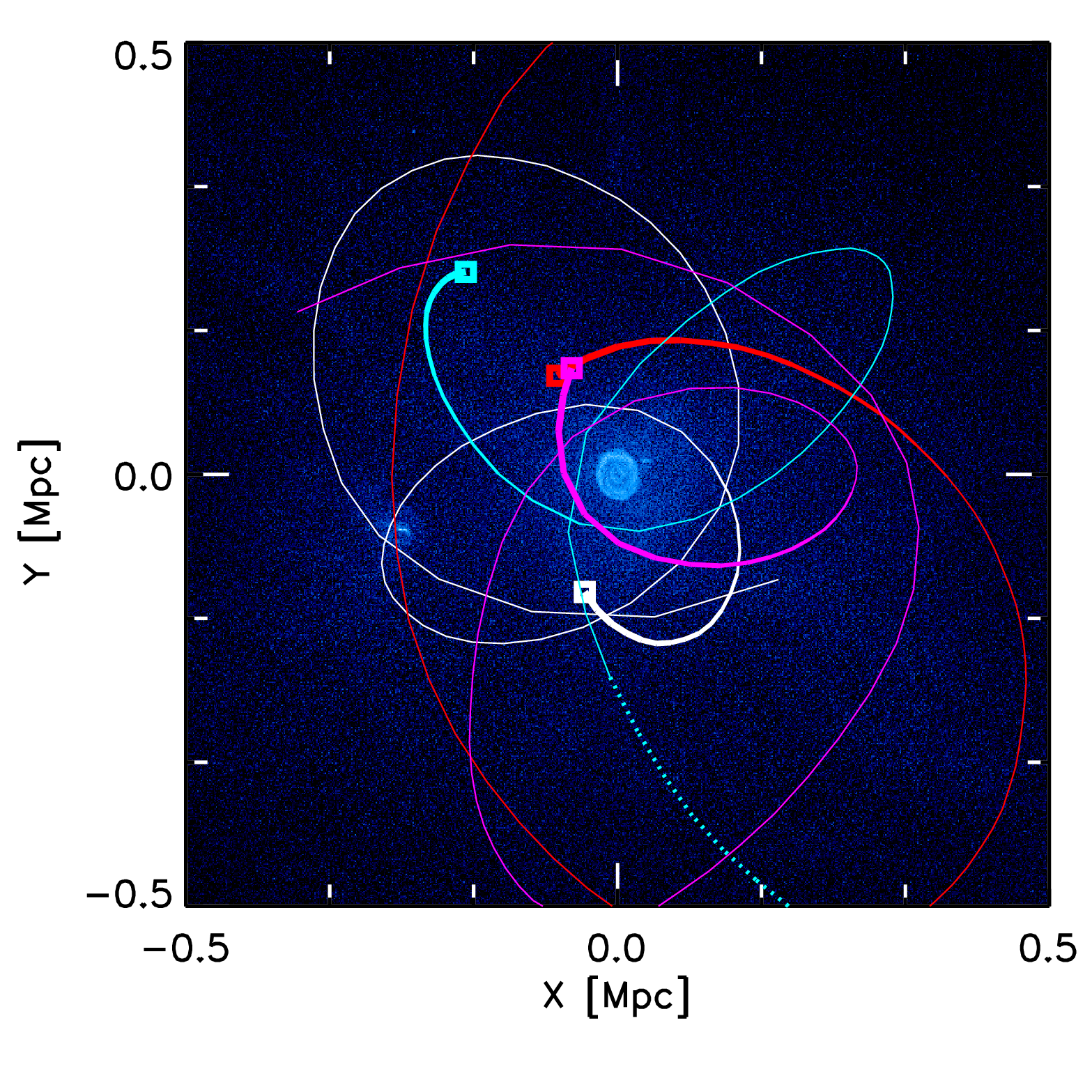} &
       \hspace{-.2in} \includegraphics*[trim = 10mm 0mm 0mm 0mm, clip, scale= .40]{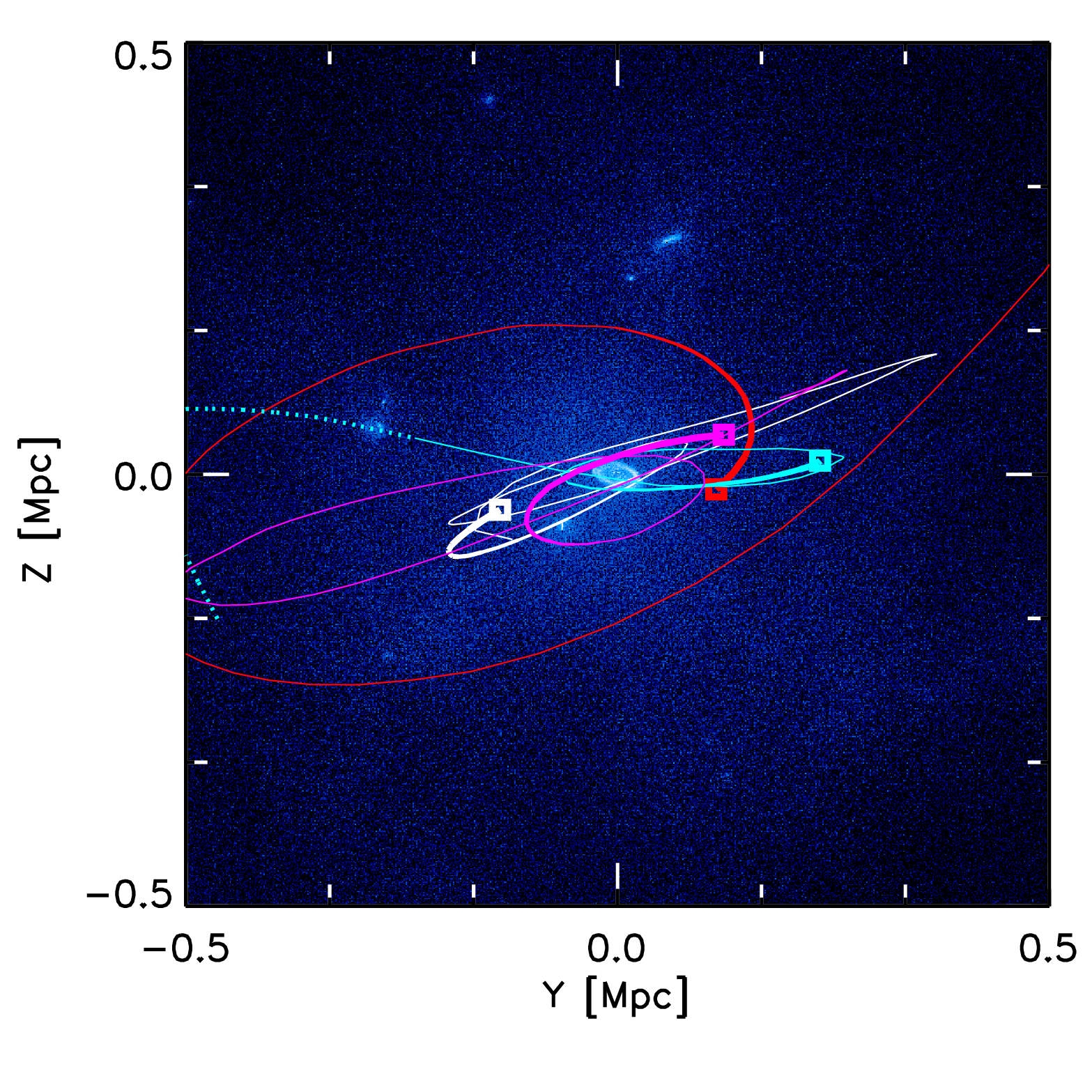} &
       \hspace{-.2in} \includegraphics*[trim = 10mm 0mm 0mm 0mm, clip, scale= .40]{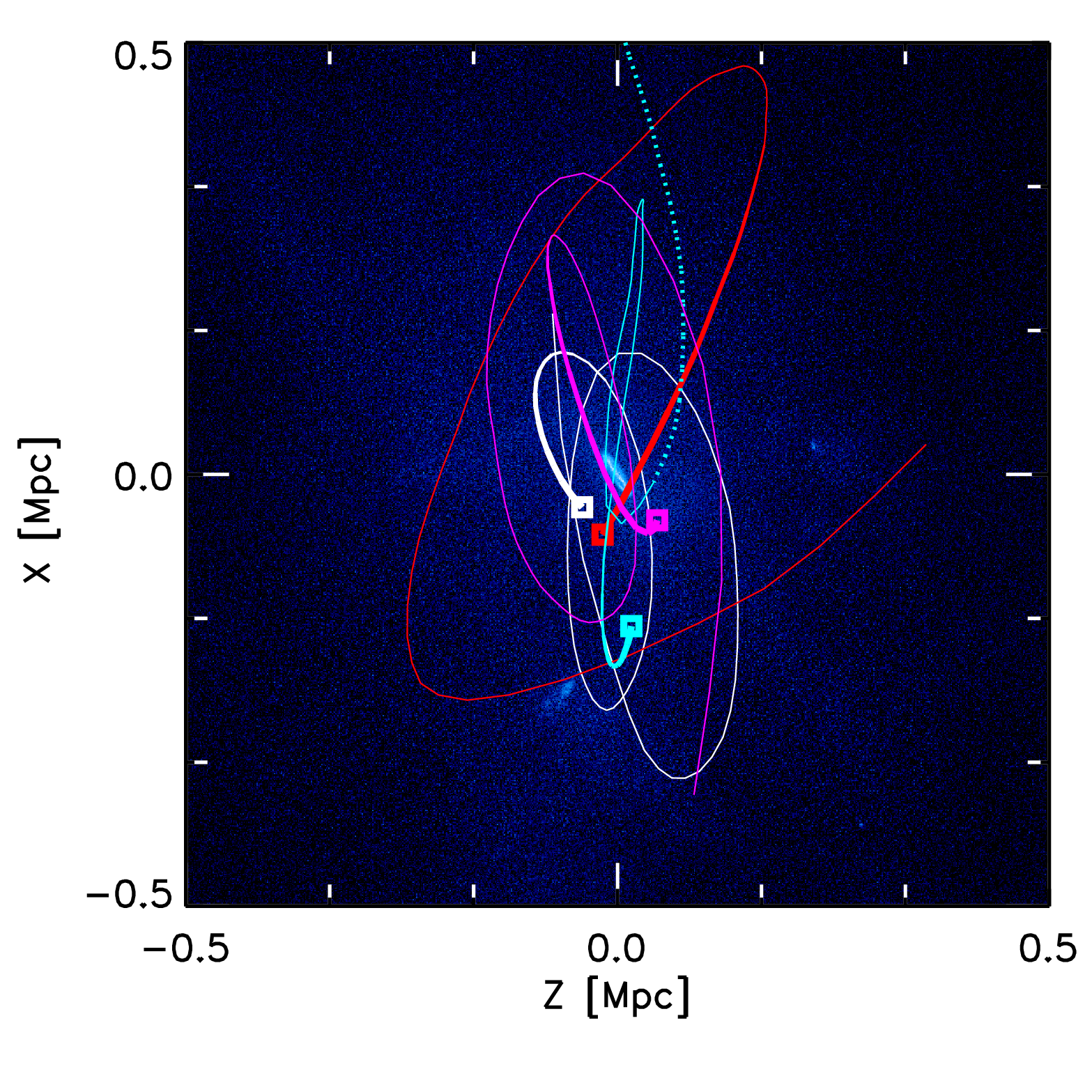} 
    \end{tabular}
  \end{center}
  \vspace{-.25in}
  \caption{Orthogonal projections of the gas distribution at $z=0$,
    identical to the middle row of
    Figure~\ref{fig:aquila_projections}, together with the orbits of
    the four subhaloes 14 (white), 23 (red), 25 (light blue) and 56
    (magenta). The square at the end of each line denotes the present
    position of the subhalo, while the tail end extends to
    $z=1.9$. All four subhaloes contain stars at $z=0$, but only halo
    25 also contains gas until $z=0.8$, as indicated by the
    corresponding dotted segment of the light blue
    orbit. \label{fig:aquila_orbits}}
\end{figure*}

In the top panel of Figure~\ref{fig:selected_evolution}, we show the
evolution of the three mass components, dark matter (black), gas
(blue) and stellar mass (green) of 4 selected subhaloes, each
representing a different evolutionary path. Also shown in the bottom
panel of Figure~\ref{fig:selected_evolution} is the distance from the
satellite centre of mass to the main halo centre as a function of
redshift.

\begin{itemize}
\item First from the left in Figure~\ref{fig:selected_evolution} is
  subhalo 14. Here, the gas is blown out by the combined effect of
  supernova feedback and UV heating while the dwarf halo is still in
  isolation. As it approaches the central halo, some gas is
  re-acquired at $z\sim3.5$, but this does not lead to renewed star
  formation. This residual gas is lost when the dwarf halo finally
  falls in to the central halo at $z=2.1$. As it spirals inwards on
  multiple orbits, the dark matter mass decreases from its peak of
  $2.4 \times 10^9\Ms$ to its $z=0$ value of $1.3\times10^9\Ms$. On
  each of the two final pericentre passages, the stellar mass also
  decreases slightly.

\item Second from the left, subhalo 23 loses its gas already at high
  redshift while still in isolation, in the same way as subhalo 14. It
  evolves passively and free of gas from $z=3$ onwards, first falling
  into the main halo at $z=1.53$. The dark matter mass and stellar
  mass are not significantly affected by tides between infall and
  $z=0$.

\item Subhalo 25, in the third column, keeps a significant amount of
  gas and continues to form stars up to its first infall at
  $z=0.96$. The ISM is stripped before it reaches pericentre for the
  first time and star formation ceases. The dark matter mass is
  subsequently reduced from its peak value of $4 \times 10^9 \Ms$ to
  $6.6 \times 10^8 \Ms$ on two close pericentre passages. The stellar
  mass also decreases, particularly during the second passage.

\item On the right, subhalo 56 is a peculiar case. Having lost some
  gas due to supernova feedback, the remainder is lost almost
  instantaneously at $z=6$, when the effect of the UV background sets
  in. As for subhalo 14, a small amount of gas gets re-accreted and is
  lost again as the subhalo approaches the central halo. The galaxy
  first becomes a satellite at $z=2.0$, but does not get close to the
  centre on its first orbit, resulting in no stripping of stars, and
  only a small reduction of the dark matter mass. The final two
  passages are much closer, and as a result, the dark matter and
  stellar mass are both significantly reduced.
\end{itemize}

Three of the four objects have peak dark matter masses between $7
\times 10^8$ and $2.5 \times 10^9 \Ms$, with subhalo 56, which loses
its gas due to UV heating, has a peak mass of $5 \times 10^8$. The
final dark matter masses lie between $5.8 \times 10^8$ and $2 \times
10^8 \Ms$, with final stellar masses in the range of $2 \times 10^6$
to 4 $\times 10^7 \Ms$, and corresponding stellar mass~-~total mass
ratios of 50-100. The four objects follow the overall scaling
relations of Figures~\ref{fig:relation_mdm-ms}
to~\ref{fig:ratio-infall-approach}, and are also similar to the
isolated dwarf galaxies of \cite{Sawala-2010}, except for higher final
stellar mass~-~total mass ratios for those objects most strongly
effected by stripping.

\begin{figure*}
  \begin{center}
    \begin{tabular}{lcc}
      \hspace{-.2in} \includegraphics*[trim = 10mm 0mm 0mm 0mm, clip, scale= .40]{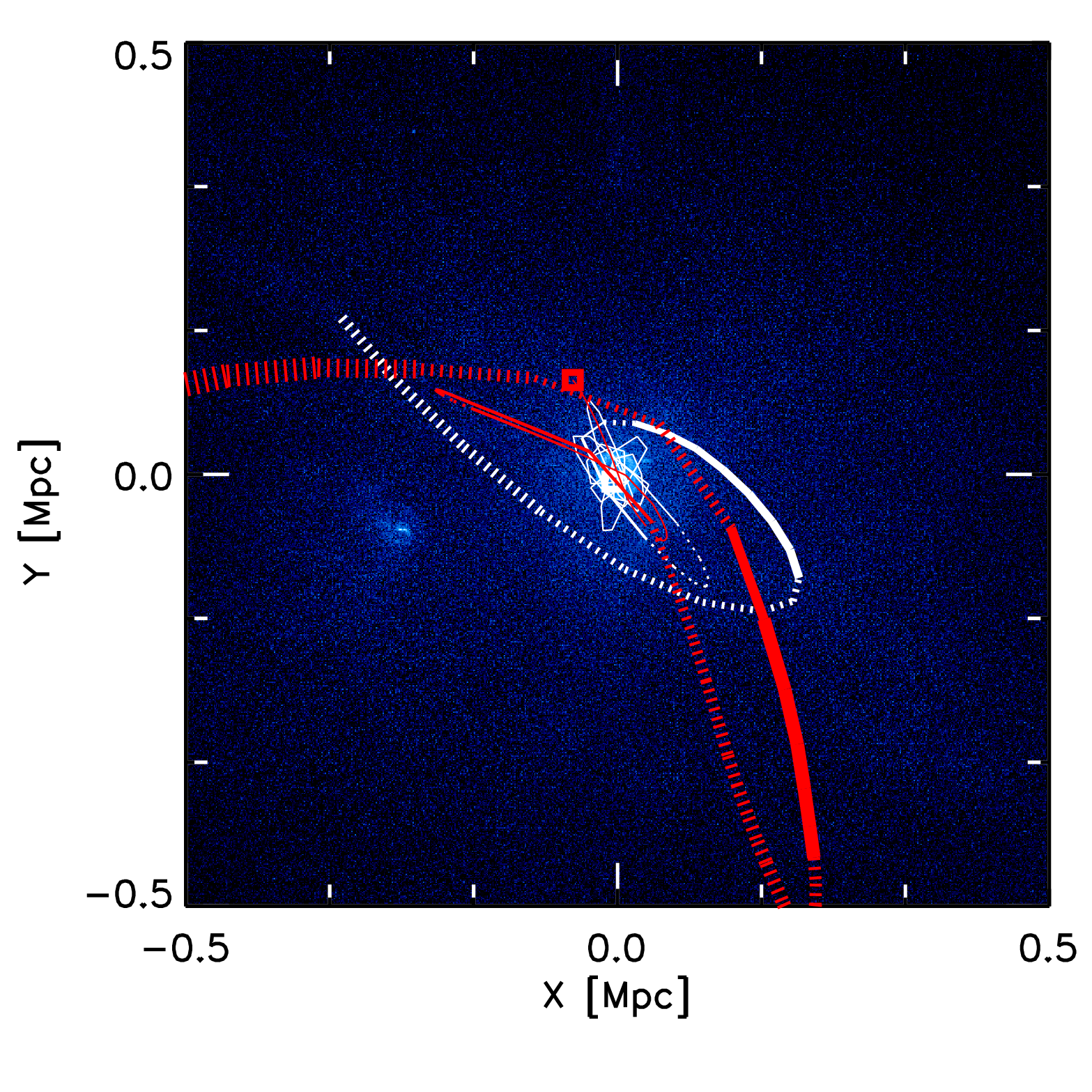} &
       \hspace{-.2in} \includegraphics*[trim = 10mm 0mm 0mm 0mm, clip, scale= .40]{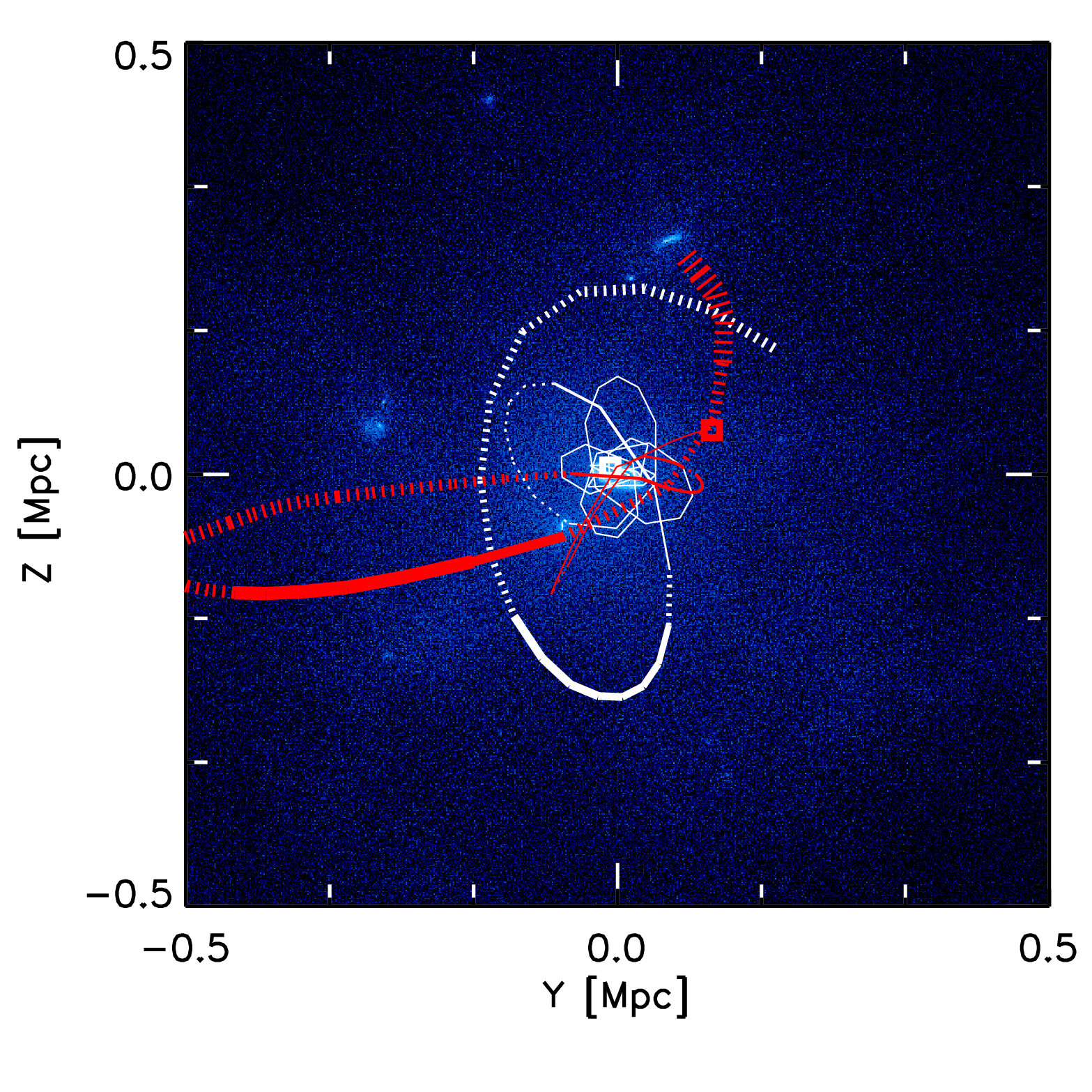} &
       \hspace{-.2in} \includegraphics*[trim = 10mm 0mm 0mm 0mm, clip, scale= .40]{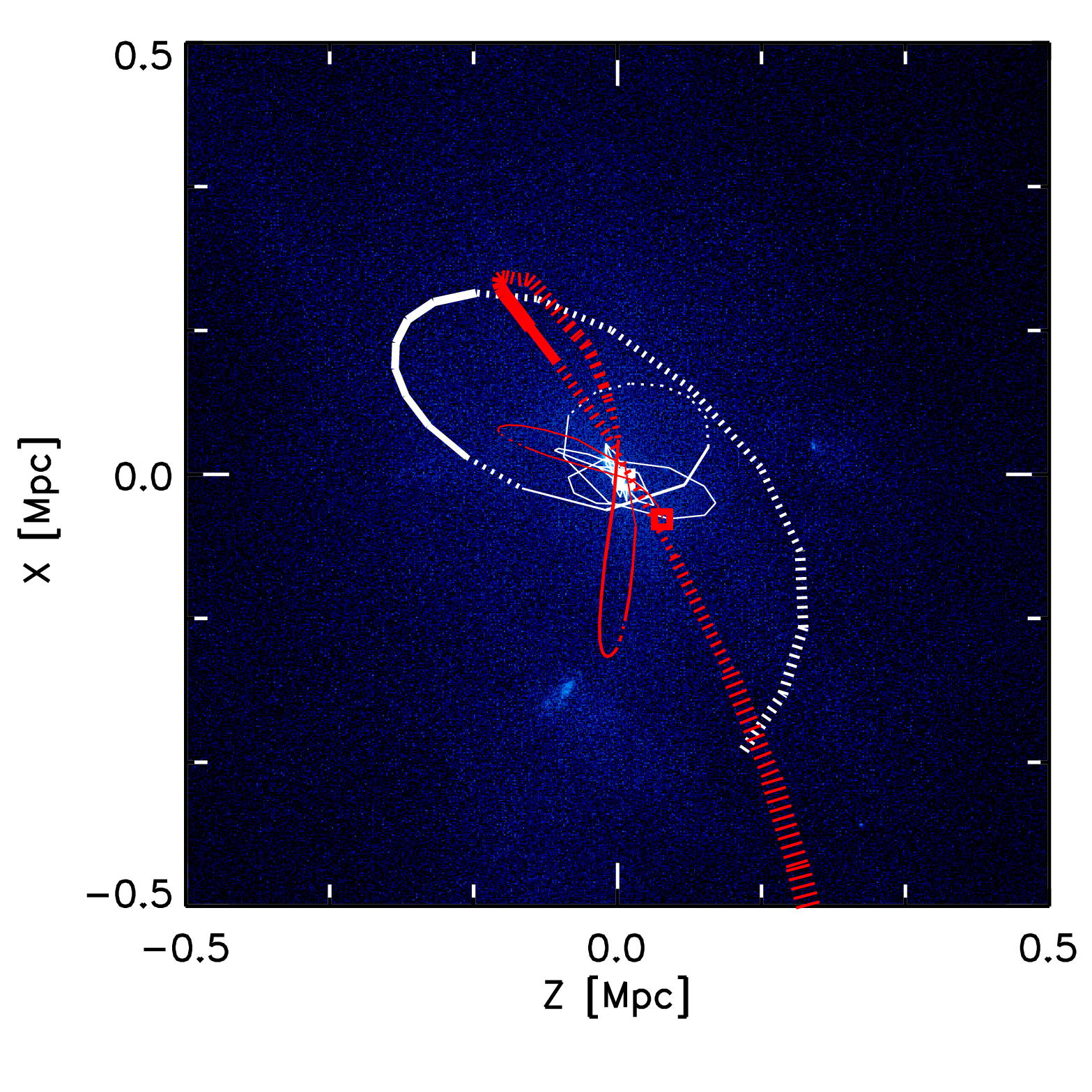} 
     \end{tabular}
  \end{center}
  \vspace{-.25in}
  \caption{Orthogonal projections of the gas distribution at $z=0$,
    identical to the middle row of
    Figure~\ref{fig:aquila_projections}, together with the orbits of
    subhaloes 3 (white) and 5 (red) after $z=2.45$. Both of the
    subhaloes lost more than 97\% of their dark matter mass since
    infall, and the thickness of the line changes along the orbit
    proportional to the dark matter mass relative to the infall
    mass. Dotted line segments denote the presence of gas, while solid
    lines indicate a gas-free phase. The squares denote the present
    positions of the satellites.\label{fig:extreme}}
\end{figure*}

Figure~\ref{fig:aquila_orbits} shows the orbits of the four subhaloes
described above in three different projections after z=1.9. While the
four satellites are on similar orbits (in contrast to the two cases
shown in Figure~\ref{fig:extreme}), it can be seen that only subhalo
25 (light blue curve), still has gas when it enters the main halo. The
other three fall in gas-free. All four subhaloes are free of gas
during most of their evolution as satellites.

\subsection{Two extreme cases} \label{aquila:extremes}
\begin{figure*}
  \begin{center}
    \begin{tabular}{ll}
       \hspace{-.2in} \includegraphics*[width = .5\textwidth]{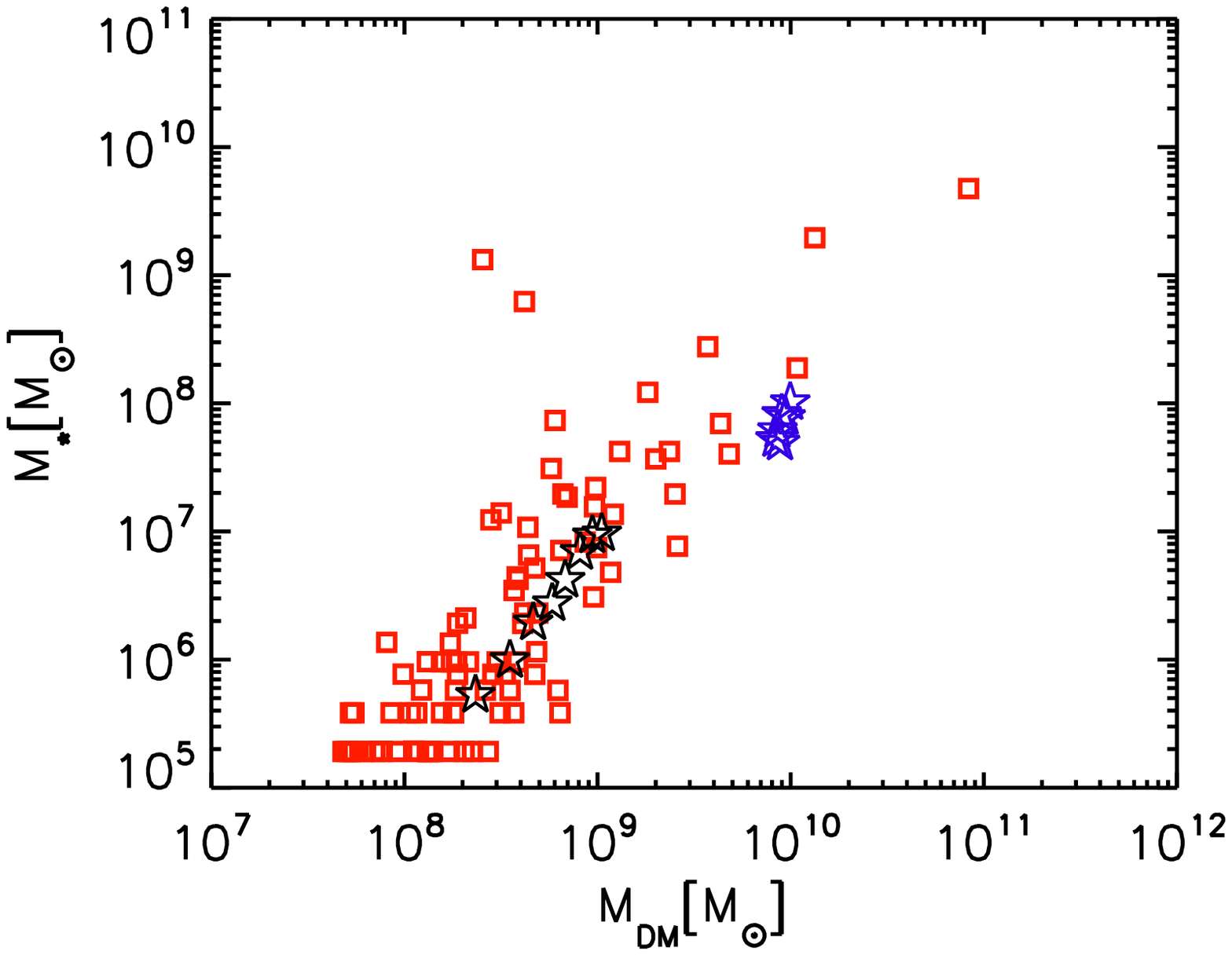} &
       \hspace{-.3in} \includegraphics*[width = .5\textwidth]{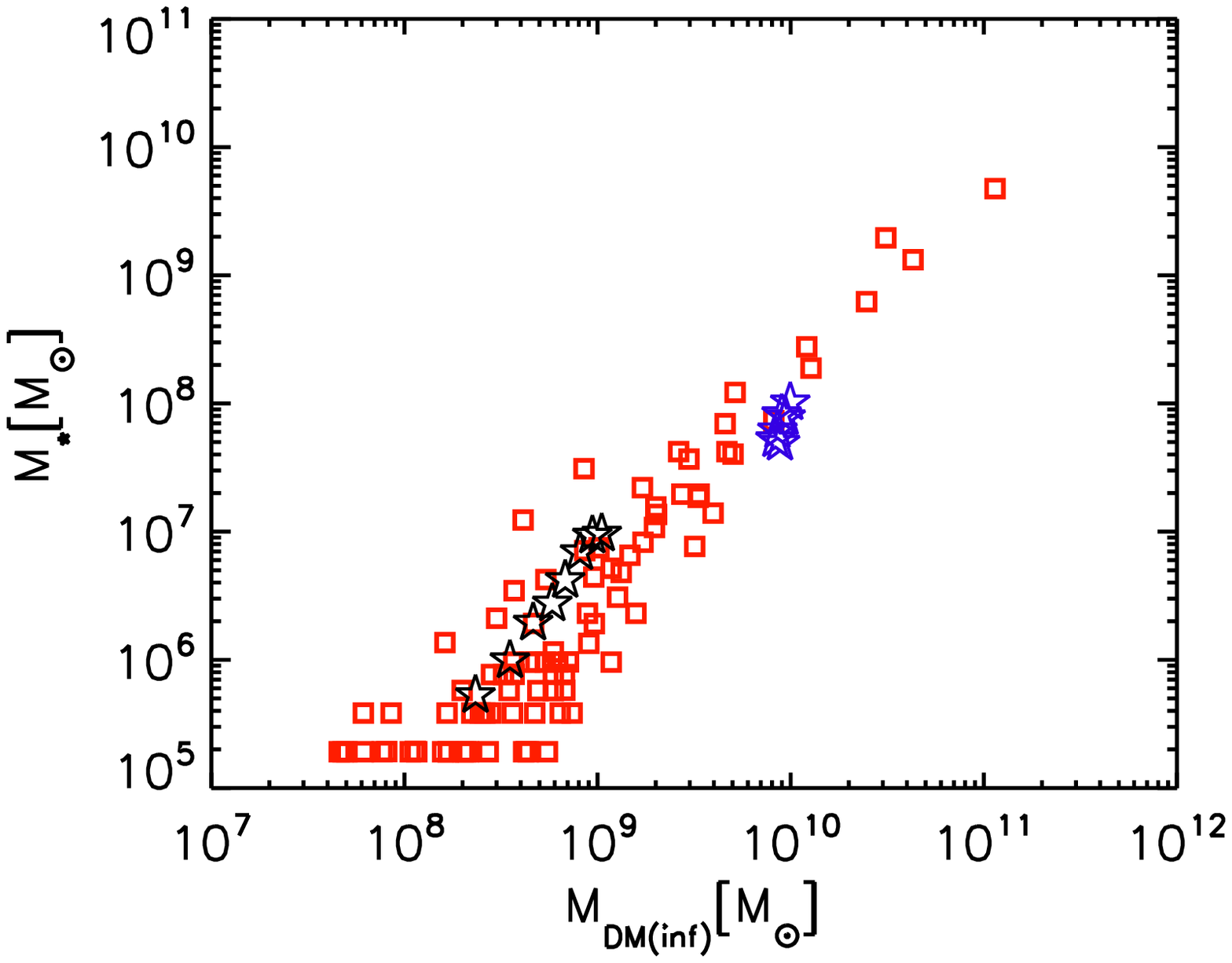}
    \end{tabular}
  \end{center}
  \vspace{-.1in}
  \caption{Stellar mass per subhalo as a function of mass in dark
    matter. In both panels, red squares denote satellite galaxies in
    the {\it Aquila} simulation. The left panel shows the dark matter
    mass at $z=0$, while the right panel shows the mass at infall. In
    both cases, stellar mass and total mass are clearly
    correlated. For comparison, the star symbols denote the results of
    high-resolution simulations of individual dwarf galaxies: black
    stars are haloes 12--20 of Sawala et al. (2010), blue stars are
    haloes 1--6 of Sawala et al. (2011). The smaller scatter in the
    relation between stellar mass and infall mass is testament to the
    fact that for most satellites, the stellar component is determined
    before infall. The scatter increases for smaller objects, and
    reaches about two orders of magnitude at an infall mass of $\sim
    10^9\Ms$. For masses below $10^9 \Ms$, an increasing fraction of
    haloes is completely dark. As haloes without stars are not
    included in these plots, this may give the false visual impression
    of a flattening relation or decreasing scatter at the low mass
    end.
 \label{fig:relation_mdm-ms}}
\end{figure*}

\begin{figure*}
  \begin{center}
    \begin{tabular}{ll}
      \includegraphics*[width = .5\textwidth]{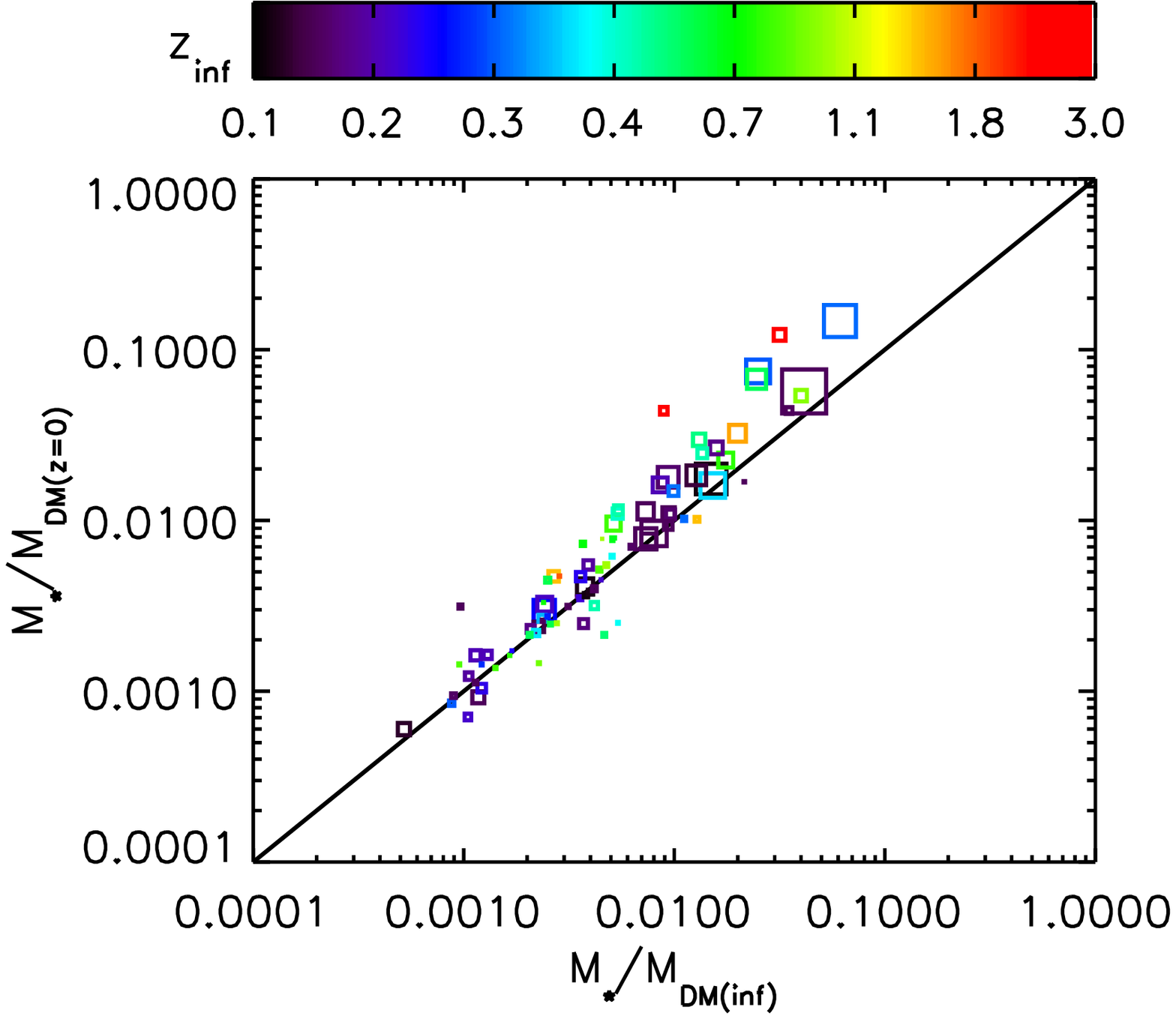}
       \hspace{-.2in} \includegraphics*[width = .5\textwidth]{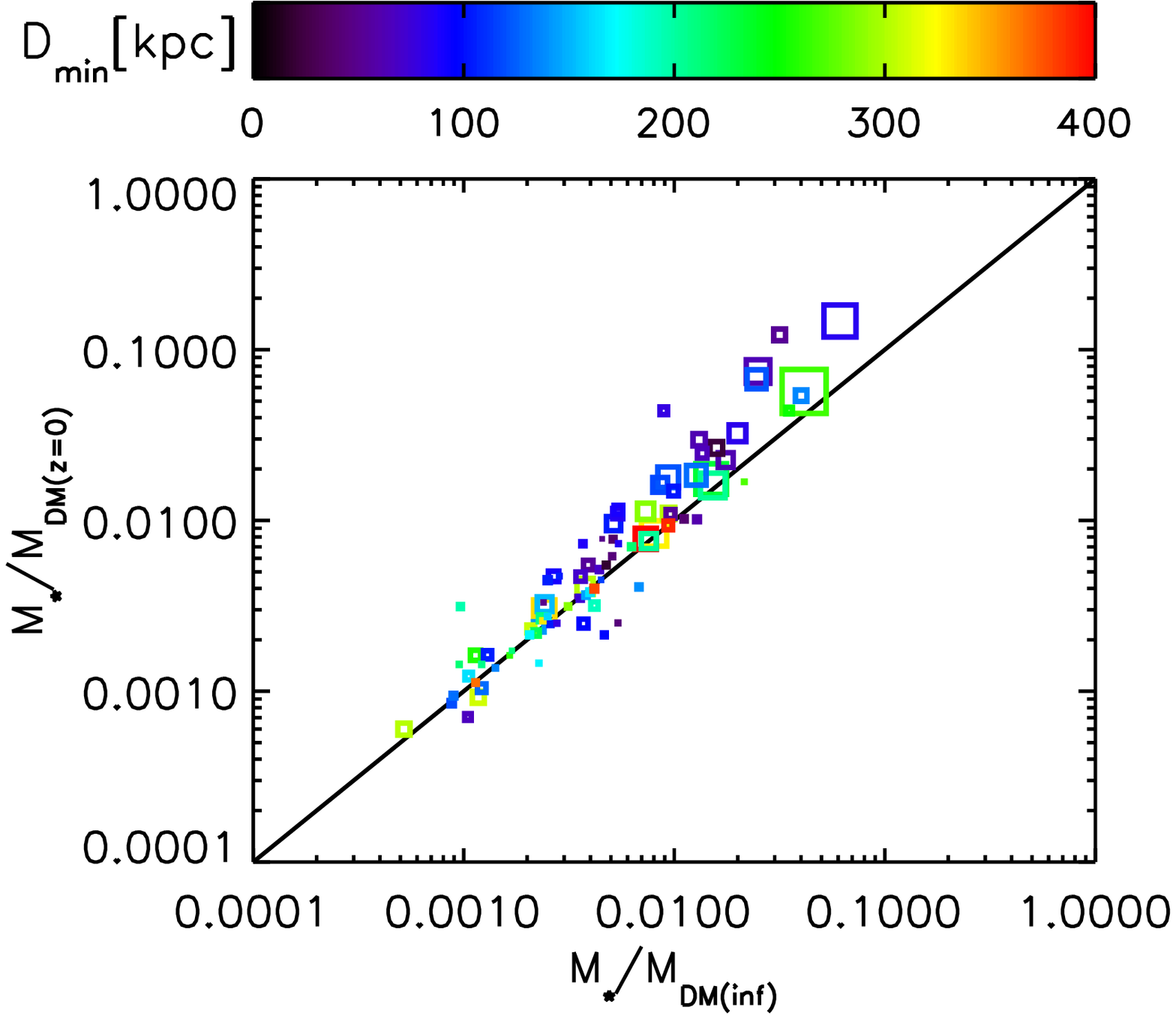} &
    \end{tabular}
  \end{center}
  \vspace{-.1in}
  \caption{Ratio of stellar mass to halo mass of each subhalo at
    present, compared to the same ratio at infall. The colour coding
    in the left panel indicates the infall redshift of each subhalo,
    while the colour coding in the right panel indicates the distance
    of closest approach between the subhalo and the centre of the host
    halo. The size of the symbol is representative of the present
    total mass of each object. Haloes above the black line increased
    their stellar mass - halo mass ratio since infall, while haloes
    below the black line decreased it.
\label{fig:ratio-infall-approach}}
\end{figure*}

Two satellites, subhaloes~3 and 5, have final stellar masses that
exceed their dark matter mass. They can be identified as outliers on
the $M_\star$--$M_{DM}$ relation shown in
Figure~\ref{fig:relation_mdm-ms}, and are off the chart in the
$M_\star$/$M_{DM,z=0}$--$M_\star$/$M_{DM,infall}$ relations of
Figure~\ref{fig:ratio-infall-approach}. This testifies to the fact
that their dark matter masses decreased by a factor of 35 in the case
of subhalo 5, and even $\sim 200$ for subhalo 3. Not surprisingly, the
orbits of the two haloes, plotted in Figure~\ref{fig:extreme}, both
include a number of recent close pericentre passages. A difference in
orbital shape is also apparent in comparison with
Figure~\ref{fig:aquila_orbits}.  The dashed and solid line segments in
Figure~\ref{fig:extreme} indicate the presence or absence of gas,
respectively. It can be seen that both objects still have gas when
they first fall into the main halo. Interestingly, they both go
through a gas-free phase on their first pericentre passage, but both
recollect gas twice during their subsequent evolution, only to lose it
again at each pericentre passage.

The two objects enter the halo with stellar mass~--~total mass ratios
of $30$ to $40$ and properties similar to dwarf irregular galaxies,
including gas to stellar mass ratios of order unity. They end up
gas-free, with significantly reduced stellar masses, but also with
final stellar mass~--~total mass ratios close to
unity. Quantitatively, this result should be taken with caution, as
the close pericentre passages make the evolution of the subhaloes also
dependent on the detailed properties of the simulated central
galaxy. In addition, some particles belonging to the outer parts of a
satellite subhalo may be misattributed to the main halo instead, even
though they remain gravitationally bound, and continue to move with
the subhalo. However, qualitatively, this result suggests that strong
tidal stripping {\it decreases}, rather than increases, the total
mass-to-light ratio of satellite galaxies, and is therefore not a
viable way to transform gas-rich, bright dwarf irregulars into the
gas-free, faint dwarf spheroidals with {\it high} mass-light ratios
observed around the Milky Way.

\subsection{Satellite Galaxies with Gas}\label{aquila:massive}
Something can also be learned about gas loss by considering the four
satellites (subhaloes 1, 2, 4 and 7) that still contain gas at
$z=0$. With halo masses between $5\times 10^9$ and $8 \times 10^{10}
\Ms$, these are among the most massive subhaloes; no subhalo less
massive than $5 \times 10^9 \Ms$ still contains gas at $z=0$. None of
these four satellite galaxies had particularly close encounters with
the central galaxy, the minimum distance range from $\sim80$ to
330~kpc. However, this does not clearly separate them from the
gas-free satellites, many of which are on even less bound orbits, or
already fell in gas-free. The present galactocentric distances are
also not significantly different among the two sub-populations.

Interestingly, two of the only three satellite galaxies which are more
massive than $5 \times 10^9 \Ms$ {\it and} gas-free, are hosted in
subhaloes 3, 5, which underwent particularly strong tidal interactions
and were discussed in more detail in
Section~\ref{aquila:extremes}. Thus, it appears that maximal total
masses of a few $\times10^9\Ms$ and orbits which avoid the inner halo
are both required to retain any gas at $z=0$. Equivalently, masses
below a few $\times10^9\Ms$ or very close orbits, are both sufficient
to produce gas-free satellite galaxies. If the majority of dwarf
spheroidals reside in subhaloes with masses below $10^9\Ms$ at present
as well as at infall, cases of orbital metamorphosis are rare.

It is also worth noting that the most massive surviving satellite
(subhalo 1), fell in as late as $z=0.13$, and did so together with
subhalo 7, another gas-rich companion. The two can easily be
identified in Figures~\ref{fig:aquila_dm} to \ref{fig:aquila_stars},
where both are visible only in the last panels. It has been noted
previously that the presence of two satellites as bright as the
Magellanic Clouds near the Milky Way is rather unusual in $\Lambda$CDM
\citep[e.g][]{Boylan-Kolchin-2010}. \cite{Tremaine-1976} showed that
dynamical friction in the halo of the Milky Way would lead to a rapid
decay of the orbits of such large satellites, which would therefore be
short-lived, surviving only a few Gyrs. Proper motions
\citep[e.g.][]{Besla-2007, Piatek-2008} suggest that the Magellanic
Clouds are indeed near their first pericentre after infall.

\begin{figure*}
  \begin{center}
    \begin{tabular}{lll}
       \hspace{-2mm}\includegraphics*[scale= .38]{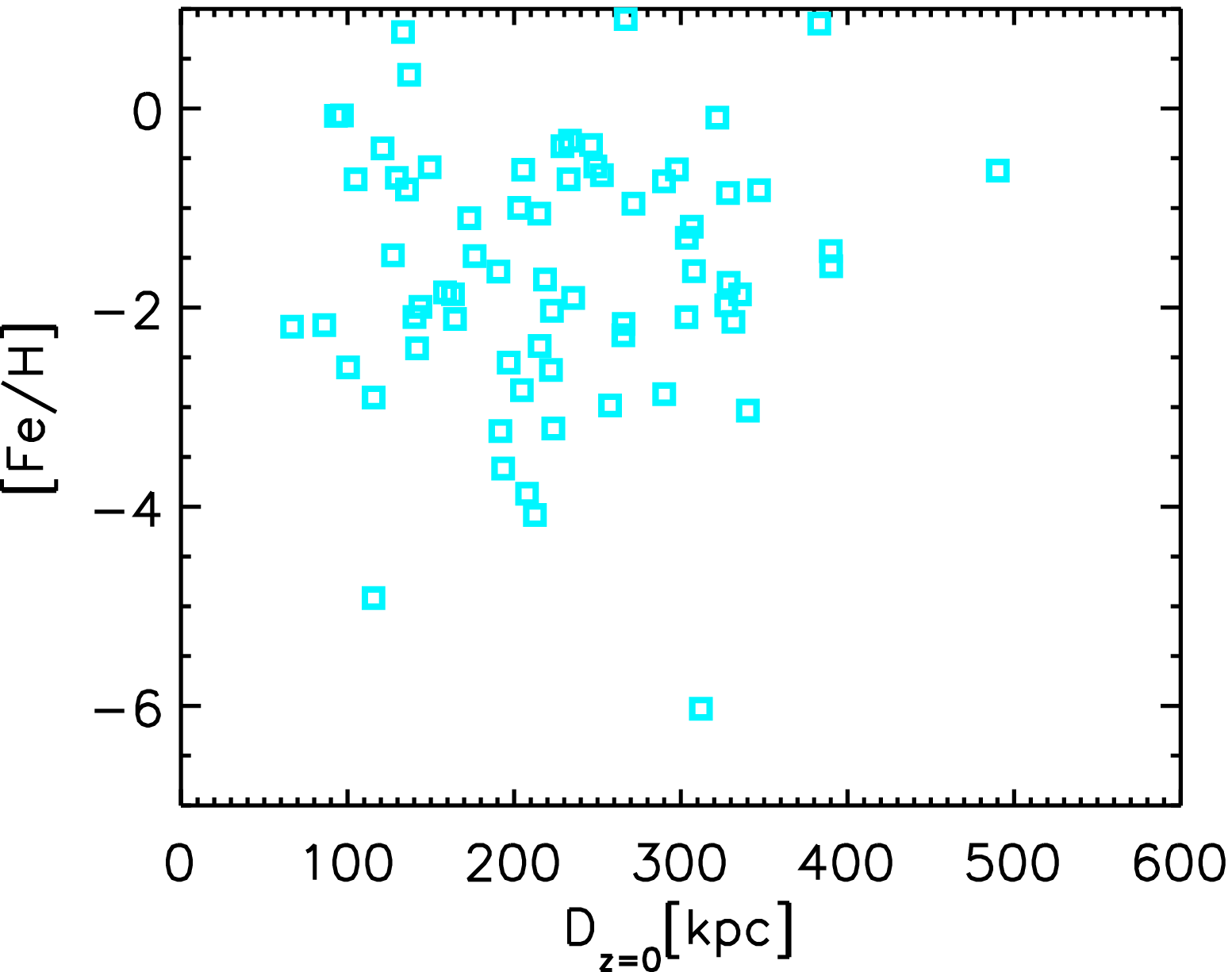} &
       \hspace{-3mm}\includegraphics*[scale= .38]{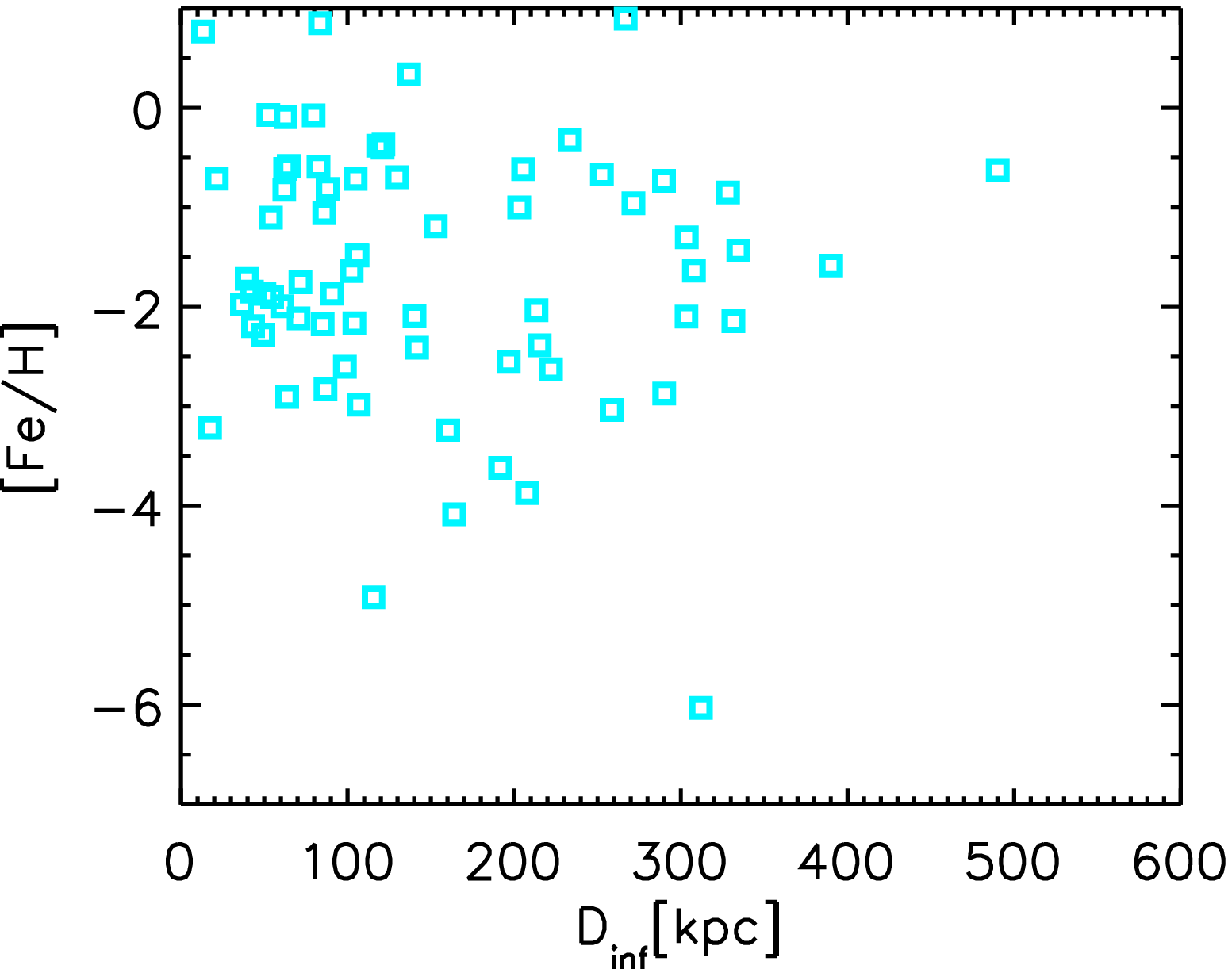} &
       \hspace{-3mm}\includegraphics[scale= .38]{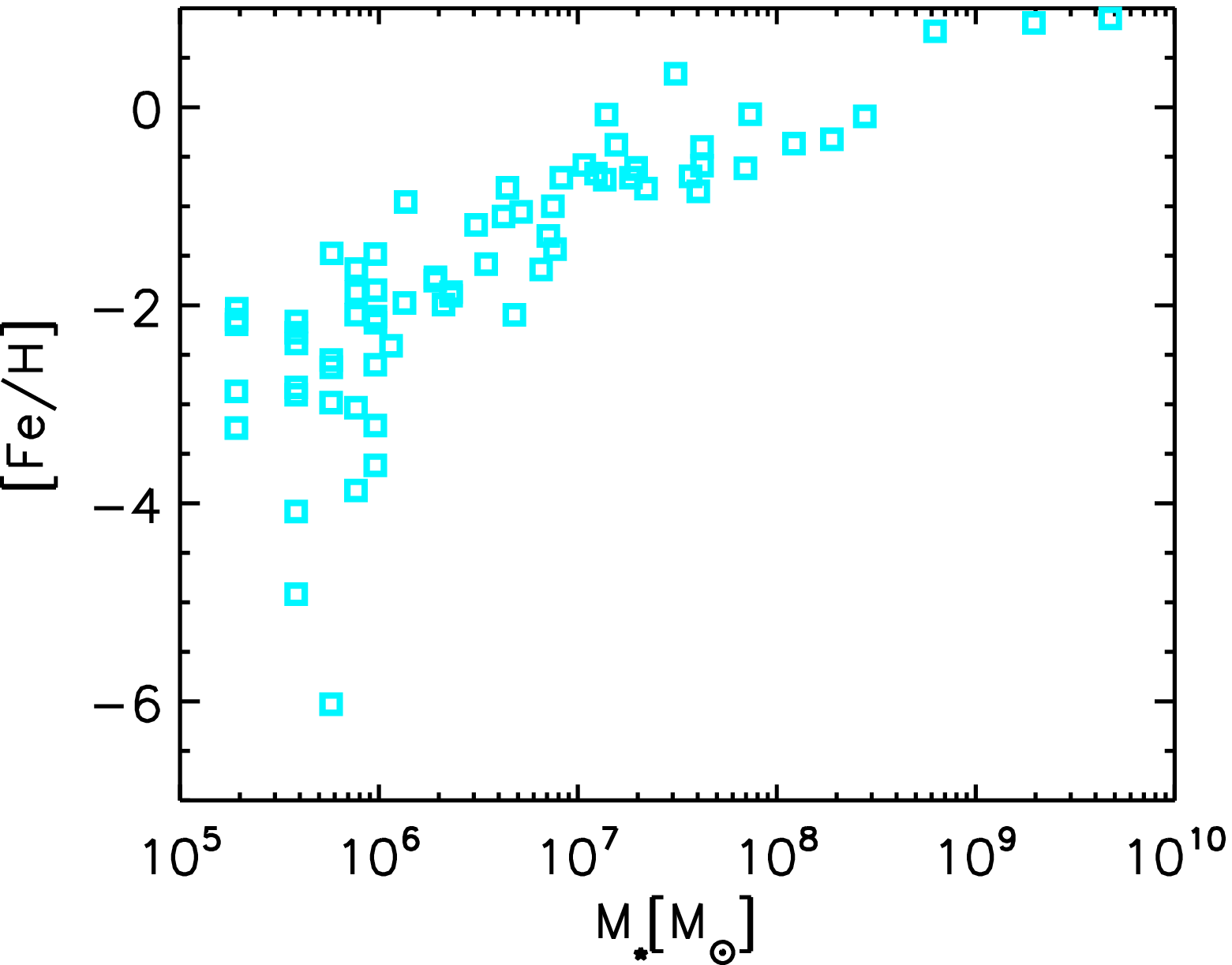} 
    \end{tabular}
  \end{center}
  \caption{Maximum stellar iron abundance within each subhalo, as a
    function of present distance (left), distance of closest approach
    (centre), and stellar mass (right). Note that those satellites
    with only primordial abundances are not included. As is observed
    in the Local Group, there is a clear correlation of metallicity
    with stellar mass, but not with present position nor distance of
    closest approach.
\label{fig:fe-relations}}
\end{figure*}

\section{Scaling Relations}\label{aquila_relations}
At redshift $z = 0$, the halo contains 199 satellite subhaloes, with
masses between $10^{11}$ and $10^{8}\Ms$. Of these, 90 have stars, and
4 also contain gas. The properties of all satellites are listed in
Table~\ref{table:aquila-satellites} of the appendix. They include the
present stellar, gas and dark matter mass, the present galactocentric
distance and the distance of closest approach, the infall redshift,
the mass at infall, the median and maximum stellar metallicity. In
this section, we explore the scaling relations among these
properties. We focus in particular on the way in which the formation
and evolution of the satellite population is linked to the subhalo
mass, and to the influence of the environment.

\subsection{Stellar Mass - Halo Mass}
Figure~\ref{fig:relation_mdm-ms} shows the relationship between
present stellar mass and dark matter mass for all satellites that
contain stars at $z=0$. In the left panel, the dark matter mass is the
current mass of each subhalo, while in the right panel, the dark
matter mass is the mass of the satellite at infall, i.e. when it first
became a subhalo of the host halo (see
Section~\ref{aquila:substructure}). In both cases, there is a clear
correlation of stellar mass and halo mass, indicating that the
processes that determine the amount of star formation per subhalo are
regulated primarily by its mass. For a halo with an infall mass of
$\sim10^9 \Ms$, the corresponding stellar mass is between a few times
$10^5$ to a a few times $10^7 \Ms$. It should be noted that the
minimum stellar mass resolved in the simulations is $2\times10^5 \Ms$.

The two subhaloes discussed in Section~\ref{aquila:extremes}, which
underwent particularly strong tidal stripping, can be identified as
outliers in the relation of stellar mass to present halo
mass. Overall, the scatter is noticeably smaller when the mass at
infall, rather than the present day mass is considered, suggesting
that the evolution of the satellite after infall also plays a role in
some cases. However, it is worth noting that environmental effects
primarily reduce the {\it halo mass}, rather than the {\it stellar
  mass}, contrary to the scenario described in
Section~\ref{introduction}, whereby faint dwarf galaxies are formed
through stripping of baryons.

Figure~\ref{fig:relation_mdm-ms} also includes a comparison with
results from our earlier simulations of isolated dwarf galaxies with
much higher resolution. In both panels, the black stars denote results
from simulations labeled 12--20, with total masses of
$2.3\times10^8$--$10^9\Ms$, presented in \cite{Sawala-2010}, with
stellar particle masses of $5.4\times 10^2$--$2.7 \times 10^3
\Ms$. Blue stars are adopted from \cite{Sawala-2011}, where six haloes
with representative merger histories and a common mass scale of
$\sim10^{10}\Ms$ were re-simulated, with a stellar particle mass
resolution of $9\times10^3\Ms$. We find that, despite the difference
in resolution of up to two orders of magnitude, the results are in
good agreement between the different sets of simulations, particularly
when the dark matter masses are corrected for the effect of stripping,
as shown in the right panel. Because the same code has been used in
all three sets of simulations, it follows that the results are not
strongly affected by resolution.

The two panels in Figure~\ref{fig:ratio-infall-approach} both show the
change in stellar mass -- halo mass ratio of each object from infall
to the present. The ratio at infall is shown on the x-axis, while the
present ratio is shown on the y-axis. Most points lie close to the
black line, which indicates a constant ratio. Notably however, the
majority of haloes are above the line, meaning that their stellar mass
fraction has increased since infall. This can be understood as a
consequence of preferential stripping of dark matter compared to
stellar matter, which is more centrally concentrated and therefore
more strongly bound to the satellite. In the left panel, the
colour-coding is by infall redshift; black and blue symbols indicate
recent accretion, yellow and red symbols indicate infall at high
redshift. In general, satellites that fell in earlier are more likely
to have changed their ratio since infall, as expected if the change is
due to continuous tidal stripping. In the right panel, the
colour-coding is done by distance of closest approach between the
subhalo and the halo of the central galaxy. As expected, haloes that
had closer encounters are also the ones that underwent a slightly
stronger change in the stellar mass to halo mass ratio since
infall. It appears that the haloes with the greatest distance
(D$_{min} > 300$kpc) have seen no change in the ratio, but these are
commonly also subhaloes that have fallen in only recently (z$_{inf} <
0.2$). In both panels, the sizes of the symbols indicate total mass;
larger satellites are typically found with higher stellar
mass~--~total mass ratios, independent of infall time or orbit.
\subsection{Stellar Populations}
Due to the small numbers of stellar particles per subhalo in the
simulation, a detailed analysis of stellar populations is not
possible. As a proxy for star formation history, we consider the
maximum iron abundance [Fe/H] of the stars in each satellite
galaxy. Because iron is formed only in the late stages of stellar
evolution and injected into the interstellar medium via supernovae,
the amount of iron observed in stars corresponds the specific degree
of reprocessing of material within each galaxy, and the intensity and
duration of star formation.

Figure~\ref{fig:fe-relations} shows the maximum stellar iron abundance
of the satellites, as a function of present distance (left), distance
of closest approach (centre), and present-day stellar mass
(right). Note that satellites with only a single generation of stars
have primordial abundances, i.e. $[\mathrm{Fe/H}]\equiv- \infty$, and
therefore do not appear on the plotted relations.

The lack of a correlation on both the left and central panels indicate
that the iron abundance does not depend strongly on either present
distance, or distance of closest approach in the past. By contrast,
there is a strong correlation with stellar mass, as observed in the
Local Group, and also reproduced in our earlier simulations of
isolated dwarf galaxies. At lower stellar mass, the scatter increases,
similar to the trend in the the relation of stellar mass and halo mass
seen in Figure~\ref{fig:relation_mdm-ms}.

\section{Isolated Dwarf galaxies} \label{aquila_outside}

\begin{figure}
  \begin{center}
    \hspace{-.3in} \includegraphics*[scale=.5]{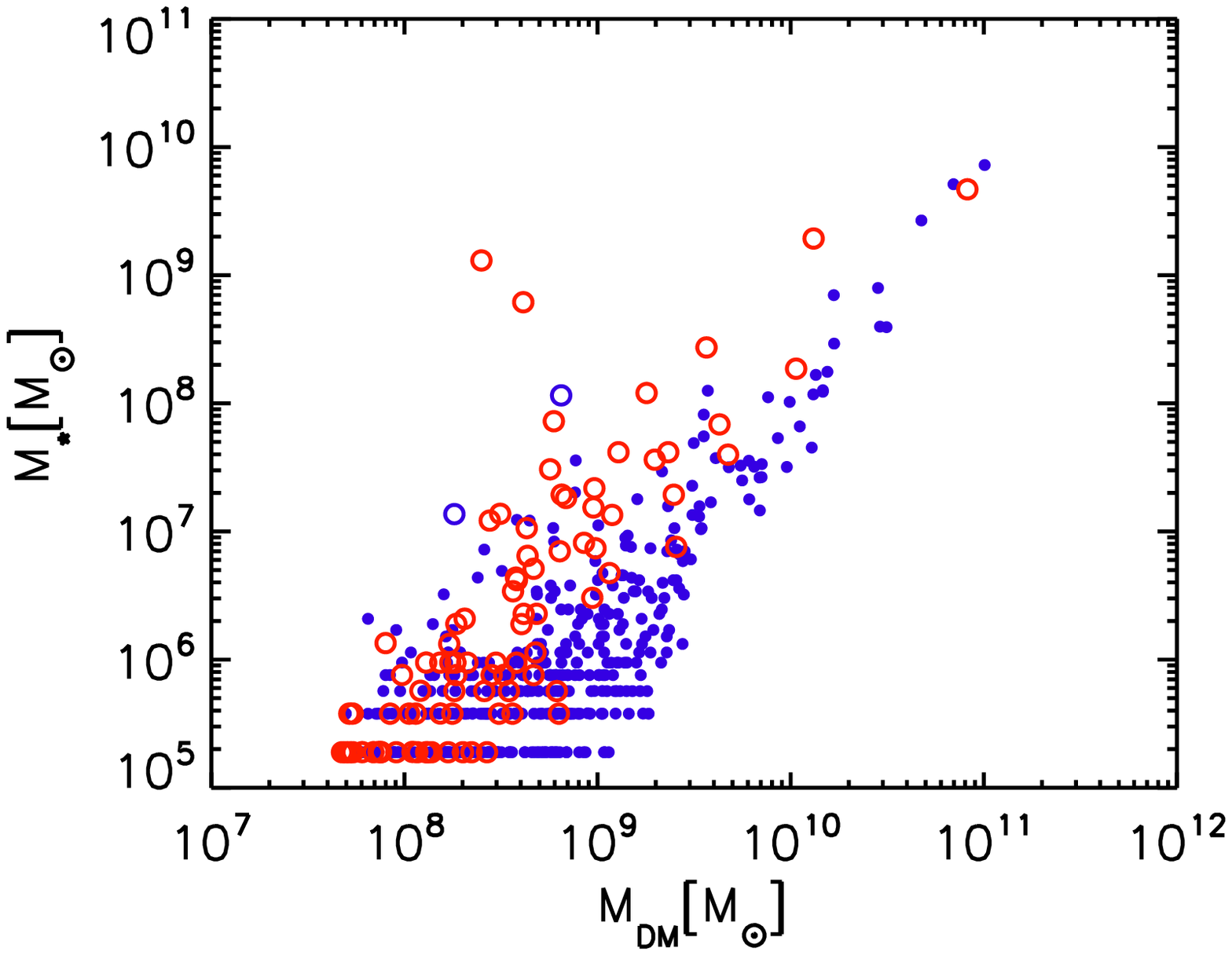} \\
    \hspace{-.3in} \includegraphics*[scale=.5]{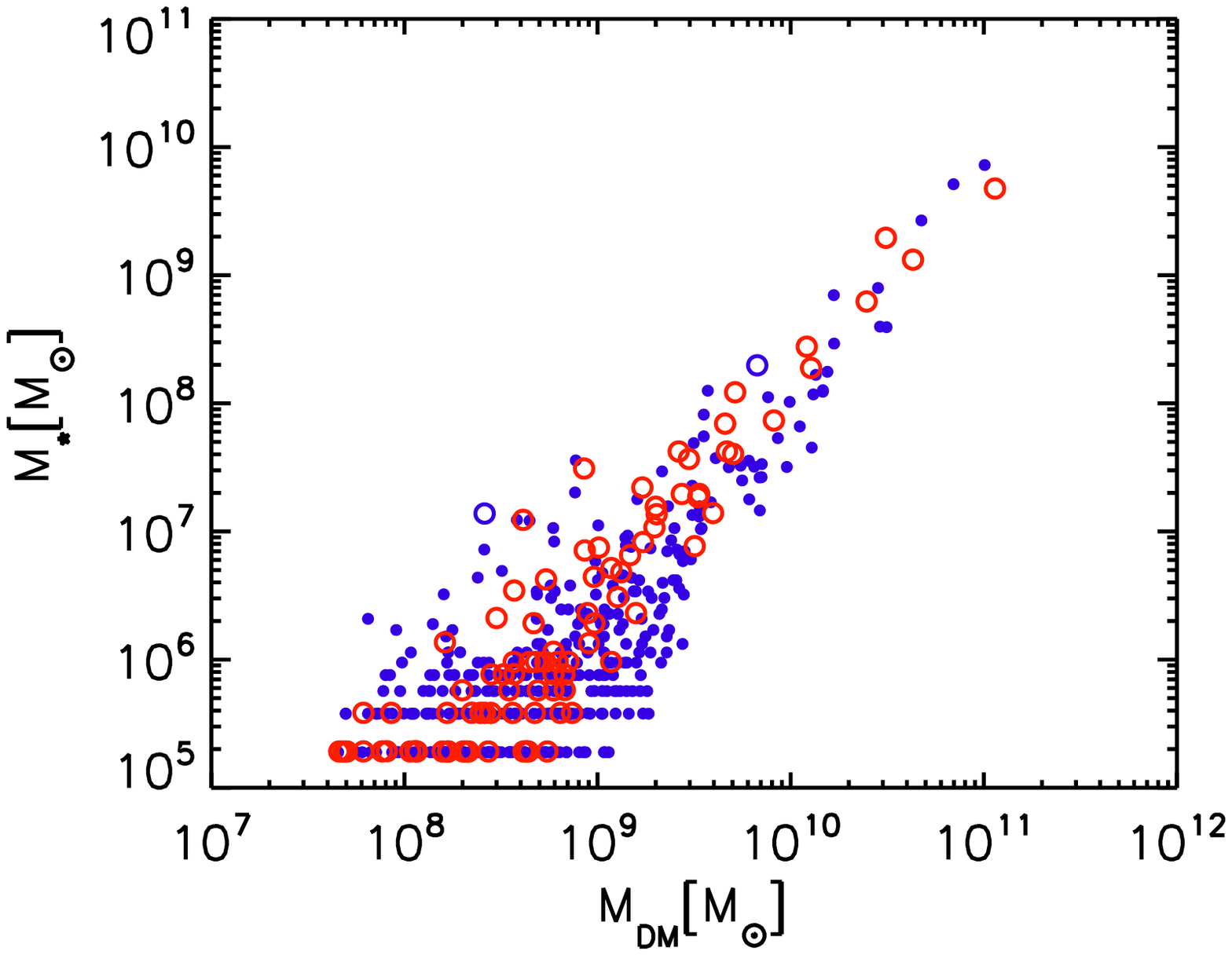}
  \end{center}
  \vspace{-.1in}
  \caption{Stellar mass per subhalo as a function of dark matter
    mass. In both panels, {\it isolated} galaxies are shown as blue
    dots, {\it satellite} galaxies are overplotted as red, open
    circles. The two blue, open circles represent dwarf galaxies
    presently isolated, but which had past interactions, as described
    in the text. In the upper panel, the dark matter mass of both the
    isolated and the satellite galaxies is taken at $z=0$, while the
    infall mass of the satellites and the interacting isolated
    galaxies is used in the lower panel, as in
    Figure~\ref{fig:relation_mdm-ms}. The fraction of subhaloes
    without stars (not shown) is $55\%$ for the satellite subhaloes of
    the Milky Way, and $77\%$ for the isolated
    subhaloes.\label{fig:relation_mdm-ms_outside}}
\end{figure}

As is evident from Figures~\ref{fig:aquila_dm}
through~\ref{fig:aquila_projections}, the high resolution volume also
contains plenty of structures outside the Milky Way halo. In this
section, we compare the population of Milky Way satellites discussed
in the previous section to the isolated dwarf galaxies that form in
the remaining volume. At $z=0$, there are a total of 2097 subhaloes in
the simulation, only $\sim 10\%$ of which are part of the most massive
halo. Not all objects outside of the main group are truly isolated, as
some of the other FoF groups also contain multiple subhaloes (see
Section~\ref{aquila:substructure}). The second largest FoF group in
the simulation hosts a galaxy with a stellar mass of $1.1 \times
10^{10} \Ms$ and contains 68 additional subhaloes. The next most
massive galaxy in the simulation has a stellar mass of $1.7 \times
10^9\Ms$.  Throughout this section, we consider as {\it isolated} all
subhaloes which are not part of the two most massive haloes. Of these
$1810$ subhaloes, 420 host stars, and 144 also contain gas. We compare
these to the {\it satellites} of the Milky Way halo, that were
discussed in the previous sections.

Figure~\ref{fig:relation_mdm-ms_outside} shows the relation of stellar
mass to halo mass for satellite galaxies (in red) and for these
isolated dwarfs (in blue). When the mass ratios are compared at $z=0$
(upper panel), the satellites contain a systematically higher stellar
mass for a given halo mass compared to the isolated galaxies. The
difference is reduced when the infall masses are considered for the
satellites (lower panel). The trend that low-mass, isolated subhaloes
still show a higher mean dark matter mass in this second relation may
be partly due to the fact that they typically grow in mass until
$z=0$, while satellites peak at $z_{inf} > 0$. However, it is also
partly attributable to the identification of substructures, which
requires a higher density if the mean background density is higher.
Interestingly, in the upper panel, the population of isolated dwarf
galaxies also shows two outliers in the $M_\star-M_{DM}$ relation,
which are denoted by blue open circles in both panels. While these two
deviate not as significantly as the satellites which were discussed in
Section~\ref{aquila:extremes}, it points to the fact that even some
dwarf galaxies which are isolated at $z=0$, may have undergone
interactions in the past. The two objects with high stellar to total
mass ratios were never satellites of the main halo, but have
interacted with smaller groups of galaxies, which also lead to tidal
stripping, mostly of dark matter. In the lower panel, we plot their
dark matter masses before their last interactions, which brings them
closer to the relation defined by both populations. It is also worth
noting that the fraction of subhaloes without stars are different
among the two populations: Whereas $45\%$ of the satellite subhaloes
contain stars at $z=0$, only 23$\%$ of the isolated subhaloes contain
stars. This indicates that the lowest mass subhaloes, which are unable
to form stars even in isolation, often do not survive to $z=0$ if they
become satellites.

\begin{figure}
  \begin{center}
       \hspace{-.3in} \includegraphics*[scale=.5]{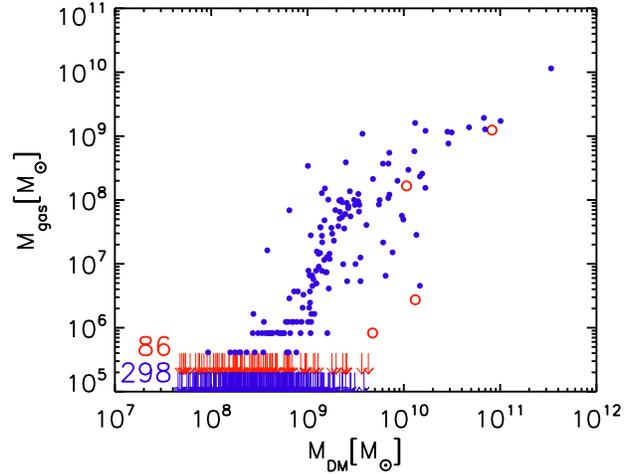}
   \end{center}
  \vspace{-.1in}
  \caption{Gas mass as a function of present subhalo mass for {\it
      isolated} dwarf galaxies (blue) and {\it satellite} dwarf
    galaxies (red). Filled (open) circles indicate isolated
    (satellite) galaxies with $M_{\mathrm{gas}} > 0$; arrows show the
    dark matter masses of galaxies that contain stars, but no gas. The
    red and blue numbers indicate the total number of gas-free
    subhaloes with stars for the two populations. The fraction of
    gas-free galaxies over the mass range shown, is $\sim95\%$ for
    satellites, and $\sim75\%$ for isolated dwarf
    galaxies. \label{fig:mdm-mgas}}
\end{figure}

In Figure~\ref{fig:mdm-mgas}, we compare the gas content of satellites
and isolated dwarf galaxies as a function of halo mass. As described
in Section~\ref{aquila:massive}, there are only four satellite
galaxies with gas at $z=0$, which are found in some of the most
massive subhaloes. Most of the isolated dwarf galaxies at $z=0$ are
also gas-free, but about 1/3rd of the isolated galaxies still contain
some gas at $z=0$. Among the isolated galaxies, there is a sharp drop
in gas mass at subhalo masses of $\sim 2 \times 10^9 \Ms$. More
massive isolated galaxies predominantly contain gas, less massive
galaxies are most often gas-free, and resemble the satellite
population. This points to a mass-threshold below which gas removal is
efficient, and mostly independent of environment, while more massive
galaxies can keep their gas in isolation. Close to this mass
threshold, the populations differ; while some isolated dwarf galaxies
still contain gas, no gas is found in the satellites. It should be
noted, however, that while there are only four satellite galaxies with
gas at $z=0$, the total number of satellite galaxies above the mass
threshold is also very low. The most massive, gas-free galaxies are
found in subhaloes of $\sim 5\times 10^9 \Ms$ in both the satellite
and the isolated populations.

\begin{figure}
  \begin{center}
    \begin{tabular}{l}
      \hspace{-.3in} \includegraphics*[scale = .5]{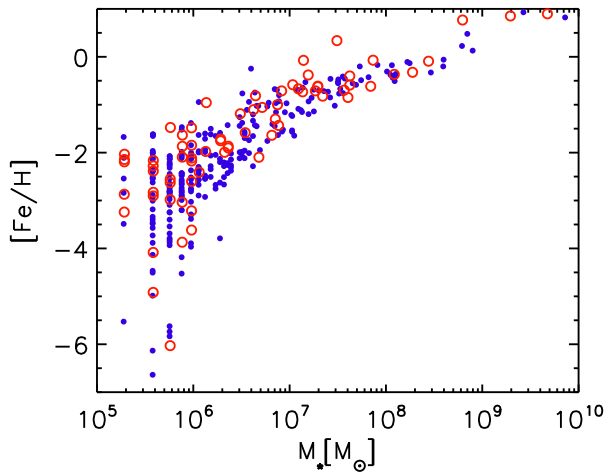} 
    \end{tabular}
  \end{center}
  \vspace{-.1in}
  \caption{Maximum stellar iron abundance within each subhalo for
    isolated subhaloes (blue) and satellite subhaloes (red). There is
    no significant difference either in the correlation of metallicity
    with stellar mass or in the amount of scatter, which in both cases
    increases significantly at the low mass end.
\label{fig:fe-relations_outside}}
\end{figure}

Figure~\ref{fig:fe-relations_outside} compares the maximum iron
enrichment [Fe/H] of stars in satellites and isolated galaxies, and is
similar to the right panel in Figure~\ref{fig:fe-relations}. In both
cases, there is a clear trend of increasing metallicity with stellar
mass, with a large scatter at the low mass end, probably attributable
in part to discreteness effects. At a given mass, the iron enrichment
and the scatter are similar in the two populations, indicating that
mass, rather than environment, is the primary determinant for the star
formation history of dwarf galaxies.

\section{Summary}  \label{aquila_summary}
Before summarizing, it should be emphasized again that some of the
results presented in this paper are scraping the resolution limit of
the {\it Aquila} simulation, which was designed to study much larger
galaxies. We have, however, been able to demonstrate that the results
are consistent with our much higher resolution simulations of
individual dwarf galaxies. The simulation is also not enough to solve
all the questions pertaining to the dwarf galaxy satellites of the
Milky Way; the so-called {\it ultra-faint} dwarfs are clearly outside
the scope of the simulation, and even some of the classical faint
satellites are only resolved with a handful of particles. Tidal
effects, believed to be the main stripping mechanism, should not
depend very strongly on force resolution. However, if ram-pressure
stripping were effective in the Local Group, it would likely be
severely underestimated, because the gas pressure is only poorly
determined, and pressure gradients are artificially smoothed in the
SPH formalism.

While the resolution is limited, it is reassuring that the properties
of these galaxies tend to agree with the higher resolution simulations
of isolated dwarf galaxies, and that the number of $\sim20-30$
dwarf-spheroidal-like galaxies with stellar masses in the range of
$\sim10^6\Ms$ qualitatively agrees with the observational counts. At
the same time, the fact that the simulation was not {\it designed} to
study dwarf galaxies may actually be an advantage: Considering that
dwarf galaxies and Milky Way sized galaxies form in the same universe,
it is only natural that they are simulated with the same realization
of the physical laws and parameters.

We have described the formation of the satellite population in a
cosmological simulation of a Milky Way sized halo and its environment
that includes hydrodynamics, cooling, star formation, supernova
feedback, enrichment and heating by the UV background. Of the 199
subhaloes with masses above $6\times10^7\Ms$, 90 contain stars and
four also contain gas at $z=0$.

We identified different mechanisms of gas-removal, both independent of
environment (supernova feedback and UV heating), and caused by
interactions with the host halo (tidal stripping). It was found that
with a few notable exceptions, the properties of the satellites as a
whole depend only weakly on environment, but very strongly on the mass
of the subhalo. The satellite galaxies that contain stars follow a
steep stellar mass~--~total mass relation, and a stellar
mass--metallicity relation which are similar to those observed, and
indistinguishable from those for isolated dwarf galaxies in the same
simulation. The relations are also similar to those obtained by high
resolution simulations of isolated dwarf galaxies
\cite[e.g.][]{Revaz-2009, Sawala-2010}.

In our simulation, tidal interactions after infall affect the dark
matter haloes of satellites more strongly than their stellar
components. The result is an average {\it increase} in total stellar
mass~--~halo mass ratio, or a corresponding {\it decrease} in total
mass~--~light ratio after infall. This is difficult to reconcile with
the transformation of luminous, late-type galaxies with moderate
mass-to-light ratios into dwarf-spheroidals with the high
mass-to-light ratios inferred from observations, and required from
abundance-matching arguments \cite[e.g.][]{Guo-2010}. Furthermore, the
trend of decreasing mass~--~light ratio is more pronounced for
satellites on closer orbits, and with earlier infall times. This
suggests that, if dwarf spheroidal galaxies of different luminosities
originated from common, gas rich and bright progenitors subject to
different levels of interaction, their final stellar mass after
stripping would scale {\it proportional} to their total mass~--~light
ratio; the opposite of which is commonly inferred from observations.

\cite{Penarrubia-2010} have shown in purely gravitational simulations
that preferential stripping of stars may be possible in a Milky Way
potential if the infalling satellite haloes are cusped, so that the
dark matter is more concentrated than the stars. Unfortunately, our
fully cosmological, hydrodynamical simulations fall short of resolving
the inner parts of satellite subhaloes to distinguish between cusped
and cored profiles by several orders of magnitude, but the softened
potential acts as an {\it effective} core. While the direct
observational evidence for cusps or cores in dwarf spheroidals is
still not clear, it should be noted that cores are not necessarily
unique to ``warm'' dark matter: as shown by \cite{Navarro-1996}, and
recently confirmed in simulations by \cite{Governato-2010}, baryonic
feedback processes may result in cores in low-mass, cold dark matter
haloes. If cores are a universal feature of dwarf galaxies, the
transformation of dwarf irregulars to dwarf spheroidals purely via
tidal effects would be difficult to reconcile with our results.

Instead, we find that satellite galaxies that end up with high total
mass-to-light ratios at $z=0$ are already faint on infall, and many of
them have already lost their gas as a result of supernova feedback and
UV radiation. While tidal interactions can remove remaining
interstellar gas, as we observe in several cases, all these results
suggest that the star formation of dwarf spheroidal galaxies is mostly
determined independent of environment, and very strongly dependent on
mass.

At a total mass of $\sim 1-3\times10^9 \Ms$, the populations of
isolated and satellite dwarf galaxies differ in the fraction of
galaxies with gas. Whereas all satellite galaxies in this mass range
are gas-free, isolated galaxies show a sharp decline in gas fraction,
but many of them still contain gas at $z=0$. Qualitatively, this is in
agreement with the HI mass--distance relation reported by
\cite{Grebel-2003} and \cite{Geha-2006}. In the mass regime of dwarf
spheroidals, however, with stellar masses below $10^7\Ms$ and inferred
dynamical masses below $10^9\Ms$, both the satellites and most of the
isolated dwarf galaxies are gas-free. Consequently, the dwarf
spheroidal galaxies formed in the simulation do not follow a clear
morphology-distance dichotomy. If such a sharp relation exists, the
galaxies that constitute this relationship would not only have to form
in a different way compared to the simulation, but also be on
different orbits, as we find that present day position is not a good
proxy for past interaction.

\section*{Acknowledgements}

All simulations were carried out at the computing centre of the
Max-Planck Society in Garching.

\bibliographystyle{mn2e} \bibliography{aquila}

\label{lastpage}

\newpage \onecolumn

\begin{longtable}{lccccccccc}

\caption[Satellites]{Data for satellite galaxies in the {\it Aquila}
  simulation}\label{table:aquila-satellites} \\

\hline \hline \\[-2ex]

Label &  M$_\star$ &  M$_\mathrm{gas}$  & M$_\mathrm{DM}$ & M$_\mathrm{inf}$ & Z$_\mathrm{inf}$ & Z$_\mathrm{form}$ & D & D$_\mathrm{min}$ & $[$Fe$/$H$]$ \\
      & [$10^6 \Ms$]   & [$10^6 \Ms$]    & [$10^8 \Ms$]  & [$10^8 \Ms$] &                &                 & [kpc] & [kpc]      &           \\  \hline
   \\[-1.8ex]
\endfirsthead

\multicolumn{10}{c}{{\tablename} \thetable{} -- Continued} \\[0.5ex]
  \hline \hline \\[-2ex]

Label &  M$_\star$ &  M$_\mathrm{gas}$  & M$_\mathrm{DM}$ & M$_\mathrm{inf}$ & Z$_\mathrm{inf}$ &  Z$_\mathrm{form}$ & D & D$_\mathrm{min}$ & $[$Fe$/$H$]$ \\
      & [$10^6 \Ms$]   & [$10^6 \Ms$]    & [$10^8 \Ms$]  & [$10^8 \Ms$] &                &                  & [kpc] & [kpc]      &           \\  \hline
  \\[-1.8ex]
\endhead

\multicolumn{10}{l}{{Continued on next page\ldots}} \\
\endfoot

\hline\\ \multicolumn{10}{p{\columnwidth}}{Notes: Col.~2: Stellar
  mass, Col.~3: Gas Mass, Col.~4: Dark matter mass (all at z=0),
  Col.~5 : Dark matter mass (at infall), Col.~6: Infall redshift ,
  Col.~7: Formation redshift, Col.~8: Distance to the centre of
  central galaxy (at z=0), Col.~9: Distance of closest approach,
  Col.~10: Maximum stellar iron abundance.}
\\ \multicolumn{10}{p{\columnwidth}}{Remarks: ** indicates that more
  than half of the stars have primordial abundances.  -- indicates
  that the mass of the component is zero. Z$_\mathrm{form}$* with an
  asterisk indicates the redshift of {\it fragmentation}, as defined
  in Section~\ref{aquila:substructure}. A total of 109 ``dark''
  subhaloes, without baryons, are omitted.}

\endlastfoot

  1 & 4745.17 & 1259.41 &  833.43 & 1142.22 & 0.13 & 9.37* &  266.9 &  266.9 & -1.06 \\
  2 & 1960.70 &    2.78 &  133.21 &  310.85 & 0.53 & 10.91 & 383.1 &   83.5 &-0.94 \\
  3 & 1322.04 &    -- &    2.54 &  429.65 & 2.32 & 12.07 &  13.2 &   13.2 & -0.63 \\
  4 &  188.92 &  168.56 &  108.16 &  128.96 & 0.07 & 5.23* &  233.5 &  233.5 & -1.82 \\
  5 &  623.97 &    -- &    4.18 &  247.15 & 2.10 & 15.47 & 133.3 &   13.2 & -0.72 \\
  6 &  276.84 &    -- &   37.15 &  121.12 & 0.50 & 14.73 & 321.9 &   62.8 & -1.49 \\
  7 &   40.32 &    0.84 &   48.05 &   50.31 & 0.13 & 13.34* &  328.3 &  328.3  & -2.21 \\
  8 &   69.45 &    -- &   43.37 &   45.73 & 0.70 & 4.67* &  205.4 &  205.4 & -1.89 \\
  9 &  122.04 &    -- &   18.19 &   51.53 & 1.07 & 12.07 & 246.1 &  121.5  & -1.84 \\
 10 &   42.12 &    -- &   23.53 &   46.69 & 0.15 &  9.86 & 121.0 &  121.0  & -2.05 \\
 11 &   19.61 &    -- &   25.15 &   27.30 & 0.13 &  9.37 & 490.2 &  490.2 & -2.36 \\
 12 &    7.68 &    -- &   25.90 &   31.83 & 0.38 & 10.38 & 390.0 &  334.5 & -2.74 \\
 13 &   36.82 &    -- &   20.05 &   29.71 & 0.09 & 8.45* &  129.6 &  129.6 & -1.95 \\
 14 &   42.11 &    -- &   13.00 &   26.31 & 2.20 & 13.34 & 149.3 &   82.6& -2.09 \\ 
 15 &   73.51 &    -- &    6.02 &   81.83 & 2.90 & 15.47 &  96.6 &   52.5 & -1.43 \\
 16 &   13.64 &    -- &   12.06 &   20.23 & 0.13 & 16.25 & 289.9 &  289.9 & -1.93 \\
 17 &   22.03 &    -- &    9.76 &   17.05 & 1.36 & 5.83* &  346.9 &   62.0 & -2.21 \\
 18 &    4.80 &    -- &   11.69 &   13.24 & 0.05 & 10.38 & 303.4 &  303.4  & ** \\
 19 &   15.60 &    -- &    9.65 &   20.01 & 0.24 & 14.73 & 228.9 &  118.4  & -2.16 \\
 20 &    7.49 &    -- &    9.89 &   10.16 & 0.13 & 12.07 & 203.0 &  203.0  & -2.90 \\
 21 &    3.08 &    -- &    9.52 &   12.70 & 0.26 &  9.86 & 306.6 &  152.9 & -2.86 \\
 22 &    8.26 &    -- &    8.63 &   17.18 & 1.41 &  7.22 & 104.9 &  104.9  & -2.23 \\
 23 &   30.98 &    -- &    5.76 &    8.52 & 1.60 & 12.69 & 137.0 &  137.0 & -1.40 \\
 24 &   18.51 &    -- &    6.96 &   33.16 & 0.18 & 12.07 & 232.6 &   21.6 & -1.98 \\
 25 &   19.63 &    -- &    6.61 &   33.58 & 0.99 & 12.69 & 297.6 &   62.6 & -1.88 \\
 26 &    7.10 &    -- &    6.46 &    8.58 & 0.13 &  8.90 & 303.6 &  303.6  & -3.05 \\
 27 &    0.38 &    -- &    6.39 &    7.38 & 0.07 &  7.62 & 304.1 &  304.1 & ** \\
 28 &    0.58 &    -- &    6.24 &    4.90 & 0.13 & 11.48 & 312.0 &  312.0 & ** \\
 29 &   10.77 &    -- &    4.36 &   19.70 & 0.92 & 11.48 & 248.7 &   64.7  & -2.32 \\
 30 &    5.19 &    -- &    4.73 &   11.78 & 0.85 & 11.48 & 215.0 &   86.0  & -2.23 \\
 31 &    2.31 &    -- &    4.90 &   15.81 & 2.10 & 10.91 & 163.3 &   90.8 & -2.99 \\
 32 &    6.54 &    -- &    4.40 &   14.73 & 0.55 &  8.02 & 190.6 &  102.5 & -2.85 \\
 33 &    1.15 &    -- &    4.84 &    5.91 & 0.10 &  7.62 & 141.7 &  141.7 & -3.05 \\
 34 &    0.77 &    -- &    4.73 &    6.74 & 0.24 &  9.37 & 340.2 &  258.5  & -4.45 \\
 35 &   13.93 &    -- &    3.17 &   39.60 & 3.09 &  9.37 &  93.1 &   79.6  & -1.09 \\
 36 &    2.31 &    -- &    4.20 &    8.87 & 0.21 &  8.02 & 235.4 &   54.7 & -2.57 \\
 37 &    4.23 &    -- &    3.87 &    5.41 & 0.15 &  0.15 & 173.0 &   54.0 & -2.19 \\ 
 38 &    4.42 &    -- &    3.83 &    9.57 & 0.92 &  8.45 & 135.6 &   88.1 & -1.77 \\
 39 &    1.92 &    -- &    4.09 &    9.63 & 0.33 &  8.90 & 328.7 &   71.7  & -3.05 \\
 41 &   12.32 &    -- &    2.80 &    4.12 & 0.13 & 10.38 & 252.6 &  252.6 & -1.74 \\
 42 &    3.46 &    -- &    3.70 &    3.70 & 0.13 &  8.02 & 390.2 &  390.2  & -3.30 \\
 43 &    0.96 &    -- &    3.85 &    6.22 & 0.17 & 10.91 & 100.3 &   98.4 & -4.82 \\
 45 &    0.38 &    -- &    3.68 &    4.73 & 0.31 &  7.62 & 447.4 &  127.2  & ** \\
 46 &    0.58 &    -- &    3.52 &    5.91 & 0.24 &  8.90 & 257.5 &  106.7 & -3.36 \\
 47 &    0.77 &    -- &    3.33 &    3.70 & 0.18 &  8.02 & 307.9 &  307.9 & -2.96 \\
 48 &    0.38 &    -- &    3.13 &    3.63 & 0.21 &  8.45 & 212.5 &  164.0 & -4.08 \\
 49 &    0.96 &    -- &    3.02 &    3.68 & 0.92 & 12.07 & 193.4 &  191.6 & ** \\
 51 &    0.77 &    -- &    2.87 &    3.26 & 0.65 &  2.74 & 207.7 &  207.7 & -4.07 \\
 52 &    0.19 &    -- &    2.71 &    5.49 & 0.28 &  6.50 & 357.1 &   66.6 & ** \\
 53 &    0.58 &    -- &    2.63 &    3.48 & 0.73 &  7.62 & 222.0 &  222.0 & -2.86 \\
 56 &    2.11 &    -- &    2.08 &    3.00 & 2.10 & 12.07 & 143.7 &   60.8 & -6.99 \\
 57 &    0.19 &    -- &    2.25 &    4.38 & 0.50 &  7.62 & 259.8 &  120.2 & ** \\
 58 &    0.96 &    -- &    2.15 &    5.36 & 1.11 &  7.62 & 176.3 &  106.0 & -2.62 \\
 59 &    1.92 &    -- &    1.88 &    4.66 & 0.53 & 10.38 & 218.5 &   39.5 & -2.36 \\
 60 &    0.19 &    -- &    2.04 &    2.15 & 0.14 &  1.36 & 129.8 &  129.8 & ** \\
 63 &    0.96 &    -- &    1.86 &    4.79 & 1.47 & 11.48 & 164.4 &   70.6 & -4.77 \\
 64 &    0.77 &    -- &    1.88 &    2.82 & 3.29 & 10.38 & 140.2 &  140.2 & -4.07 \\
 65 &    0.58 &    -- &    1.84 &    1.99 & 0.13 &  7.22 & 196.6 &  196.6 & -2.89 \\
 66 &    1.34 &    -- &    1.73 &    9.02 & 1.20 & 8.45* &  327.3 &   36.7 & -2.98 \\
 67 &    0.96 &    -- &    1.75 &    4.44 & 1.67 & 10.38 & 223.5 &   17.5  & ** \\
 68 &    0.38 &    -- &    1.80 &    2.47 & 1.03 & 10.91 & 267.2 &   97.9  & ** \\
 70 &    0.38 &    -- &    1.80 &    2.80 & 1.07 &  8.02 & 235.1 &  172.1  & ** \\
 71 &    0.19 &    -- &    1.71 &    1.69 & 0.13 &  4.42 & 367.7 &  367.7  & ** \\
 73 &    0.96 &    -- &    1.55 &   11.76 & 0.79 & 10.91 & 158.5 &   42.6 & -2.38 \\
 74 &    0.38 &    -- &    1.55 &    2.23 & 1.11 & 1.11* &  265.7 &  104.1 & -2.16 \\
 78 &    0.96 &    -- &    1.31 &    7.07 & 3.71 & 15.47 &  86.0 &   85.1 & -2.58 \\
 79 &    0.19 &    -- &    1.40 &    2.71 & 1.41 & 10.38 & 136.0 &  136.0  & ** \\
 83 &    0.19 &    -- &    1.34 &    1.58 & 0.44 & 0.44* &  207.3 &  207.3 & ** \\
 84 &    0.19 &    -- &    1.34 &    2.01 & 1.47 &  8.02 & 251.0 &  214.5  & ** \\
 86 &    0.19 &    -- &    1.31 &    1.69 & 1.53 &  7.22 & 307.0 &  170.2  & ** \\
 87 &    0.58 &    -- &    1.23 &    6.76 & 2.45 &  7.22 & 127.3 &  105.6 & -2.13 \\
 94 &    0.38 &    -- &    1.16 &    6.41 & 1.30 & 10.38 & 265.5 &   49.5 & -2.28 \\
 95 &    0.19 &    -- &    1.18 &    1.16 & 1.36 & 1.36* &  261.9 &  261.9 & ** \\
 97 &    0.19 &    -- &    1.12 &    1.14 & 0.35 &  0.60 & 222.6 &  213.3 & -2.03 \\
 98 &    0.38 &    -- &    1.07 &    1.66 & 0.18 &  4.17 & 115.7 &   63.7 & -2.90 \\
101 &    0.77 &    -- &    0.99 &    5.91 & 1.67 &  8.45 & 335.5 &   50.1 & -2.06 \\
105 &    1.36 &    -- &    0.81 &    1.62 & 0.13 & 11.48 & 271.6 &  271.6 & -1.68 \\
110 &    0.19 &    -- &    0.92 &    4.14 & 2.58 &  6.85 & 188.1 &  145.8 & ** \\
113 &    0.38 &    -- &    0.85 &    0.85 & 0.28 & 0.28* &  115.6 &  115.6 & -4.92 \\
119 &    0.19 &    -- &    0.77 &    4.27 & 0.79 & 10.91 &  66.7 &   43.3 & -2.19 \\
120 &    0.19 &    -- &    0.77 &    2.08 & 1.67 &  1.82 &  78.8 &   78.8  & ** \\
123 &    0.19 &    -- &    0.74 &    0.81 & 0.13 & 10.91 & 331.6 &  331.6 & -2.14 \\
128 &    0.19 &    -- &    0.70 &    0.77 & 0.42 & 0.42* &  191.7 &  160.4 & -3.24 \\
133 &    0.19 &    -- &    0.61 &    0.61 & 0.13 &  0.28 & 290.0 &  290.0 & -2.87 \\
147 &    0.38 &    -- &    0.55 &    0.61 & 0.14 &  0.42 & 215.3 &  215.3 & -2.39 \\
152 &    0.38 &    -- &    0.53 &    2.61 & 1.25 & 1.25* &  204.6 &   86.8 & -2.83 \\
159 &    0.19 &    -- &    0.55 &    1.07 & 0.29 & 0.29* &  271.5 &   74.8 & ** \\
161 &    0.19 &    -- &    0.53 &    0.50 & 0.04 & 0.04* &  144.7 &  144.7 & ** \\
167 &    0.19 &    -- &    0.50 &    0.48 & 0.05 & 0.05* &  165.1 &  165.1 & ** \\
177 &    0.19 &    -- &    0.48 &    0.46 & 0.13 & 0.33* &  379.8 &  379.8 & ** \\

\hline

\end{longtable}

\end{document}